\documentclass[12pt,a4paper]{article}
\usepackage{epsf}
\usepackage[backref,citecolor=blue]{hyperref}

\begin{document}
%
% Local font definitions
%
% `Blackboard Bold' font
%
\font\twelvebb=msbm10 scaled 1200
\font\tenbb=msbm10
  \newfam\bbfam
  \def\bb{\fam\bbfam\twelvebb}
  \textfont\bbfam=\twelvebb
  \scriptfont\bbfam=\tenbb
  \scriptscriptfont\bbfam=\scriptfont\bbfam
%
% `Euler' font
%
\font\twelveeusm=eusm10 scaled 1200
\font\teneusm=eusm10
  \newfam\eusmfam
  \def\eusm{\fam\eusmfam\twelveeusm}
  \textfont\eusmfam=\twelveeusm
  \scriptfont\eusmfam=\teneusm
  \scriptscriptfont\eusmfam=\scriptfont\eusmfam
%
% `Fraktur' font
%
\font\twelvefk=eufm10 scaled 1200
\font\tenfk=eufm10
  \newfam\fkfam
  \def\fk{\fam\fkfam\twelvefk}
  \textfont\fkfam=\twelvefk
  \scriptfont\fkfam=\tenfk
  \scriptscriptfont\fkfam=\scriptfont\fkfam
%
% Draft stuff
%
\def\draftversion{N}                % Y for draft, N for final version
\def\note[#1]#2{\message{(#1)}\if\draftversion Y{\noindent\em[#2]\/}\fi}
%
% Mark all printed pages for draft version using dvips \specials
%
\if \draftversion Y
\fi
%
% Functions
%
\def\Tr{\mathop{\rm tr}}            % Trace
\def\Det{\mathop{\rm det}}          % Determinant
\def\normedTr{\mathop{\sigma}}      % Normalised trace, (1/N) Tr
\def\Re{\mathop{\rm Re}}            % Real Part
\def\Im{\mathop{\rm Im}}            % Imaginary Part
\def\sgn{\mathop{\rm sgn}}          % Signum
\def\identity{{\bb I}}              % Identity matrix
\def\implies{\Rightarrow}           % =>
\def\J#1{{\mathop{\rm J_{#1}}}}     % Bessel function
\def\erf{\mathop{\rm erf}}          % Error function
\def\erfc{\mathop{\rm erfc}}        % Complementary error function
\def\FT{{\cal F}}                   % Fourier transform
\def\LT{{\cal L}}                   % Laplace transform
%
% Fractions
%
\def\rational#1#2{{\mathchoice{\textstyle{#1\over#2}}%
  {\scriptstyle{#1\over#2}}{\scriptscriptstyle{#1\over#2}}{#1/#2}}}
\def\half{\rational12}                      % One half
\def\third{\rational13}                     % One third
\def\quarter{\rational14}                   % One quarter
%
% Sets
%
\def\N{{\bb N}}                     % Set of natural numbers
\def\Q{{\bb Q}}                     % Set of rational numbers
\def\R{{\bb R}}                     % Set of real numbers
\def\Z{{\bb Z}}                     % Set of integers
\def\C{{\bb C}}                     % Set of complex numbers
\def\S{{\bb S}}                     % Unit circle
%
% Symbols and variable names
%
\def\opt#1{{{#1}_{\hbox{\rm\tiny opt}}}}  % `opt' subscript
\def\expo#1{{{#1}_{\hbox{\rm\tiny exp}}}} % `exp' subscript
\def\acc#1{{{#1}_{\hbox{\rm\tiny acc}}}}  % `acc' subscript
\def\n{N}                           % # MC samples
\def\t{t}                           % MD time
\def\dt{{\delta\tau}}               % MD integration time step
\def\pmomtheta{\vartheta}           % Partial momentum mixing angle
\def\pmomthetaopt{\opt\pmomtheta}   % Optimal noise
\def\trjlen{\tau}                   % Trajectory length
\def\trjlenav{{\bar\trjlen}}        % Average trajectory length
\def\trjlenavopt{\opt\trjlenav}     % Optimal average trajectory length
\def\texp{\expo\t}                  % Exponential autocorrelation time (time)
\def\nexp{\expo\n}                  % Exponential autocorrelation time (steps)
\def\betaexp{\expo\beta}            % Corresponding pole in Laplace transform
\def\dH{{\delta H}}                 % Energy change over trajectory
\def\eps{\varepsilon}               % Generic small step size parameter
\def\H{{\hat H}}                    % Liouvillian, Hamiltonian vector field,...
\def\M{{\eusm M}}                   % Hamiltonian kernel
\def\dM{{\delta\M}}                 % Kernel of \dH
\def\I{\identity}                   % Unit matrix
\def\pacc{\acc P}                   % Acceptance probability
\def\paccav{\acc{\bar P}}           % Average acceptance probability
\def\asymp{\sim}                    % Physicists' asymptopia
\def\update{{\fk A}}                % GHMC single trajectory update
\def\updateav{{\bar\update}}        % LT of single trajectory update
\def\cost{{\fk C}}                  % Cost or computer time
\def\xivv{\varphi}                  % Phase in section 5.5
%
% Local abbreviations and commonly used names
%
\def\hmc{Hybrid Monte Carlo}        % Algorithm's name
\def\lmc{Langevin Monte Carlo}      % Algorithm's name
\def\qcd{{\sc qcd}}                 % Quantum Chromodynamics
\def\lgt{Lattice Gauge Theory}      % LGT
\def\mc{Monte Carlo}                % MC
\def\md{Molecular Dynamics}         % MD
\def\mcmd{Monte Carlo Molecular Dynamics}   % MCMD
\def\cg{Conjugate Gradient}         % CG
\def\steps{T}                       % number of Markov steps
\def\gap{K}                         % number of steps per measurement
%
% Constructions
%
\def\twovector[#1,#2]{\left(\begin{array}{c} #1 \\ #2 \end{array}\right)}
\def\twomatrix[#1,#2;#3,#4]%
  {\left(\begin{array}{cc}
    #1 & #2 \\
    #3 & #4
  \end{array}\right)}
\def\threevector[#1,#2,#3]%
  {\left(\begin{array}{c} #1 \\ #2 \\ #3\end{array}\right)}
\def\sixvector[#1,#2,#3,#4,#5,#6]%
  {\left(\begin{array}{c} #1 \\ #2 \\ #3\\ #4\\ #5\\ #6\end{array}\right)}
\def\threematrix[#1,#2,#3;#4,#5,#6;#7,#8,#9]%
  {\left(\begin{array}{ccc}
    #1 & #2 & #3 \\
    #4 & #5 & #6 \\
    #7 & #8 & #9
  \end{array}\right)}
%
% Algebraic numbers, radicals, and the like
%
\def\cuberoot#1{\root3\of{#1}}      % \cuberoot{x} = x^{1/3}
\def\fifthroot#1{\root5\of{#1}}     % \fifthroot{x} = x^{1/5}
\def\stripsign#1{\if+#1\else#1\fi}  % remove leading + signs
\def\crte(#1,#2,#3){\left(\stripsign#3#2\cuberoot2#1\cuberoot4\right)}
\def\frte(#1,#2,#3,#4,#5){{\left(
  \begin{array}{c}
z    \scriptstyle{\stripsign#5} \\
    \scriptstyle{#4\fifthroot2} \\
    \scriptstyle{#3\fifthroot4} \\
    \scriptstyle{#2\fifthroot8} \\
    \scriptstyle{#1\fifthroot{16}}
  \end{array}
\right)}}
%
% Auxilliary integrals, e.g., \GCC_{p,q}(\beta). Avoid clutter by suppressing
% the argument when it is \(\beta\).
%
\def\notbeta#1{\if\beta#1\else(#1)\fi}
\def\G(#1){G\notbeta#1}
\def\GC_#1(#2){{G^{(C)}_{#1}\notbeta#2}}
\def\GS_#1(#2){{G^{(S)}_{#1}\notbeta#2}}
\def\GCC_#1(#2){{G^{(CC)}_{#1}\notbeta#2}}
%
% LaTeX environments and suchlike
%
\newtheorem{example}{Example}[subsection]
\newtheorem{assumption}{Assumption}[subsection]
\renewcommand{\theenumi}{\roman{enumi}}   % Roman enumeration
%%
%% Titlepage
%%
\title{Cost of the \\
  Generalised Hybrid Monte Carlo Algorithm \\
  for Free Field Theory}

\author{A.~D.~Kennedy\thanks{adk@ph.ed.ac.uk}
  \hskip1.3em and\hskip1.3em
  Brian Pendleton\thanks{bjp@ph.ed.ac.uk}\\[1ex]
  Department of Physics and Astronomy, \\
  The University of Edinburgh, The King's Buildings, \\
  Edinburgh, EH9~3JZ, Scotland}
  
\date{}

\maketitle

\begin{abstract}
  \noindent We study analytically the computational cost of the
  Generalised Hybrid Monte Carlo (GHMC) algorithm for free field theory. We
  calculate the Metropolis acceptance probability for leapfrog and
  higher-order discretisations of the Molecular Dynamics (MD) equations of
  motion. We show how to calculate autocorrelation functions of arbitrary
  polynomial operators, and use these to optimise the GHMC momentum mixing
  angle, the trajectory length, and the integration stepsize for the special
  cases of linear and quadratic operators. We show that long trajectories are
  optimal for GHMC, and that standard HMC is more efficient than algorithms
  based on Second Order Langevin Monte Carlo (L2MC), sometimes known as
  Kramers Equation. We show that contrary to naive expectations HMC and L2MC
  have the same volume dependence, but their dynamical critical exponents are
  \(z = 1\) and \(z = 3/2\) respectively.{\parfillskip=0pt\par}\vskip2ex


  \vskip2ex

\end{abstract}

\newpage
\section{Introduction}
\label{sec:introduction}

Hybrid Monte Carlo (HMC) \cite{duane87a} remains the most popular algorithm
for simulation of Quantum Chromodynamics (QCD) including the dynamical effects
of fermions. It is therefore imperative that we have a detailed understanding
of how the computational cost of HMC varies as we change simulation parameters
such as the lattice volume and the fermion mass. In this paper we present
detailed analytical results for the computational cost of the generalised HMC
algorithm for free field theory. We expect many of the results obtained to
also be applicable to more general field theories where most of the ``modes''
are weakly interacting, for example the high momentum modes of asymptotically
free theories such as QCD.

In this paper we give detailed descriptions of the techniques we have
developed to enable us to calculate fictitious-time evolution operators,
Metropolis acceptance probabilities, and autocorrelation functions for
arbitrary polynomial operators in GHMC simulations of free field theory. This
allows us to minimise the computational cost of such simulations, and to
compare the efficiency of various popular limits of GHMC. This paper brings
together and extends many results that have been presented previously
\cite{kennedy90a,kennedy91b,kennedy91c,kennedy95a,Kennedy:1999di,kennedy99a,%
kennedy00a}. The techniques developed for this paper have also been used in
several other papers \cite{horvath96a,kennedy96c,kennedy97a,Joo:2000dh}.

The structure of this paper is as follows: in Section~\ref{sec:ghmc} we
describe the Generalised Hybrid Monte Carlo (GHMC) Algorithm and discuss
various limiting cases. Section~\ref{free-field-theory} demonstrates that GHMC
computations for free field theory are equivalent to GHMC computations for a
set of uncoupled harmonic operators with judiciously chosen frequencies. In
Section~\ref{leapfrog-evolution} we describe the generalised leapfrog
discretisation schemes for the classical equations of motion in fictitious
time, and introduce specific parameterisations of the evolution operators in
order to facilitate the calculation of the Metropolis acceptance probability
in Section~\ref{sec:acceptance-rates}. In
Section~\ref{sec:autocorrelation-functions} we introduce general techniques
for calculating Laplace transforms of autocorrelation functions of polynomial
operators in free field theory, and present explicit results for arbitrary
linear and quadratic operators for several of the limiting cases of GHMC, both
for fixed- and exponentially-distributed GHMC trajectory lengths.
Section~\ref{sec:costs-for-pacc-equals-1} presents a detailed analysis of the
cost of HMC simulations in the approximation that the Metropolis acceptance
rate is unity, whilst Section~\ref{sec:costs-for-pacc-neq-1} analyses the
relative cost of GHMC, HMC, and the second order Langevin algorithm (L2MC) for
non-trivial acceptance rates under some very weak assumptions.  Finally
Section~\ref{sec:autocorrelations-and-frequency-of-measurement} discusses the
somewhat related topic of how to optimise relative frequencies of updates and
measurements in general Monte Carlo simulations. Some concluding remarks are
contained in Section~\ref{sec:conclusions}. Several detailed technical results
are relegated to Appendices.

\section{Generalised Hybrid Monte Carlo Algorithm}
\label{sec:ghmc}

For simplicity, we shall describe the Generalised Hybrid Monte Carlo (GHMC)
algorithm for a theory of scalar fields, denoted generically by \(\phi\), with
action \(S(\phi)\).

In order to generate the distribution
\begin{displaymath}
  \frac1Z e^{-S(\phi)} \, [d\phi]
\end{displaymath}
we introduce a set of fictitious momenta \(\pi\) and generate the fields
\(\phi\) and \(\pi\) according to the distribution
\begin{equation}
  \frac1{Z'} e^{-H(\phi,\pi)} \, [d\phi][d\pi]
  \label{eq:phase-space-distribution}
\end{equation}
where
\begin{displaymath}
  H(\phi,\pi) = \half \sum_x \pi_x^2 + S(\phi),
\end{displaymath}
and ignore \(\pi\).

We begin by recalling \cite{kennedy86b} that a Markov Process will converge to
some distribution of configurations if it is constructed out of update steps
each of which has the desired distribution as a fixed point, and which taken
together are ergodic. The generalised HMC algorithm
\cite{horowitz90a,kennedy95a,Kennedy:1999di} is constructed out of two such
steps.

\subsection{Molecular Dynamics Monte Carlo}

This in turn consists of three parts:
\begin{enumerate}
\item \emph{Molecular Dynamics (MD):} an approximate integration of Hamilton's
  equations on phase space which is exactly area-preserving and reversible;
  this is a map \(U(\tau): (\phi, \pi) \mapsto (\phi', \pi')\) with
  \(\det\frac{\partial(\phi',\pi')}{\partial(\phi,\pi)}=1\) and \(U(\trjlen) =
  U^{-1}(-\trjlen)\).
\item A \emph{momentum flip} \(F: \pi \mapsto -\pi\).
\item \emph{Monte Carlo (MC):} a Metropolis accept/reject test
  \begin{displaymath}
    (\phi',\pi') = \left\{
      \begin{array}{rl}
        F\cdot U(\phi,\pi)
        & \mbox{with probability \(\min(1,e^{-\delta H})\)} \\
        (\phi,\pi) & \mbox{otherwise.}
      \end{array}
    \right.
  \end{displaymath}
  This may be implemented by generating a uniformly distributed random number
  \(y\in [0,1]\), so that
  \begin{displaymath}
    \int_0^1 dy\, \theta\left(e^{-\delta H} - y\right)
    = \min(1, e^{-\delta H}).
  \end{displaymath}
\end{enumerate}
The composition of these gives the MDMC update step
\begin{equation}
  \twovector[\phi',\pi'] =
    \left[F\cdot U(\trjlen) \, \theta(e^{-\dH} - y)
      + \identity \, \theta(y - e^{-\dH})\right] \twovector[\phi,\pi].
  \label{eq:mdmc-update}
\end{equation}
This satisfies detailed balance because \((F\cdot U)^2 = \identity\).

\subsection{Partial Momentum Refreshment}
\label{partial-momentum-refreshment}

This mixes the Gaussian-distributed momenta \(\pi\) with Gaussian noise
\(\xi\)
\begin{equation}
  \twovector[\pi',\xi'] = \twomatrix[ \cos\pmomtheta, \sin\pmomtheta;
  -\sin\pmomtheta, \cos\pmomtheta] \cdot F \twovector[\pi,\xi].
  \label{eq:momentum-refresh}
\end{equation}
If \(\pi\) and \(\xi\) are Gaussian distributed, \(P_G(\pi)[d\pi] \propto
e^{-\pi^2/2}[d\pi]\), then so is \(P_G(\pi')\):
\begin{displaymath}
  P_G(\pi') = \int_\infty^\infty d\xi \int_\infty^\infty d\pi P_G(\xi)
  P_G(\pi) \delta(\pi'+\pi\cos\pmomtheta -\xi\sin\pmomtheta).
\end{displaymath}
The extra momentum flip \(F\) is included so that the trajectory is reversed
upon an MC rejection instead of on an acceptance.

We may combine the MDMC update of equation~(\ref{eq:mdmc-update}) with the
partial momentum refreshment of equation~(\ref{eq:momentum-refresh}) to obtain
the following expression for their combined effect
\begin{equation}
  \threevector[\phi',\pi',\xi'] = \threematrix[ 1, 0, 0; 0, \cos\pmomtheta,
  \sin\pmomtheta; 0, -\sin\pmomtheta, \cos\pmomtheta] \left[U(\trjlen) \,
    \theta(e^{-\dH} - y) + F \, \theta(y - e^{-\dH})\right]
  \threevector[\phi,\pi,\xi].
  \label{eq:ghmc-update}
\end{equation}

\subsection{Special Cases of GHMC}

Several well-known algorithms are special cases of GHMC:
\begin{itemize}
\item The usual Hybrid Monte Carlo (HMC) algorithm is the special case where
  \(\pmomtheta = \pi/2\), i.e., the fictitious momenta \(p\) are replaced by
  the Gaussian distributed \(\xi\), the old momenta being discarded
  completely.  The momentum flips may be ignored in this case as long as MCMD
  and momentum refreshment steps are strictly alternated.
  
\item \(\pmomtheta = 0\) corresponds to pure MDMC, which is an exact version
  of the MD or microcanonical algorithm. It is in general non-ergodic.
  
\item The second order Langevin Monte Carlo (L2MC) algorithm of Horowitz
  \cite{horowitz90a,beccaria94a} corresponds to choosing arbitrary
  \(\pmomtheta\) but MDMC trajectories of a single leapfrog integration step,
  \(\trjlen=\dt\). This is an exact version of the second order Langevin
  algorithm \cite{klein22,kuti88b,horowitz85,horowitz87}.
  
\item The Langevin Monte Carlo (LMC)~\cite{sugar} algorithm has
  \(\pmomtheta=\pi/2\) and \(\trjlen=\dt\). This is an exact version of the
  Langevin algorithm.
\end{itemize}

\subsection{Tunable Parameters}
\label{sec:tunable-parameters}

The GHMC algorithm has three free parameters, the trajectory length
\(\trjlen\) the momentum mixing angle \(\pmomtheta\), and the integration step
size~\(\dt\). These may be chosen arbitrarily without affecting the validity
of the method, except for some special values for which the algorithm ceases
to be ergodic. We may adjust these parameters to minimise the cost of a Monte
Carlo computation, and the main goal of this work is to carry out this
optimisation procedure for free field theory.

Horowitz~\cite{horowitz90a} pointed out that the L2MC algorithm has the
advantage of having a higher acceptance rate than HMC for a given step size,
but he did not take in to account that it also requires a higher acceptance
rate to get the same autocorrelations because the trajectory is reversed at
each MC rejection. It is not obvious a priori which of these effects
dominates.

The parameters \(\trjlen\) and \(\pmomtheta\) may be chosen independently from
some distributions \(P_R(\trjlen)\) and \(P_M(\pmomtheta)\) for each
trajectory (but of course they cannot be chosen in a way which is correlated
with the starting point in phase space). In the following we shall consider
various choices for the momentum refreshment distribution \(P_M\), but we
shall always take a fixed value for \(\pmomtheta\), \(P_M(\pmomtheta) =
\delta(\pmomtheta-\pmomtheta_0)\). The generalisation of our results to the
case of other distributions of \(\pmomtheta\) values is trivial, but it is not
immediately obvious that it is useful. We choose each trajectory \(\trjlen\)
length independently from some distribution \(P_R(\trjlen)\), as this avoids
the lack of ergodicity caused by choosing a fixed trajectory length which is a
rational multiple of the period of any mode of the system \cite{mackenzie89b}.
This is a disease of free field theory which in interacting models is removed
to some extent by mode coupling.

\section{Free Field Theory}
\label{free-field-theory}

\subsection{Complex Fields on a Finite Lattice}

Consider a \(d\) dimensional free scalar field theory on a \(V\equiv L^d\)
lattice for which the expectation of an arbitrary operator \(\Omega(\phi)\) is
defined by the functional integral
\begin{displaymath}
  \left<\Omega\right> \equiv \frac1Z \int [d\phi]\, e^{-S(\phi)} \Omega(\phi)
\end{displaymath}
where \(Z\) is chosen such that \(\langle1\rangle=1\). For a complex field
\(\phi:\, \Z_L^d\to\C\) the action is
\begin{displaymath}
  S(\phi) = \half \sum_{x\in\Z_L^d}
    \phi^*_x \left(-\partial^2 + m^2\right) \phi_x
\end{displaymath}
where \(\partial^2\) is the lattice Laplacian
\begin{displaymath}
  \partial^2\phi_x \equiv \sum_{\mu=1}^d
    (\phi_{x+\hat\mu} - 2\phi_x + \phi_{x-\hat\mu}).
\end{displaymath}
For the Generalised Hybrid Monte Carlo (GHMC) algorithm discussed in
section~\ref{sec:ghmc} we introduce a set of fictitious momentum fields
\(\pi\), and the corresponding Hamiltonian
\begin{displaymath}
  H(\phi,\pi) \equiv \half \sum_{x\in\Z_L^d} \pi_x^* \pi_x + S(\phi),
\end{displaymath}
and we evaluate the functional integral
\begin{displaymath}
  \langle\Omega\rangle \equiv \frac1{Z'}
    \int [d\phi][d\pi]\, e^{-H(\phi,\pi)} \Omega(\phi).
\end{displaymath}

For theoretical analysis it is useful to diagonalise the Hamiltonian by
Fourier transformation to ``real'' (as opposed to ``fictitious'') momentum
space. From the identity of example~(\ref{ZN-ZN-completeness}) we have
\begin{equation}
  \left.
  \begin{array}{rcccl}
    \tilde{\phi}_p & \equiv & (\FT\phi)_p & \equiv &
      L^{-d/2} \displaystyle \sum_{x\in\Z_L^d}
        e^{-2\pi i p\cdot x/L} \: \phi_x \\[4ex]
    \phi_x  & = & (\FT^{-1}\tilde{\phi})_x & = &
      L^{-d/2} \displaystyle \sum_{p\in\Z_L^d}
        e^{ 2\pi i p\cdot x/L} \: \tilde{\phi}_p
  \end{array}
  \right\}
  \label{eq:complex-finite-fft}
\end{equation}
and similarly for the fictitious momenta. We obtain
\begin{equation}
  H(\phi,\pi)= \tilde{H}(\tilde{\phi},\tilde{\pi})      
    = \half \sum_{p\in\Z_L^d}
      \left(\tilde{\pi}_p^* \tilde{\pi}_p
        + \omega^2_p \tilde{\phi}_p^\ast \tilde{\phi}_p \right),
  \label{eq:hamiltonian-fft}
\end{equation}
where the frequencies are
\begin{equation}
  \omega_p^2 \equiv m^2 + 4 \sum_{\mu=1}^d \sin^2 \frac{\pi p_\mu}{L}.
  \label{eq:frequency-definition}
\end{equation}
The phase space configuration \((\phi,\pi)\) is generated by GHMC with
probability density proportional to
\begin{eqnarray*}
  e^{-H(\phi,\pi)} \: [d\phi][d\pi]
    & = & e^{-\tilde{H}(\tilde{\phi},\tilde{\pi})} \;
      \det\left( \frac{\delta\phi}{\delta\tilde{\phi}} \right)
      \det\left( \frac{\delta\pi} {\delta\tilde{\pi}}  \right)
      [d\tilde{\phi}][d\tilde{\pi}] \\
  & \propto & e^{-\tilde{H}(\tilde{\phi},\tilde{\pi})} \:
      [d\tilde{\phi}][d\tilde{\pi}],
\end{eqnarray*}
where the last relation follows because the Jacobian is a (real) constant. We
thus see that free field theory corresponds to a set of harmonic oscillators
with a specific choice of frequency spectrum.

\subsection{Infinite Lattices}

For an infinite lattice \(\Z^d\) (instead of \(\Z_L^d\)) the corresponding
momentum space is the torus \(S_1^d\) (instead of \(\Z_L^d\)), and
equations~(\ref{eq:complex-finite-fft}), (\ref{eq:hamiltonian-fft}), and
(\ref{eq:frequency-definition}) get replaced by
\begin{displaymath}
  \left.
  \begin{array}{rcccl}
    \tilde{\phi}_p & \equiv & (\FT\phi)_p & \equiv &
      (2\pi)^{-d/2} \displaystyle \sum_{x\in\Z^d}
        e^{-i p\cdot x} \: \phi_x \\[4ex]
    \phi_x  & = & (\FT^{-1}\tilde{\phi})_x & = &
      (2\pi)^{-d/2} \displaystyle \int_{S_1^d} dp\,
        e^{ i p\cdot x} \: \tilde{\phi}_p
  \end{array}
  \right\},
\end{displaymath}
\begin{displaymath}
  H(\phi,\pi) = \tilde{H}(\tilde{\phi},\tilde{\pi})     
    = \half \int_{S_1^d} dp\,
      \left( \tilde{\pi}_p^* \tilde{\pi}_p
        + \omega^2_p \tilde{\phi}_p^* \tilde{\phi}_p \right),
\end{displaymath}
and
\begin{displaymath}
  \omega(p)^2 \equiv m^2 + 4 \sum_{\mu=1}^d \sin^2 \frac{p_\mu}{2}.
\end{displaymath}
where we made use of examples~(\ref{Z-S1-completeness})
and~(\ref{S1-Z-completeness}).

\subsection{Real Fields}

For real (as opposed to complex) fields \(\phi:\,\Z_L^d \to \R\) we define
\begin{eqnarray*}
  \tilde{\phi}_p^{(e)} & \equiv & \Re \tilde{\phi}_p \\
  \tilde{\phi}_p^{(o)} & \equiv & \Im \tilde{\phi}_p
\end{eqnarray*}

%%% In more detail ...
%%%
%%% \begin{displaymath}
%%%   \left.
%%%   \begin{array}{rcl}
%%%     \tilde{\phi}_p^{(e)} &\equiv & L^{-d/2} \displaystyle \sum_{x\in\Z_L^d}
%%%       \cos \frac{2\pi i p\cdot x}{L} \: \phi_x \\[4ex]
%%%     \tilde{\phi}_p^{(o)} &\equiv & L^{-d/2} \displaystyle \sum_{x\in\Z_L^d}
%%%       \sin \frac{2\pi i p\cdot x}{L} \: \phi_x
%%%   \end{array}
%%%   \right\}
%%%   \end{displaymath}

Since \(\tilde{\phi}_{-p}^{\,(e)} = \tilde{\phi}_p^{\,(e)}\) and
\(\tilde{\phi}_{-p}^{\,(o)} = -\tilde{\phi}_p^{\,(o)}\) these two fields are
independent degrees of freedom only on a subset of the momentum space \(p\in
\Z_L^d\). We may choose to define the real momentum field
\(\tilde{\phi}:\,\Z_L^d \to \R\) as
\begin{displaymath}
  \tilde{\phi}_p = \sqrt2 \;
  \left\{
  \begin{array}{lcl}
    \tilde{\phi}_p^{(e)} && \mbox{if \(p\ge-p\)} \\[1ex]
    \tilde{\phi}_p^{(o)} && \mbox{if \(p<-p\)}
  \end{array}
  \right.
\end{displaymath}
where the ordering is arbitrary (\emph{e.g.}, lexicographic). The Hamiltonian
is then
\begin{eqnarray*}
  \tilde{H}(\tilde{\phi},\tilde{\pi})
    & = & \half \sum_{p\in\Z_L^d}
      \left(
        \left| \tilde{\pi}_p^{(e)} + i\tilde{\pi}_p^{(o)} \right|^2
      + \omega_p^2
        \left| \tilde{\phi}_p^{(e)} + i\tilde{\phi}_p^{(o)} \right|^2
      \right) \\
  & = & \half \sum_{p\in\Z_L^d}
    \left(
      \tilde{\pi}_p^{(e)\,2} + \tilde{\pi}_p^{(o)\,2}
      + \omega_p^2 (\tilde{\pi}_p^{(e)\,2} + \tilde{\pi}_p^{(o)\,2})
    \right) \\
  & = & \half \sum_{p\in\Z_L^d}
    \left( \tilde{\pi}_p^2 + \omega_p^2 \tilde{\phi}_p^2 \right)
\end{eqnarray*}
because \(\omega_{-p}^2 = \omega_p^2\).

\subsection{Spectral Averages}

Results obtained for a general set of harmonic oscillators will be expressed
in terms of ``spectral averages,'' which may be evaluated explicitly for free
field theory. We use the notation \(\normedTr(f)\) to denote the spectral
average of some function \(f:S_1^d\to\R\), \(\normedTr(f) \equiv
L^{-d}\sum_{p\in\Z_L^d} f\left(\frac{2\pi p}L\right)\).

The finite lattice results may further be expanded about the infinite lattice
results to obtain a large volume expansion by means of the Poisson resummation
formula~(\ref{poisson-resummation-formula}) of
section~(\ref{subsec-a-poisson-resummation-formula}),
\begin{displaymath}
  \normedTr(f) = \sum_{k\in\Z} \left[\prod_{\mu=1}^d
    \left({1\over2\pi}\right) \int_0^{2\pi} dp_\mu\,\cos(p_\mu kL) \right]
    \; f(p).
\end{displaymath}

\begin{example}
  Consider the massless spectral average in one dimension (where the volume
  \(V\equiv L^d=L\))
\begin{eqnarray*}
  \normedTr\Bigl(\omega(p)^2\Bigr)
  &=& \frac1V \sum_{p\in\Z_V} \omega\left(\frac{2\pi p}V\right)^{2\alpha}
    = \frac1V \sum_{p\in\Z_V} \omega_p^{2\alpha}
    = \frac{4^\alpha}{V} \sum_{p\in\Z_V} \sin^{2\alpha} \frac{\pi p}{V} \\
  &=& \frac{4^\alpha}{2\pi} \int_0^{2\pi} dp \sin^{2\alpha} \frac{p}{2}
    + \sum_{k=1}^\infty \frac{4^\alpha}\pi
      \int_0^{2\pi} dp \sin^{2\alpha} \frac{p}{2} \cos pkV \\
  &=& \frac{4^\alpha}{\pi} B(\half, \alpha+\half)
    = \frac1\pi {2\alpha\choose \alpha},
\end{eqnarray*}
where all the higher order terms vanish for large \(V\) because
\(\int_0^{2\pi} dp \sin^{2\alpha}\frac{p}2\cos pkV=0\) for \(kV>\alpha\) upon
integration by parts. This result is to be compared with the exact answer
given in equation~(\ref{eq:A}) of Appendix~\ref{sec:spectral-sums}.
\end{example}

\section{Leapfrog Evolution}
\label{leapfrog-evolution}

\subsection{Lowest Order Leapfrog}

We wish to consider a system of \(V\) uncoupled harmonic oscillators
\(\{\phi_p\}\) for \(p\in\Z_V\). The \hmc\ algorithm requires us to introduce
a corresponding set of ``fictitious'' momenta \(\{\pi_p\}\), and the dynamics
on this ``fictitious'' phase space is described by the Hamiltonian
\begin{equation}
  H = {1\over2} \sum_{p=1}^V \left( \pi_p^2 + \omega_p^2 \phi_p^2 \right).
\label{hamiltonian}
\end{equation}
The classical trajectory through phase space must obey Hamilton's equations
(using the symplectic 2-form \(\sum_{i=1}^N d\pi_p \wedge d\phi_p\))
\begin{displaymath}
  \dot\phi_p = {\partial H\over\partial\pi_p} = \pi_p, \qquad
  \dot\pi_p = - {\partial H\over\partial\phi_p} = - \omega_p^2 \phi_p.
\end{displaymath}
The leapfrog equations are the simplest discretisation of these which are
exactly reversible and area-preserving,
\begin{eqnarray*}
  \pi_p(\half\dt) =
    & \pi_p(0) + \dot\pi_p(0) \,\half\dt
    &= \pi_p(0) - \omega_p^2 \phi_p(0) \,\half\dt \\
  \phi_p(\dt) =
    & \phi_p(0) + \dot\phi_p(\half\dt) \,\dt
    &= \phi_p(0) + \pi_p(\half\dt) \,\dt \\
  \pi_p(\dt) =
    & \pi_p(\half\dt) + \dot\pi_p(\dt) \,\half\dt
    &= \pi_p(\half\dt) - \omega_p^2 \phi_p(\dt) \,\half\dt.
\end{eqnarray*}
This leapfrog integration scheme is a linear mapping on phase
space\footnote{For the rest of this section we shall consider only a single
  oscillator with \(\omega_p=1\), as everything will be diagonal in~\(p\) and
  the frequency dependence can be recovered by dimensional analysis.}
\begin{displaymath}
  \twovector[\phi(\trjlen+\dt),\pi(\trjlen+\dt)] =
    U_0(\dt) \twovector[\phi(\trjlen),\pi(\trjlen)]
\end{displaymath}
where the matrix
\begin{equation}
  U_0(\dt) \equiv \left(
  \begin{array}{cc}
    1 - \half\dt^2 & \dt \\
    -\dt + \quarter\dt^3 & 1 - \half\dt^2
  \end{array}
  \right)
\label{delta0}
\end{equation}
satisfies \(\Det U_0=1\), as it must because the mapping is area-preserving.
This lowest-order leapfrog integration agrees with the exact Hamiltonian
evolution up to errors of \(O(\dt^3)\),
\begin{equation}
  U_0(\dt) = e^{\H\dt} \left[\identity + U_{0,1}\,\dt^3
    + O(\dt^5)\right]
\label{leapfrog-error}
\end{equation}
where
\begin{equation}
  \H \twovector[\phi,\pi]
  \equiv \left[{\partial H\over\partial\pi}{\partial\over\partial\phi}
    - {\partial H\over\partial\phi}{\partial\over\partial\pi}\right]
    \twovector[\phi,\pi]
  = \left(
  \begin{array}{cc}
    0 & 1 \\ -1 & 0
  \end{array}
  \right) \twovector[\phi,\pi]
\label{hamiltonian-operator}
\end{equation}
is the generator of a translation through fictitious time, and \(U_{0,1}\) is
some operator on phase space. The error must be an odd function of \(\dt\)
because leapfrog integration is reversible to all orders in~\(\dt\).

\subsection{Higher Order Leapfrog}

Following Campostrini \emph{et al.} \cite{campostrini89a,creutz89a} we can
easily construct a higher-order leapfrog integration schemes with errors of
\(O(\dt^5)\) by defining a ``wiggle''
\begin{displaymath}
  U_1(\dt) \equiv
    U_0\left(\dt\over\alpha_1\right)
    U_0\left(-{\sigma_1\dt\over\alpha_1}\right)
    U_0\left(\dt\over\alpha_1\right).
\end{displaymath}
Clearly this is area-preserving and reversible because \(U_0\) is. Using
equation~(\ref{leapfrog-error}), we find that
\begin{displaymath}
  U_1(\dt) = \exp\left({2-\sigma_1\over\alpha_1}\H\,\dt\right)
    \left[\identity +
    U_{0,1}\left(2-\sigma_1^3\right) \left(\dt\over\alpha_1\right)^3 +
    O(\dt^5)\right];
\end{displaymath}
if we choose \(\sigma_1=\root3\of2\) then the coefficient of \(\dt^3\)
vanishes, and we can arrange the step size to equal that of the lowest-order
leapfrog scheme by taking \(\alpha_1=2-\sigma_1\). The explicit form for the
first-order wiggle \(U_1(\dt)\) is
\begin{displaymath}
  \left(
  \begin{array}{cc}
    {1 - {1\over2}\,\dt^2 + {1\over24}\,\dt^4
      + {6 + 5\cuberoot2 + 4\cuberoot4\over288}\,\dt^6} &
    {\dt - {1\over6}\,\dt^3
      - {4 + 4\cuberoot2 + 3\cuberoot4\over144}\,\dt^5} \\
    {- \dt + {1\over6}\,\dt^3
      + {\cuberoot2 + \cuberoot4\over144}\,\dt^5
      - {25 + 20\cuberoot2 + 16\cuberoot4\over1728}\,\dt^7} &
    {1 - {1\over2}\,\dt^2 + {1\over24}\,\dt^4
      + {6 + 5\cuberoot2 + 4\cuberoot4\over288}\,\dt^6}
  \end{array}
  \right).
\end{displaymath}
This construction can be iterated by defining
\begin{equation}
  U_n(\dt) \equiv
    U_{n-1}\left(\dt\over\alpha_n\right)
    U_{n-1}\left(-{\sigma_n\dt\over\alpha_n}\right)
    U_{n-1}\left(\dt\over\alpha_n\right),
\label{general-wiggle}
\end{equation}
which again clearly is area-preserving and reversible. This may be shown to
have errors of the form
\begin{equation}
  U_n(\dt) = e^{\H\dt} \left[\identity + U_{n,1}\dt^{2n+3} +
      O(\dt^{2n+5})\right]
\label{general-error}
\end{equation}
by induction on \(n\): Assume equation~(\ref{general-wiggle}) holds \(\forall
n<N\), then from equations~(\ref{general-wiggle}) and~(\ref{general-error}) we
find that
\begin{displaymath}
  U_N(\dt) = \exp\left({2-\sigma_N\over\alpha_N}\H\,\dt\right)
    \left[\identity + U_{N-1,1}\left(2-\sigma_N^{2N+1}\right)
      \left(\dt\over\alpha_N\right)^{2N+1} + O(\dt^{2N+3})\right],
\end{displaymath}
which gives us equation~(\ref{general-error}) for \(n=N\) upon setting
\(\sigma_N = \root{2N+1}\of2\) and \(\alpha_N = 2-\sigma_N\).

\subsection{Parameterisation of Leapfrog Evolution Operators}

In order to calculate the explicit form for a \md\ trajectory consisting of
\(\trjlen/\dt\) leapfrog steps it is useful to parameterise the leapfrog
matrices \(U_n(\dt)\) in the following way. Reversibility requires that for
any leapfrog matrix \(U_n(-\dt) = U_n(\dt)^{-1}\) and area-preservation
requires that \(\Det U_n(\dt)=1\), hence
\begin{displaymath}
  U_n(-\dt) = \left(
  \begin{array}{cc}
    U_{n_{1,1}}(-\dt) & U_{n_{1,2}}(-\dt) \\
    U_{n_{2,1}}(-\dt) & U_{n_{2,2}}(-\dt)
  \end{array}
  \right) = \left(
  \begin{array}{cc}
    U_{n_{2,2}}(\dt) & -U_{n_{1,2}}(\dt) \\
    -U_{n_{2,1}}(\dt) & U_{n_{1,1}}(\dt)
  \end{array}
  \right).
\end{displaymath}
For the lowest-order leapfrog matrix of equation~(\ref{delta0}) we observe
that the diagonal elements are even functions of \(\dt\) and the off-diagonal
elements are odd functions of \(\dt\); this property also holds for all the
higher-order leapfrog matrices defined by equation~(\ref{general-wiggle}). We
may therefore parameterise \(U_n\) in terms of two even functions
\begin{equation}
  \begin{array}{rcl}
    \kappa_n(\dt) &=& 1 + \kappa_{n,1}\,\dt^{2n+2} + O(\dt^{2n+4}) \\
    \rho_n(\dt) &=& 1 + \rho_{n,1}\,\dt^{2n+2} + O(\dt^{2n+4}).
  \end{array}
  \label{eq:kappa-rho}
\end{equation}
as
\begin{equation}
  U_n(\dt) = \left(
  \begin{array}{cc}
    \cos[\kappa_n(\dt)\,\dt] &
    {\displaystyle\sin[\kappa_n(\dt)\,\dt] \over \displaystyle\rho_n(\dt)} \\
    -\rho_n(\dt) \sin[\kappa_n(\dt)\,\dt] & \cos[\kappa_n(\dt)\,\dt]
  \end{array}
  \right).
\label{eq:u-parameterisation}
\end{equation}
We may easily compute the leading terms in the Taylor expansions of these
functions. For the lowest-order leapfrog scheme we obtain
\begin{eqnarray*}
  \kappa_0 &=& 1 + {1\over24}\,\dt^2 + {3\over640}\,\dt^4 + O(\dt^6)
    = 1 + 0.041\dot6\,\dt^2 + O(\dt^4) \\
  \rho_0 &=& 1 - {1\over8}\,\dt^2 - {1\over128}\,\dt^4 + O(\dt^6)
    = 1 - 0.125\,\dt^2 + O(\dt^4);
\end{eqnarray*}
for the Campostrini ``wiggle''
\begin{eqnarray*}
  \kappa_1 &=& 1 - {\crte(+20,+25,+32)\over1440}\,\dt^4 + O(\dt^6)
    \approx 1 - 0.06614\,\dt^4 + O(\dt^6) \\
  \rho_1 &=& 1 + {\crte(+2,+3,+4)\over288}\,\dt^4 + O(\dt^6)
    \approx 1 + 0.03804\,\dt^4 + O(\dt^6);
\end{eqnarray*}
and for the second-order ``wiggle''
\begin{eqnarray*}
  \kappa_2 &=& 1 + {
    \frte(%
      +{\crte(+457072,+571340,+726418)},%
      +{\crte(+510776,+638470,+811769)},%
      +{\crte(+602896,+753620,+958174)},%
      +{\crte(+721280,+901600,+1146320)},%
      +{\crte(+831376,+1039220,+1321294)})
    \over\scriptstyle881798400}\,\dt^6 + O(\dt^8)
    \approx 1 + 0.02169\,\dt^6 + O(\dt^8) \\
  \rho_2 &=& 1 - {
    \frte(%
      +{\crte(+134352,+168880,+211570)},%
      +{\crte(+153684,+194042,+243521)},%
      +{\crte(+184248,+233308,+293134)},%
      +{\crte(+219456,+277664,+348752)},%
      +{\crte(+247752,+312244,+391582)})
    \over\scriptstyle125971200}\,\dt^6 + O(\dt^8)
    \approx 1 - 0.04573\,\dt^6 + O(\dt^8).
\end{eqnarray*}
Approximate values for \(\kappa_{n,1}\) and \(\rho_{n,1}\) are listed in
Table~\ref{tbl:kappa-rho}. We note in passing that the magnitude of these
leading non-trivial coefficients do not grow rapidly with increasing \(n\).

\begin{table}
  \begin{center}
    \begin{tabular}{|r|r|r|r|}
    \hline
    \(n\) & \multicolumn{1}{c|}{\(\kappa_{n,1}\)} &
      \multicolumn{1}{c|}{\(\rho_{n,1}\)} &
      \multicolumn{1}{c|}{\(\log_{10}\langle\dH_n\rangle/x\)} \\
    \hline
    \(0\) & \( 0.0417\) & \(-0.1250\) & \(-1.0280\) \\
    \(1\) & \(-0.0661\) & \( 0.0380\) & \(-0.9945\) \\
    \(2\) & \( 0.0217\) & \(-0.0457\) & \( 0.2861\) \\
    \(3\) & \(-0.0204\) & \( 0.0038\) & \(-0.7422\) \\
    \(4\) & \(-0.0258\) & \( 0.0118\) & \( 1.4128\) \\
    \(5\) & \( 0.0437\) & \(-0.0483\) & \\
    \(6\) & \(-0.0137\) & \( 0.0371\) & \\
    \(7\) & \(-0.0528\) & \( 0.1178\) & \\
    \(8\) & \( 0.1545\) & \(-0.0813\) & \\
    \hline
    \end{tabular}
  \end{center}
  \caption[kappa-rho]{The coeffcients \(\kappa_{n,1}\) and \(\rho_{n,1}\) of
    equation~(\ref{eq:kappa-rho}) for \(n\) up to 8. The limiting values of
    \(\log_{10}\langle\dH_n\rangle/x\) for \(\trjlen\to\infty\) with \(m=0\)
    are also given for the values of \(n\) appearing in
    Figure~\ref{fig:acc-logdh}.}
  \label{tbl:kappa-rho}
\end{table}

\subsection{Time Evolution Operators}

The parameterisation given in equation (\ref{eq:u-parameterisation})
facilitates the computation of the time evolution operator \(U_n(\trjlen,\dt)
\equiv U_n(\dt)^{\trjlen/\dt}\) for trajectories of length \(\trjlen\) where,
in general, \(\trjlen\gg\dt\).
We assume as an induction hypothesis that
\begin{displaymath}
  U_n(k\dt,\dt) = U_n(\dt)^k = \left(
  \begin{array}{cc}
    \cos[\kappa_n(\dt)\,k\dt] &
    {\displaystyle\sin[\kappa_n(\dt)\,k\dt] \over \displaystyle\rho_n(\dt)} \\
    -\rho_n(\dt) \sin[\kappa_n(\dt)\,k\dt] & \cos[\kappa_n(\dt)\,k\dt]
  \end{array}
  \right);
\end{displaymath}
from this it immediately follows that \(U_n(\dt)^j U_n(\dt)^k =
U_n(\dt)^{j+k}\) using simple trigonometric identities. Expressing the result
in terms of the trajectory length rather than the number of integration steps,
\(\trjlen\equiv k\dt\), we obtain
\begin{equation}
  U_n(\trjlen,\dt) = \left(
  \begin{array}{cc}
    \cos[\kappa_n(\dt)\,\trjlen] &
    {\displaystyle\sin[\kappa_n(\dt)\,\trjlen]\over\displaystyle\rho_n(\dt)} \\
    -\rho_n(\dt) \sin[\kappa_n(\dt)\,\trjlen] & \cos[\kappa_n(\dt)\,\trjlen]
  \end{array}
  \right).
\label{Un}
\end{equation}
This time evolution matrix may be expanded about the exact Hamiltonian
evolution as a Taylor series in~\(\dt\),
\begin{equation}
  U_n(\trjlen,\dt) = e^{\H\trjlen} \left[\I + U_{n,1}(\trjlen)\,\dt^{2n+2}
    + O(\dt^{2n+4})\right],
\label{U-Taylor}
\end{equation}
where from equation~(\ref{hamiltonian-operator})
\begin{equation}
  e^{\H\trjlen} = \left(
  \begin{array}{cc}
    \cos\trjlen & \sin\trjlen \\
    -\sin\trjlen & \cos\trjlen
  \end{array}
  \right),
  \label{eq:exact-evolution}
\end{equation}
and
\begin{displaymath}
  U_{n,1}(\trjlen) = - \rho_{n,1}\sin\trjlen\left(
  \begin{array}{cc}
    - \sin\trjlen & \cos\trjlen \\
    \cos\trjlen & \sin\trjlen
  \end{array}
  \right) + \kappa_{n,1}\trjlen\left(
  \begin{array}{cc}
    0 & 1 \\
    -1 & 0
  \end{array}
  \right).
\end{displaymath}
It is instructive to compare equations~(\ref{general-error})
and~(\ref{U-Taylor}).

\section{Acceptance Rates}
\label{sec:acceptance-rates}

In this section we describe the calculation of the acceptance rate for MDMC
(and thus for GHMC) for a system of \(V\) uncoupled harmonic oscillators
\cite{gausterer89a,gupta90a}. The method of calculation is the same for both
\lmc\ and \hmc, and is independent of whether one uses lowest order leapfrog
or a higher order discretisation scheme; the various algorithms differ only in
the explicit form of the time evolution matrix \(U_n(\trjlen,\dt)\) given in
equation~(\ref{Un}).

The Hamiltonian of equation~(\ref{hamiltonian}) is a quadratic form
\begin{displaymath}
  H \equiv {1\over2} \sum_{p=1}^V
    \twovector[\omega_p\phi_p,\pi_p]^T \twovector[\omega_p\phi_p,\pi_p],
\end{displaymath}
so the change in ``fictitious energy'' over a trajectory is
\begin{displaymath}
  \dH_n \equiv H\Bigl(\phi(\trjlen),\pi(\trjlen)\Bigr) - H\Bigl(\phi,\pi\Bigr)
    = {1\over2} \sum_{p=1}^V
      \twovector[\omega_p\phi_p,\pi_p]^T \dM_n
      \twovector[\omega_p\phi_p,\pi_p],
\end{displaymath}
where \(\dM_n\equiv U_n^TU_n - \I\), and we have abbreviated \(\phi_p(0)\) and
\(\pi_p(0)\) to \(\phi_p\) and \(\pi_p\). Inserting the Taylor expansion of
equation~(\ref{U-Taylor}) and noting that \((e^{\H\trjlen})^T
e^{\H\trjlen}=1\), corresponding to exact energy conservation for \(\dt=0\),
we find that
\begin{eqnarray*}
  \dM_n &\equiv& \dM_{n,1}\,\dt^{2n+2} + \dM_{n,2}\,\dt^{2n+4}
    + O(\dt^{2n+6}) \\
  &=& \left[\I + U_{n,1}\dt^{2n+2} + O(\dt^{2n+4})\right]^T
    \left[\I + U_{n,1}\,\dt^{2n+2} + O(\dt^{2n+4})\right] - \I \\
  &=& \left(U_{n,1}^T+U_{n,1}\right)\,\dt^{2n+2} + O(\dt^{2n+4})
\end{eqnarray*}

The probability of the change in energy \(\dH_n\) having the value \(\xi\)
when averaged over the equilibrium distribution of starting points on phase
space is
\begin{displaymath}
  P_{\dH_n}(\xi) = {1\over Z} \int[d\phi][d\pi] e^{-H} \delta(\xi-\dH_n),
\end{displaymath}
where as usual the ``partition function'' \(Z\) is just the normalisation
constant required to ensure that \(\int d\xi\,P_{\dH_n}(\xi)=1\). It is
convenient to choose an integral representation for the \(\delta\)-function as
a superposition of plane waves, as then
\begin{eqnarray}
  P_{\dH_n}(\xi) &=& {1\over Z} \int[d\phi][d\pi] e^{-H}
      \int_{-\infty+i0}^{\infty+i0} {d\eta\over2\pi} e^{i\eta(\xi-\dH_n)}
    = \int {d\eta\over2\pi} e^{i\eta\xi}
      \,{1\over Z} \int[d\phi][d\pi] e^{-H-i\eta\dH_n} \nonumber \\
  &=& \int {d\eta\over2\pi} e^{i\eta\xi}
    {1\over Z'} \int[d\phi][d\pi] \exp\left[ - {1\over2} \sum_{p,q=1}^V
      \twovector[\omega_p\phi_p,\pi_p]^T \left\{\I + i\eta\dM_n\right\}_{p,q}
        \twovector[\omega_q\phi_q,\pi_q] \right] \nonumber \\
  &=& \int {d\eta\over2\pi}
    {e^{i\eta\xi} \over \sqrt{\Det(\I+i\eta\dM_n)}}
  = \int {d\eta\over2\pi} e^{i\eta\xi} e^{-\half\Tr\ln(\I+i\eta\dM_n)}.
  \label{p-dh-1}
\end{eqnarray}

It will prove useful to observe that the exact area-preservation property of
the time evolution operator ensures that \(\Det U_n=1\), and thus that
\begin{displaymath}
  \exp\Tr\ln(\I+\dM_n) = \Det U_n^TU_n = \Det U_n^T \Det U_n = 1,
\end{displaymath}
and hence \(\Tr\ln(\I+\dM_n)=0\). This implies that the quantity
\(\Tr\ln(\I+i\eta\dM_n)\) must vanish not only for \(i\eta=0\) but also for
\(i\eta=1\). Expanding the logarithm, and making use of this fact, we find
that
\begin{equation}
  \Tr\ln(\I+i\eta\dM_n) = i\eta(1-i\eta)
    {1\over2} \Tr\left(\dM_{n,1}^2\right) \,\dt^{4n+4} + O(\dt^{4n+6}).
  \label{tr-ln-expansion}
\end{equation}

We may perform an asymptotic expansion of the integral over \(\eta\) using
Laplace's method. First we recast equation~(\ref{tr-ln-expansion}) into a form
where the \(V\) dependence is explicit
\begin{equation}
  \Tr\ln(\I+i\eta\dM_n) = i\eta(1-i\eta)
    {1\over2} \normedTr\left(\dM_{n,1}^2\right) x
      + O\left(1,\eta^3\over\root2n+2\of V\right),
  \label{tr-ln-x}
\end{equation}
where we have introduced the variable \(x\equiv V\dt^{4n+4}\), and the
spectral average \(\normedTr(f)={1\over V}\Tr f\), which has a finite limit as
\(V\to\infty\). In equation~(\ref{tr-ln-x}) the correction terms are
negligible if the only important contributions come from the regions where
\(\eta^3\ll\root2n+2\of V\). Laplace's method shows that such contributions
are exponentially suppressed.
Using equation~(\ref{tr-ln-x}) in equation~(\ref{p-dh-1}) we obtain
\begin{displaymath}
  P_{\dH_n}(\xi) \asymp \int {d\eta\over2\pi} e^{i\eta\xi}
    e^{-{1\over4} i\eta(1-i\eta) \normedTr\left(\dM_{n,1}^2\right)x}.
\end{displaymath}
Completing the square gives
\begin{displaymath}
  P_{\dH_n}(\xi) \asymp {1\over\sqrt{\pi\normedTr\left(\dM_{n,1}^2\right)x}}
    \exp\left[-{1\over\normedTr\left(\dM_{n,1}^2\right)x}
      \left\{ \xi - {\normedTr\left(\dM_{n,1}^2\right)x\over4}
      \right\}^2 \right],
\end{displaymath}
and as \(\langle\dH_n\rangle=\int d\xi\,P_{\dH_n}(\xi)\xi={1\over4}
\normedTr\left(\dM_{n,1}^2\right)x\) this may also be written
as~\cite{gupta90a}
\begin{displaymath}
  P_\dH(\xi) \asymp {1\over\sqrt{4\pi\langle\dH\rangle}} \exp\left[ - {(\xi -
      \langle\dH\rangle)^2 \over 4\langle\dH\rangle}\right],
\end{displaymath}
where we have supressed the index~\(n\).

The average Metropolis acceptance rate is now easily found,
\begin{eqnarray}
  \pacc &\equiv& \left\langle\min\left(1,e^{-\dH}\right)\right\rangle
    = \int_{-\infty}^\infty d\xi\,P_\dH(\xi) \min\left(1,e^{-\xi}\right)
                                                \nonumber \\
  &=& \int_{-\infty}^0 d\xi\,P_\dH(\xi)
    + \int_0^\infty d\xi\,P_\dH(\xi)e^{-\xi}    \nonumber \\
  &=& {2\over\sqrt\pi}
    \int_{{1\over2}\sqrt{\langle\dH\rangle}}^\infty d\xi\,e^{-\xi^2}
    = \erfc\left({1\over2}\sqrt{\langle\dH\rangle}\right)
    = \erfc\left(\sqrt{{1\over8}\langle\dH^2\rangle}\right).
    \label{eq:pacc-formula}
\end{eqnarray}
The explicit form for
\begin{eqnarray}
  \langle\dH_n\rangle
    &=& {1\over4} \normedTr\left(\dM_{n,1}^2 \right)x
      = 2x\rho_{n,1}^2
        \normedTr\Bigl((\sin\omega\trjlen)^2 \omega^{4n+4}\Bigr)
                                                        \nonumber \\
    &=& 2x\rho_{n,1}^2\times\frac1V\sum_{p=1}^V
      (\sin\omega_p\trjlen)^2 \omega_p^{4n+4},
      \label{eq:dh-formula}
\end{eqnarray}
where
\begin{displaymath}
  2\rho_{0,1}^2 = {1\over32} = 0.03125
\end{displaymath}
for the lowest-order leapfrog integration scheme, and
\begin{eqnarray*}
  2\rho_{1,1}^2 &=& {\crte(+25,+32,+40)\over41472} \approx 0.002894, \\
  2\rho_{2,1}^2 &=&
    \begin{array}{l}
      {\frte(%
        +{\crte(+79980474344,+100768983952,+126961318646)},%
        +{\crte(+91967883784,+115872242816,+145990198426)},%
        +{\crte(+105742133216,+133226114824,+167855113064)},%
        +{\crte(+121414378108,+152971330676,+192732972517)},%
        +{\crte(+139299694184,+175505620552,+221124374726)})
      \over\scriptstyle793437161472000}
    \end{array}
    \approx 0.004183
\end{eqnarray*}
for higher-order ``wiggles.''

\subsection{One Dimensional Free Field Theory}
\def\1#1{{\mathop{\rm J_{#1}}}(4\trjlen)}

For the case of free field theory we can compute the spectral averages using
the explicit form for the frequency spectrum. Using the methods of
appendix~\ref{sec:spectral-sums} we find that the spectral average
\begin{eqnarray}  %% acc-sigma2.tex
  \bar\sigma^{(2)}_{m^2}
  &\equiv& \frac1V \sum_{p\in\Z_V} (\sin\omega_p\trjlen)^2 \omega_p^4
                                                        \nonumber \\
  &=& \Bigl[3 - 3\1{0} + 4\1{2} - \1{4}\Bigr]           \nonumber \\
  &+& \Bigl[2 - 2\1{0} + 4t\1{1} + \1{2}\Bigr] m^2      \nonumber \\
  &+& \frac12 \Bigl[1 + (2t^2-1)\1{0} + 3t\1{1}\Bigr] m^4 + O(m^6)
  \label{eq:hmc0-sigma}
\end{eqnarray}
up to terms which vanish as \(V\to\infty\). The finite volume corrections are
given explicitly in appendix~\ref{sec:spectral-sums}. For the higher order
integration schemes the corresponding spectral averages are also given in
Appendix~\ref{sec:spectral-sums}.

\begin{figure}
  \begin{center}
    \epsfxsize=0.9\textwidth \leavevmode\epsffile{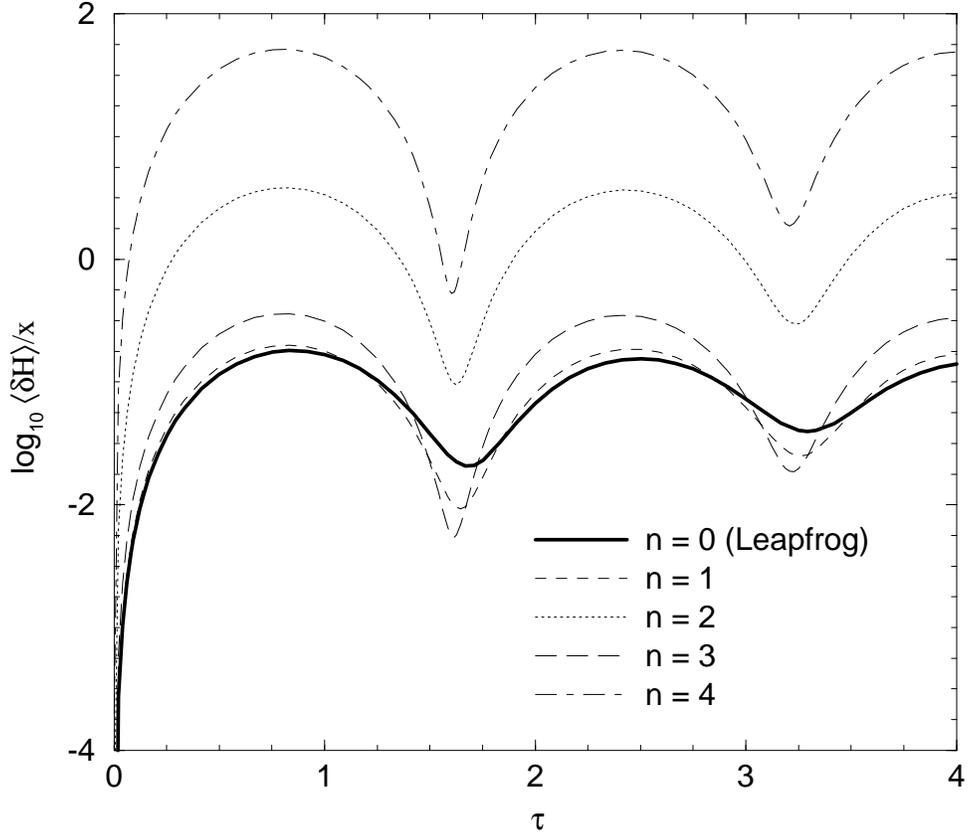}
  \end{center}
  \caption[acc-logdh]{The logarithm of \(\langle\dH_n\rangle/x\), where
    \(x\equiv V\dt^{4n+4}\), for one dimensional free field theory in an
    infinite volume and with a mass \(m=0.01\) is shown for various orders of
    Campostrini wiggles (\(n=0\) corresponds to leapfrog integration). See
    Table~\ref{tbl:kappa-rho} on page~\pageref{tbl:kappa-rho} for the limiting
    values of the curves as \(\trjlen\to\infty\) with \(m=0\).}
  \label{fig:acc-logdh}
\end{figure}

The values for the logarithm of \(\langle\dH\rangle/x\) are shown in
Figure~\ref{fig:acc-logdh}. The limiting values for \(\trjlen\to\infty\) for
the massless case (\(m=0\)) are given in Table~\ref{tbl:kappa-rho}; for
\(m>0\) the corresponding quantities diverge.

\section{Autocorrelation Functions}
\label{sec:autocorrelation-functions}

\subsection{Simple Markov Processes}
\label{sec:simple-markov-processes}

Let \((\phi_0,\phi_2,\ldots,\phi_{\n-1})\) be a sequence of field
configurations generated by an equilibrated ergodic Markov process, and let
\(\Bigl\langle \Omega(\phi)\Bigr\rangle\) denote the expectation value of some
operator \(\Omega\) for \(\phi\) distributed according to the fixed point
distribution of this Markov process. For simplicity we shall assume that
\(\langle\Omega\rangle = 0\) in subsections~\ref{sec:simple-markov-processes}
and~\ref{sec:hybrid-stochastic-processes}. We may define an {\em unbiased
  estimator\/} \(\bar\Omega\) over the finite sequence of configurations by
\begin{displaymath}
  \bar\Omega \equiv {1\over\n} \sum_{j=1}^\n \Omega(\phi_j),
\end{displaymath}
so \(\langle\bar\Omega\rangle = {1\over\n} \sum_{j=1}^\n \Bigl\langle
\Omega(\phi_j)\Bigr\rangle = \langle\Omega\rangle=0\). The variance of this
estimator is
\begin{eqnarray*}
  \left\langle \left(\bar\Omega-\langle\bar\Omega\rangle\right)^2
  \right\rangle &=& \Bigl\langle{\bar\Omega}^2\Bigl\rangle = {1\over\n^2}
  \sum_{j=0}^{\n-1} \sum_{j'=0}^{\n-1}
  \Bigl\langle \Omega(\phi_{j'}) \Omega(\phi_j) \Bigr\rangle \\
  &=& {1\over\n^2} \left\{ \sum_{j=0}^{\n-1} \Bigl\langle \Omega(\phi_j)^2
    \Bigr\rangle + 2\sum_{j=0}^{\n-2} \sum_{j'=j+1}^{\n-1}
    \Bigl\langle \Omega(\phi_{j'}) \Omega(\phi_j) \Bigr\rangle \right\} \\
  &=& {1\over\n} \Bigl\langle \Omega(\phi)^2 \Bigr\rangle \left\{1 +
    {2\over\n}\sum_{\ell=1}^{\n-1} (\n-\ell) \,C_\Omega(\ell)\, \right\},
\end{eqnarray*}
where
\begin{equation}
  C_\Omega(\ell) \equiv {\Bigl\langle \Omega(\phi_\ell)\Omega(\phi_0)
    \Bigr\rangle \over \Bigl\langle \Omega(\phi)^2 \Bigr\rangle}
  \label{autocorrelation-function-definition}
\end{equation}
is the {\em autocorrelation function\/} for~\(\Omega\). If the Markov process
is ergodic, then for large \(\ell\),
\begin{equation}
  \bigl|C_\Omega(\ell)\bigr| \leq \lambda_{\max}^\ell \equiv e^{-\ell/\nexp},
  \label{step-exp-autocorrelation-time}
\end{equation}
where \(\lambda_{\max}\) is the second-largest eigenvalue of the Markov matrix
and \(\nexp\) is the {\em exponential autocorrelation time\/}. If
\(\n\gg\nexp\) then
\begin{eqnarray}
  \langle{\bar\Omega}^2\rangle
  &=& \left\{1 + 2 \sum_{\ell=1}^\infty C_\Omega(\ell) \right\}
  {\Bigl\langle \Omega(\phi)^2 \Bigr\rangle \over\n}
  \left[1 + O\left({\nexp\over\n}\right)\right] \nonumber \\
  &\equiv& \{1 + 2A_\Omega\}
  {\Bigl\langle \Omega(\phi)^2 \Bigr\rangle \over\n}
  \left[1 + O\left({\nexp\over\n}\right)\right],
  \label{eq:correlated-variance}
\end{eqnarray}
where \(A_\Omega\equiv\sum_{\ell=1}^\infty C_\Omega(\ell)\) is the {\em
  integrated autocorrelation function\/} for the operator~\(\Omega\).
 
This result tells us that on average \(1+2A_\Omega\) correlated measurements
are needed to reduce the variance by the same amount as a single truly
independent measurement.

\subsection{Hybrid Stochastic Processes}
\label{sec:hybrid-stochastic-processes}

Suppose that a sequence of configurations \(\bigl(\phi(\t_0), \phi(\t_1),
\phi(\t_2), \ldots, \phi(\t_\n)\bigr)\) is generated from \(\phi(\t_0)\) as
follows: the configuration \(\phi(\t_j)\) is generated from \(\phi(\t_{j-1})\)
by choosing momenta \(\pi(\t_{j-1})\) as described in
Section~(\ref{partial-momentum-refreshment}), and integrating Hamilton's
equations for a time interval \(\trjlen_j \equiv \t_j-\t_{j-1}\), where each
trajectory length \(\trjlen_j\) is chosen randomly from the
distribution~\(P_R(\trjlen_j)\). The autocorrelation function \(C_\Omega\)
defined by equation~(\ref{autocorrelation-function-definition}) may be
expressed in terms of the autocorrelation function
\begin{displaymath}
  C_\Omega(\trjlen_1,\ldots,\trjlen_\ell) \equiv
  {\Bigl\langle\Omega\bigl(\phi(\t_\ell)\bigr)
    \Omega\bigl(\phi(\t_0)\bigr)\Bigr\rangle
    \over \Bigl\langle\Omega^2\Bigl\rangle}
\end{displaymath}
by averaging it over the refresh distribution
\begin{equation}
  C_\Omega(\ell) = \int_0^\infty d\trjlen_1\ldots d\trjlen_\ell\,
  P_R(\trjlen_1)\ldots P_R(\trjlen_\ell)
  \,C_\Omega(\trjlen_1,\ldots,\trjlen_\ell).
  \label{step-autocorrelation}
\end{equation}
The integrated autocorrelation function then becomes
\begin{displaymath}
  A_\Omega = \sum_{\ell=1}^\infty \int_0^\infty d\trjlen_1\ldots
    d\trjlen_\ell\, P_R(\trjlen_1)\ldots P_R(\trjlen_\ell)
    \,C_\Omega(\trjlen_1,\ldots,\trjlen_\ell).
\end{displaymath}

If we wish to determine autocorrelations in terms of the total elapsed
fictitious time \(\t=\t_\ell-\t_0=\sum_{k=1}^\ell \trjlen_k\) of the sequence
of trajectories we may introduce yet another autocorrelation function by
\begin{equation}
  C_\Omega(\t) \equiv \sum_{\ell=1}^\infty \int_0^\infty d\trjlen_1\ldots
    d\trjlen_\ell\, P_R(\trjlen_1)\ldots P_R(\trjlen_\ell)\,
    \delta\Bigl(\t-\sum_{j=0}^\ell \trjlen_j\Bigr)\,
    C_\Omega(\trjlen_1,\ldots,\trjlen_\ell),
  \label{t-autocorrelation}
\end{equation}
and in terms of this we find that
\begin{equation}
  A_\Omega = \int_0^\infty d\t\,C_\Omega(\t).
  \label{t-integrated-autocorrelation-function}
\end{equation}
The function \(C_\Omega(\t)\) has the advantage of giving the autocorrelations
as a function of MD time which is approximately proportional to computer time.

If we make the reasonable assumption that the cost of the computation is
proportional to the total fictitious (MD) time for which we have to integrate
Hamilton's equations, and to the volume \(V\) of the lattice\footnote{For free
  field theory the volume \(V\) is just the number of uncoupled harmonic
  oscillators.}, then the cost \(\cost\) per independent configuration is
proportional to \((1+2A_\Omega) V \trjlenav/\dt\) with \(\trjlenav\) denoting
the average length of a trajectory. The meaning of ``independent
configuration'' was discussed in section~\ref{sec:simple-markov-processes},
and depends on the particular operator \(\Omega\) under consideration. The
optimal trajectory length is obtained by minimising the cost, that is by
choosing \(\trjlenav\) so as to satisfy
\begin{eqnarray}
  {d\cost\over d\trjlenav} = 0 &\implies&
  1 + 2A_\Omega + 2\trjlenav{dA_\Omega\over d\trjlenav} = 0
  \nonumber \\
  {d\cost\over d\pmomtheta} = 0 &\implies& {dA_\Omega\over d\pmomtheta} = 0
  \label{eq:minimise-cost}
\end{eqnarray}

\subsection{Autocorrelation Functions for Polynomial Operators}

In order to carry out these calculations we make a simplifying assumption:
\begin{assumption} \label{enum:dh-correlations}
  The acceptance probability \(\min\bigl(1,e^{-\dH}\bigr)\) for each
  trajectory may be replaced by its value averaged over phase space
  \(\pacc\equiv \left\langle \min\bigl(1, e^{-\dH}\bigr) \right\rangle\); we
  neglect correlations between successive trajectories. Including such
  correlations leads to seemingly intractable complications. It is not obvious
  that our assumption corresponds to any systematic approximation except, of
  course, that it is valid when \(\dH=0\).
\end{assumption}
 
The action of a generalised HMC update on the fields \(\phi\), their conjugate
fictitious momenta \(\pi\), and the Gaussian noise \(\xi\) for the trajectory
under consideration from Equation~(\ref{eq:ghmc-update}) is
\begin{displaymath}
  \threevector[\omega\phi', \pi', \xi'] =
    \update(\trjlen, \pmomtheta, y, \dH) \threevector[\omega\phi, \pi, \xi],
\end{displaymath}
where the matrix \(\update\) depends on the trajectory length \(\trjlen\), the
noise rotation angle \(\pmomtheta\) (which can be chosen independently for
each trajectory if we so wished), the uniform random number \(y\) used in the
Metropolis accept/reject test, and the value of \(\dH\).

We may ignore the corrections of non-leading order in \(\dt\) to the MD
evolution operator because for any given value of \(\langle\pacc\rangle\) of
\(O(1)\) there is a corresponding value of \(x\) which is also \(O(1)\), and
thus \(\dt\) is of order \(V^{-1/(4n+4)}\). These corrections therefore only
contribute to the autocorrelations through the acceptance rate itself at
leading order in the large volume expansion.

From the leading order contribution (\ref{eq:exact-evolution}) we obtain
\begin{displaymath}
  \def\3{\cos\pmomtheta} \def\4{\sin\pmomtheta}
  \def\5{\cos\omega\trjlen} \def\6{\sin\omega\trjlen}
  \update =
  \threematrix[\5, \6, 0; -\6\3, \5\3, \4; \6\4, -\5\4, \3] \theta_{+} +
  \threematrix[1, 0, 0; 0, -\3, \4; 0, \4, \3] \theta_{-}
\end{displaymath}
with \(\theta_{\pm} \equiv \theta\left[\pm\left(e^{-\dH} - y\right)\right]\),
and \(\omega\) being the appropriate frequency for each mode.

\subsubsection{Linear Operators}
\label{sec:linear-operators}

We are interested in calculating the autocorrelation function for a general
linear operator
\begin{equation}
  \Omega=\sum_p\Omega_p\phi_p
  \label{eq:linear-op-definition}
\end{equation}
for a set of uncoupled harmonic oscillators \(\{\phi_p\}\). Such an operator
is of course connected, meaning that \(\langle\Omega\rangle =
\sum_p\Omega_p\langle\phi_p\rangle = 0\). Its autocorrelation function may be
expressed in terms of the autocorrelation functions for the individual
harmonic oscillators themselves, for
\begin{displaymath}
  C_\Omega = \frac{\langle\Omega(\t)\Omega(0)\rangle}{\langle\Omega^2\rangle}
  = \sum_{p,p'} \Omega_p\Omega_{p'}
  \frac{\langle\phi_p(\t)\phi_{p'}(0)\rangle}{\langle\Omega^2\rangle},
\end{displaymath}
and as the oscillators are uncoupled
\begin{displaymath}
  \Bigl\langle\phi_p(\t)\phi_{p'}(0)\Bigr\rangle
    = \delta_{p,p'} \Bigl\langle\phi_p(\t)\phi_p(0)\Bigr\rangle
    + (1-\delta_{p,p'}) \langle\phi_p\rangle \langle\phi_{p'}\rangle
     = \delta_{p,p'} \Bigl\langle\phi_p(\t)\phi_p(0)\Bigr\rangle,
\end{displaymath}
thus
\begin{displaymath}
  C_\Omega(\t) = \frac{\sum_p \Omega_p^2 C_{\phi_p}(\t)
    \langle\phi_p^2\rangle} {\sum_p \Omega_p^2 \langle\phi_p^2\rangle},
\end{displaymath}
where
\begin{displaymath}
  C_{\phi_p}(\t)
    = \frac{\langle\phi_p(\t)\phi_p(0)\rangle}{\langle\phi_p^2\rangle}.
\end{displaymath}
Let us proceed to calculate these single mode autocorrelation functions; while
doing so we can drop the subscript \(p\) for notational simplicity.

The average over the Gaussian distribution of \(\xi\) gives
\(\langle\xi\rangle = 0\), so we can drop the last column of \(\update\), and
since a new \(\xi\) is taken from a heatbath at the start of each trajectory
we may drop the last row also. This leaves us with the basis \(\Phi^{(1)}
\equiv (\omega\phi, \pi)\), which is updated by \({\Phi^{(1)}}' =
\update^{(1)} \Phi^{(1)}\) where the matrix
\begin{equation}
  \def\1{\theta_{+}}
  \def\2{\theta_{-}}
  \def\3{\cos\pmomtheta} \def\4{\sin\pmomtheta}
  \def\5{\cos\omega\trjlen} \def\6{\sin\omega\trjlen}
  \update^{(1)}
    \equiv \twomatrix[\5, \6; -\6\3, \5\3] \1 + \twomatrix[1, 0; 0, -\3] \2.
    \label{eq:linear-update-matrix}
\end{equation}

\subsubsection{Quadratic Operators}
\label{sec:quadratic-operators}

Let \(\Omega\) be a generic quadratic operator for the set of uncoupled
harmonic oscillators \(\{\phi_p\}\),
\begin{equation}
  \Omega \equiv \sum_{p\ge q} \Omega_{pq} \phi_p \phi_q
  \label{eq:quadratic-op-definition}
\end{equation}
whose connected part is \(\Omega_c = \Omega - \left<\Omega\right>\). The
autocorrelation function for \(\Omega_c\) is\footnote{The integrated
  autocorrelation function for a disconnected operator diverges in general.}
\begin{displaymath}
  C_{\Omega_c}(\t) =
    \frac{\left<\Omega_c(\t)\Omega_c(0)\right>}{\left<\Omega_c^2\right>}
\end{displaymath}
If we define the elementary autocorrelation functions for linear and quadratic
modes to be \(C_{\phi_p}(\t)\) and \(C_{{(\phi^2_p)}_c}(\t)\), then we may
express \(C_{\Omega_c}(\t)\) in terms of them:
\begin{displaymath}
  C_{\Omega_c}(\t) = \frac1{\left<\Omega_c^2\right>}
    \left\{\sum_p \Omega^2_{pp} C_{{(\phi^2_p)}_c}(\t)
      \left\langle\left({(\phi^2_p)}_c\right)^2\right\rangle +
      2 \sum_{p>q} \Omega^2_{pq} C_{\phi_p}(\t) C_{\phi_q}(\t)
      \left\langle{(\phi^2_p)}_c\right> \left<{(\phi^2_q)}_c\right\rangle
    \right\}
\end{displaymath}
Higher degree polynomial operators may be treated in the same way. As before
we calculate the single mode autocorrelation functions and again drop the
subscript \(p\) while doing so.

We cannot directly average the GHMC update matrix \(\update\) over \(\xi\) as
we did in the linear case, but we can linearise the problem by considering a
homogeneous quadratic operator as a linear combination of the quadratic
monomials \(((\omega\phi)^2, \pi\omega\phi, \pi^2, \xi\omega\phi, \xi\pi,
\xi^2)\). These quadratic monomials are updated by the symmetrised tensor
product of the update matrix~\(\update\),
\begin{displaymath}
  \left(
  \begin{array}{c}
    (\omega\phi')^2 \\ \pi'\omega\phi' \\ \pi'^2 \\ \xi'\omega\phi' \\
      \xi'\pi' \\ \xi'^2
  \end{array}
  \right) = \update\otimes_s\update \left(
  \begin{array}{c}
    (\omega\phi)^2 \\ \pi\omega\phi \\ \pi^2 \\ \xi\omega\phi \\
      \xi\pi \\ \xi^2
  \end{array}
  \right),
\end{displaymath}
where the update matrix \(\update\otimes_s\update\) is explicitly
\begin{displaymath}
  \def\1{\scriptstyle\phi} \def\2{\scriptstyle\pi} \def\3{\scriptstyle\xi}
  \def\4{\scriptstyle\update} \def\5{\scriptstyle2} \left(
   \begin{array}{ccc|cc|c}
     \4_{\1\1}^2 & \5\4_{\1\2} \4_{\1\1} & \4_{\1\2}^2 &
     \5\4_{\1\3} \4_{\1\1} & \5\4_{\1\3} \4_{\1\2} & \4_{\1\3}^2 \\
     \4_{\2\1} \4_{\1\1} & \4_{\2\2} \4_{\1\1} + \4_{\2\1} \4_{\1\2} &
     \4_{\2\2} \4_{\1\2} & \4_{\2\3} \4_{\1\1} + \4_{\2\1} \4_{\1\3} &
     \4_{\2\3} \4_{\1\2} + \4_{\2\2} \4_{\1\3} & \4_{\2\3} \4_{\1\3} \\
     \4_{\2\1}^2 & \5\4_{\2\2} \4_{\2\1} & \4_{\2\2}^2 & \5\4_{\2\3} \4_{\2\1}
     &
     \5\4_{\2\3} \4_{\2\2} & \4_{\2\3}^2 \\ [0.5ex]
     \hline \4_{\3\1} \4_{\1\1} & \4_{\3\2} \4_{\1\1} + \4_{\3\1} \4_{\1\2} &
     \4_{\3\2} \4_{\1\2} & \4_{\3\3} \4_{\1\1} + \4_{\3\1} \4_{\1\3} &
     \4_{\3\3} \4_{\1\2} + \4_{\3\2} \4_{\1\3} & \4_{\3\3} \4_{\1\3} \\
     \4_{\3\1} \4_{\2\1} & \4_{\3\2} \4_{\2\1} + \4_{\3\1} \4_{\2\2} &
     \4_{\3\2} \4_{\2\2} & \4_{\3\3} \4_{\2\1} + \4_{\3\1} \4_{\2\3} &
     \4_{\3\3} \4_{\2\2} + \4_{\3\2} \4_{\2\3} & \4_{\3\3} \4_{\2\3} \\ [0.5ex]
     \hline \4_{\3\1}^2 & \5\4_{\3\2} \4_{\3\1} & \4_{\3\2}^2 & \5\4_{\3\3}
     \4_{\3\1} & \5\4_{\3\3} \4_{\3\2} & \4_{\3\3}^2
   \end{array}
 \right).
\end{displaymath}
As the update is now linear we can average over the Gaussian distribution of
\(\xi\) as before. The basis monomials become \(((\omega\phi)^2,
\pi\omega\phi, \pi^2, 0, 0, \langle\xi^2\rangle=1)\), and this leads to three
simplifications:
\begin{itemize}
\item The fourth and fifth columns are multiplied by zero, and can thus be
  dropped.
  
\item The fourth and fifth rows are not of interest and may be dropped too, as
  \(\xi\) is chosen from a heathbath before the next trajectory, and we
  already know how the linear monomials \((\phi, \pi)\) are updated.
  
\item The last row may be replaced by \((0,0,0,0,0,1)\) as we know that
  \(\xi'\) is Gaussian distributed and thus \(\langle\xi'^2\rangle=1\).
\end{itemize}
We are thus led to consider the update of the inhomogenous monomials of even
degree in \(\omega\phi\) and \(\pi\) alone, namely \(\Phi^{(2)} \equiv
((\omega\phi)^2, \pi\omega\phi, \pi^2, 1)\), which is updated by
\({\Phi^{(2)}}' = \update^{(2)} \Phi^{(2)}\) where
\begin{eqnarray}
  \update^{(2)} &\equiv& \left(
    \def\1{\phi} \def\2{\pi} \def\3{\xi} \def\4{\update}
    \begin{array}{cccc}
      \4_{\1\1}^2 & 2 \4_{\1\2} \4_{\1\1} & \4_{\1\2}^2 & \4_{\1\3}^2 \\
      \4_{\2\1} \4_{\1\1} & \4_{\2\2} \4_{\1\1} + \4_{\2\1} \4_{\1\2} &
      \4_{\2\2} \4_{\1\2} & \4_{\2\3} \4_{\1\3} \\
      \4_{\2\1}^2 & 2 \4_{\2\2} \4_{\2\1} & \4_{\2\2}^2 & \4_{\2\3}^2 \\
      0 & 0 & 0 & 1
    \end{array}
  \right) \nonumber \\
  &=& 
  \def\1{\cos\pmomtheta} \def\2{\sin\pmomtheta}
  \def\3{\cos^2\pmomtheta} \def\4{\sin^2\pmomtheta}
  \def\5{\cos^2\omega\trjlen} \def\6{\cos\omega\trjlen\sin\omega\trjlen}
  \def\7{\sin^2\omega\trjlen}
  \left(
    \begin{array}{cccc}
      \5 & 2\6 & \7 & 0 \\
      -\6\1 & (\5-\7)\1 & \6\1 & 0 \\
      \7\3 & -2\6\3 & \5\3 & \4 \\
      0 & 0 & 0 & 1
    \end{array}
  \right) \theta_{+} \nonumber \\
  && \qquad + \left(
    \def\1{\cos\pmomtheta} \def\3{\cos^2\pmomtheta} \def\4{\sin^2\pmomtheta}
    \begin{array}{cccc}
      1 & 0 & 0 & 0 \\
      0 & -\1 & 0 & 0 \\
      0 & 0 & \3 & \4 \\
      0 & 0 & 0 & 1
    \end{array}
  \right) \theta_{-}.
  \label{eq:quadratic-update-matrix}
\end{eqnarray}

\subsection{Laplace Transforms of Autocorrelation Functions}

We can extract more information about the autocorrelation function
\(C_\Omega(t)\) than just the integrated autocorrelation function
\(A_\Omega\).  We shall discuss this further in
Section~\ref{sec:laplace-meaning}. In order to do this it is convenient to
compute the Laplace transform of the autocorrelation function
\begin{eqnarray}
  F_\Omega(\beta) &\equiv& \LT C_\Omega(\t)
  \equiv \int_0^\infty dt\, C_\Omega(\t) e^{-\beta\t} \label{eq:laplace-def} \\
  &=& \sum_{\ell=1}^\infty \int_0^\infty d\trjlen_1 \ldots d\trjlen_\ell \,
  P_R(\trjlen_1)\ldots P_R(\trjlen_\ell) \,
  C_\Omega(\trjlen_1,\ldots,\trjlen_\ell) e^{-\beta \trjlen_1}\ldots e^{-\beta
    \trjlen_\ell}. \nonumber
\end{eqnarray}

We may generalise the results of sections \ref{sec:linear-operators} and
\ref{sec:quadratic-operators} and observe that the update step for the vector
of inhomogeneous monomials in \(\omega\phi\) and \(\pi\), \(\Phi^{(d)}\), is
of the form\footnote{After averaging over the distributions of \(\pmomtheta\)
  and \(y\) which are chosen independently for each trajectory.}
\begin{displaymath}
  \Phi^{(d)}(\trjlen) = \update^{(d)}(\trjlen) \Phi^{(d)}(0)
\end{displaymath}
The matrix \(\update^{(d)}\) is block upper triangular with with first block
corresponding to the \(d+1\) homogeneous monomials of degree \(d\), the next
block to the \(d-1\) homogeneous monomials of degree \(d-2\), and so forth.
The dimension of the matrix, which is the number of even or odd inhomogeneous
monomials of degree \(d\) or less, is \(\quarter(d+2)^2\) for \(d\) even and
\(\quarter(d+3)(d+1)\) for \(d\) odd.

The connected operator \((\phi^d)_c \equiv \phi^d - \langle\phi^d\rangle =
v^T\cdot\Phi^{(d)}\) where \(v = (1/\omega^d,0,\ldots,0,
\langle\phi^d\rangle)\), and the \(\ell\)-trajectory autocorrelation function
is thus
\begin{displaymath}
  C_{(\phi^d)_c}(\trjlen_1,\ldots,\trjlen_\ell)
  = \frac{\left\langle v^T \cdot \Bigl[\update^{(d)}(\trjlen_1)) \ldots
      \update^{(d)}(\trjlen_\ell)\Phi^{(d)}(0)\Bigr]
      \; v^T \cdot\Phi^{(d)}(0)\right\rangle}
  {\Bigl\langle\bigl(v^T\cdot\Phi^{(d)}\bigr)^2\Bigr\rangle}.
\end{displaymath}
By virtue of assumption~\ref{enum:dh-correlations} we may replace each matrix
\(\update^{(d)}\) by its phase space average,
\begin{displaymath}
  C_{(\phi^d)_c}(\trjlen_1,\ldots,\trjlen_\ell)
  = \frac{v^T \cdot
    \prod_{j=1}^\ell \bigl\langle\update^{(d)}(\trjlen_j)\bigr\rangle
    \cdot \Bigl\langle\Phi^{(d)}
    \cdot \left.\Phi^{(d)}\right.^T\Bigr\rangle \cdot v}
  {v^T \cdot \Bigl\langle\Phi^{(d)}
    \cdot \left.\Phi^{(d)}\right.^T\Bigr\rangle \cdot v},
\end{displaymath}
and the Laplace transform of the degree \(d\) single mode autocorrelation
function is therefore
\begin{displaymath}
  F_{(\phi^d)_c}(\beta) = \LT C_{(\phi^d)_c}(\t)
  = \frac{v^T \cdot
    \Bigl[\sum_{\ell=1}^\infty \left.\updateav^{(d)}\right.^\ell\Bigr]
    \cdot \Bigl\langle\Phi^{(d)}
    \cdot \left.\Phi^{(d)}\right.^T\Bigr\rangle \cdot v}
  {v^T \cdot \Bigl\langle\Phi^{(d)}
    \cdot \left.\Phi^{(d)}\right.^T\Bigr\rangle \cdot v},
\end{displaymath}
where
\begin{equation}
  \updateav^{(d)} \equiv \int_0^\infty d\trjlen\, P_R(\trjlen)
  e^{-\beta\trjlen} \Bigl\langle\update^{(d)}(\trjlen)\Bigr\rangle.
  \label{eq:updateav-def}
\end{equation}
Summing the geometric series we obtain a simple expression for the Laplace
transformed autocorrelation function,
\begin{equation}
  1 + F_{(\phi^d)_c}(\beta)
  = \frac{v^T \cdot
    \left[\identity - \updateav^{(d)}\right]^{-1}
    \cdot \Bigl\langle\Phi^{(d)}
    \cdot \left.\Phi^{(d)}\right.^T\Bigr\rangle \cdot v}
  {v^T \cdot \Bigl\langle\Phi^{(d)}
    \cdot \left.\Phi^{(d)}\right.^T\Bigr\rangle \cdot v},
  \label{eq:ltac-matrix}
\end{equation}

In order to evaluate the expectation values, we need the Gaussian averages for
the equilibrium probability distribution of
equation~(\ref{eq:phase-space-distribution}) for the Hamiltonian given in
equation~(\ref{hamiltonian}):
\begin{displaymath}
  \left.
    \begin{array}{rcl}
      \left\langle \phi_p^{2k+1} \right\rangle =
      & \left\langle \pi_p^{2k+1} \right\rangle & = 0 \\ [1.5ex]
      \left\langle (\omega_p\phi_p)^{2k} \right\rangle =
      & \left\langle \pi_p^{2k} \right\rangle
      & = {1\over\sqrt\pi} 2^k \Gamma(k+\half)
    \end{array}
  \right\}
\end{displaymath}
and
\begin{displaymath}
  \Bigl\langle
  \phi_p^\alpha \phi_q^\beta \pi_r^\gamma\pi_s^\delta
  \Bigr\rangle =
  \Bigl\langle \phi_p^\alpha \Bigr\rangle
  \Bigl\langle \phi_q^\beta \Bigr\rangle
  \Bigl\langle \pi_r^\gamma \Bigr\rangle
  \Bigl\langle \pi_s^\delta \Bigr\rangle
\end{displaymath}
when \(p\neq q\) and \(r\neq s\).

\subsubsection{Linear Operators}

For linear operators the matrix of expectation values in
equation~(\ref{eq:ltac-matrix}) is
\begin{displaymath}
  \Bigl\langle\Phi^{(1)} \left.\Phi^{(1)}\right.^T\Bigr\rangle = \left(
    \begin{array}{cc}
      1 & 0 \\ 0 & 1
    \end{array}
  \right).
\end{displaymath}

The explicit form of the Laplace transformed update matrix is obtained from
equation~(\ref{eq:linear-update-matrix}) by first averaging over phase space,
which replaces \(\theta_{+}\) by \(\pacc(\omega\trjlen)\) and \(\theta_{-}\)
by \(1-\pacc(\omega\trjlen)\), and then equation~(\ref{eq:updateav-def}) gives
\begin{equation}
  \updateav^{(1)} = \left(
    \def\1{\cos\pmomtheta}
    \def\2#1#2#3{G^#1_{#2,#3}}
    \begin{array}{cc}
      \2+10+\2-00 & \2+01 \\
      -\2+01\1 & \left(\2+10-\2-00\right)\1
    \end{array}
  \right)
  \label{eq:updateav-linear}
\end{equation}
where
\begin{displaymath}
  G^{+}_{j,k} \equiv \int_0^\infty d\trjlen\, P_R(\trjlen)
  e^{-\beta\trjlen} \pacc(\trjlen) (\cos\omega\trjlen)^j (\sin\omega\trjlen)^k
\end{displaymath}
and
\begin{displaymath}
  G^{-}_{j,k} \equiv \int_0^\infty d\trjlen\, P_R(\trjlen)
  e^{-\beta\trjlen} \bigl(1-\pacc(\trjlen)\bigr)
  (\cos\omega\trjlen)^j (\sin\omega\trjlen)^k.
\end{displaymath}

The Laplace transform of the linear single mode autocorrelation function is
\begin{equation}
  F_\phi(\beta) = \frac
  {G^{+}_{1,0} + G^{-}_{0,0}
    - \left({G^{+}_{1,0}}^2 - {G^{-}_{0,0}}^2 + {G^{+}_{0,1}}^2\right)
    \cos\pmomtheta}
  {1 - G^{+}_{1,0} - G^{-}_{0,0}
    + \left(- G^{+}_{1,0} + G^{-}_{0,0} + {G^{+}_{1,0}}^2 - {G^{-}_{0,0}}^2
      + {G^{+}_{0,1}}^2\right) \cos\pmomtheta},
  \label{eq:F-linear}
\end{equation}
which for the special case of \(\pmomtheta=\pi/2\) (HMC) simplifies to
\begin{displaymath}
  F_\phi(\beta; \pmomtheta=\pi/2) =
  \frac{G^{+}_{1,0} + G^{-}_{0,0}}{1 - G^{+}_{1,0} - G^{-}_{0,0}}.
\end{displaymath}

\subsubsection{Quadratic Operators}

For quadratic operators the matrix of expectation values is
\begin{displaymath}
  \Bigl\langle\Phi^{(2)} \left.\Phi^{(2)}\right.^T\Bigr\rangle = \left(
    \begin{array}{cccc}
      3 & 0 & 1 & 1 \\
      0 & 1 & 0 & 0 \\
      1 & 0 & 3 & 1 \\
      1 & 0 & 1 & 1
    \end{array}
  \right);
\end{displaymath}
and from equations~(\ref{eq:quadratic-update-matrix})
and~(\ref{eq:updateav-def})
\begin{displaymath}
  \def\1{\cos\pmomtheta}
  \def\2{\cos^2\pmomtheta}
  \def\3{\sin^2\pmomtheta}
  \def\4#1#2#3{G^#1_{#2,#3}}
  \def\5#1#2{G_{#1,#2}}
  \updateav^{(2)} = \left(
    \begin{array}{cccc}
      \4+20 + \4-00 & 2\4+11 & \4+02 & 0 \\
      -\4+11\1 & \left(\4+20-\4+02-\4-00\right)\1 & \4+11\1 & 0 \\
      \4+02\2 & -2\4+11\2 & \left(\4+20+\4-00\right)\2 & \500\3 \\
      0 & 0 & 0 & \500
    \end{array}
  \right),
\end{displaymath}
where
\begin{equation}
  G_{j,k} \equiv G^{+}_{j,k} + G^{-}_{j,k}
  = \int_0^\infty d\trjlen\, P_R(\trjlen)
  e^{-\beta\trjlen} (\cos\omega\trjlen)^j (\sin\omega\trjlen)^k.
  \label{eq:g-def}
\end{equation}

The Laplace transform of the quadratic single mode autocorrelation function
\(F_{(\phi^2)_c}(\beta)\) is
\begin{equation}
  \frac
  {\left[
      \begin{array}{c}
        G^{+}_{2,0} + G^{-}_{0,0} + \Bigl( - {G^{+}_{2,0}}^2 -
        2{G^{+}_{1,1}}^2 + G^{+}_{0,2}G^{-}_{0,0} +
        G^{+}_{0,2}G^{+}_{2,0} + {G^{-}_{0,0}}^2\Bigr) \cos\pmomtheta \\
        - \Bigl(G^{+}_{2,0} - G^{+}_{0,2} + G^{-}_{0,0}\Bigr)
        \Bigl(G^{-}_{0,0} + G^{+}_{2,0} + G^{+}_{0,2}\Bigr) (\cos\pmomtheta)^2
        \\
        + \Bigl(G^{-}_{0,0} + G^{+}_{2,0} + G^{+}_{{0,2}}\Bigr)
        \Bigl({G^{+}_{2,0}}^2 - 2G^{+}_{0,2}G^{+}_{2,0} + {G^{+}_{0,2}}^2 +
        4{G^{+}_{1,1}}^2 - {G^{-}_{0,0}}^2\Bigr) (\cos\pmomtheta)^3
      \end{array}
    \right]}
  {\left[
      \begin{array}{c}
        1 - G^{-}_{0,0} - G^{+}_{2,0} \\
        - \bigl(G^{+}_{0,2}G^{-}_{0,0} - 2{G^{+}_{1,1}}^2 + {G^{-}_{0,0}}^2 +
        G^{+}_{0,2}G^{+}_{2,0} - G^{-}_{0,0} + G^{+}_{2,0} - {G^{+}_{2,0}}^2
        - G^{+}_{0,2} \bigr) \cos\pmomtheta \\
        - \bigl( - {G^{+}_{2,0}}^2 + G^{-}_{0,0} - 2G^{+}_{2,0}G^{-}_{0,0} +
        {G^{+}_{0,2}}^2 - {G^{-}_{0,0}}^2 + G^{+}_{2,0}\bigr)
        (\cos\pmomtheta)^2 \\
        + \left(
          \begin{array}{c}
            2{G^{+}_{1,1}}^2 - {G^{+}_{0,2}}^3 - 4{G^{+}_{1,1}}^2G^{-}_{0,0}
            + G^{+}_{0,2}{G^{+}_{2,0}}^2 - G^{+}_{0,2}G^{+}_{2,0} \\
            + {G^{+}_{0,2}}^2G^{+}_{2,0} - G^{+}_{0,2}G^{-}_{0,0}
            + G^{+}_{2,0}{G^{-}_{0,0}}^2 - {G^{+}_{2,0}}^3 + {G^{+}_{2,0}}^2 \\
            - {G^{-}_{0,0}}^2 - 4G^{+}_{2,0}{G^{+}_{1,1}}^2
            - {G^{+}_{2,0}}^2G^{-}_{0,0} - 4{G^{+}_{1,1}}^2G^{+}_{0,2} \\
            + G^{+}_{0,2}{G^{-}_{0,0}}^2 - {G^{+}_{0,2}}^2G^{-}_{0,0} +
            {G^{-}_{0,0}}^3 + 2G^{+}_{0,2}G^{+}_{2,0}G^{-}_{0,0}
          \end{array}
        \right) (\cos\pmomtheta)^3
      \end{array}
    \right]},
  \label{eq:F-quadratic}
\end{equation}
which for the special case of \(\pmomtheta=\pi/2\) (HMC) simplifies to
\begin{displaymath}
  F_{(\phi^2)_c}(\beta; \pmomtheta=\pi/2) =
  \frac{G^{+}_{2,0} + G^{-}_{0,0}}{1 - G^{+}_{2,0} - G^{-}_{0,0}}.
\end{displaymath}

\subsection{Exponentially Distributed Trajectory Lengths}

To proceed further we need to choose a specific form for the momentum refresh
distribution. In this section we will analyse the case of exponentially
distributed trajectory lengths, \(P_R(\trjlen) = re^{-r\trjlen}\) where the
parameter is just the inverse average trajectory length \(r=1/\trjlenav\). In
section~\ref{sec:pr-fixed} we will consider fixed length trajectories.

We make the approximation that \(\dH\) and thus \(\pacc\) are independent of
\(\trjlen\) (c.f., Figure~\ref{fig:acc-logdh}), so
\begin{displaymath}
  G^{+}_{j,k} \approx \paccav G_{j,k}, \qquad
  G^{-}_{j,k} \approx (1 - \paccav) G_{j,k};
\end{displaymath}
where \(\paccav\equiv\pacc(\trjlenav)\). This approximation is only made in
order to avoid unpleasant integrals which cannot be evaluated in closed form.
The integral in equation~(\ref{eq:g-def}) may be evaluated, and we find
\begin{equation}
  G^{\hbox{\tiny exp}}_{j,k} =
  \sum_{\mu=0}^j \sum_{\nu=0}^k {j\choose\mu} {k\choose\nu}
  \frac{(\half)^{j+k} (-1)^{\nu + \lfloor{\half k}\rfloor}B}
  {(\beta\trjlenav + 1)^2 + \xivv^2 (j+k-2\mu-2\nu)^2},
  \label{eq:g-result}
\end{equation}
where we have introduced the dimensionless quantity
\(\xivv\equiv\omega\trjlenav\) and
\begin{displaymath}
  B = \left\{
    \begin{array}{cl}
      \beta\trjlenav + 1 & \mbox{if \(\half k\in\N\)}, \\ [1ex]
      (j+k-2\mu-2\nu)\xivv & \mbox{if \(\half k\not\in\N\).}
    \end{array}
  \right.
\end{displaymath}

\subsubsection{Linear Operators}

The Laplace transform of the linear single mode autocorrelation function for
exponentially distributed trajectories is obtained by substituting the
explicit form for \(G^{\pm}\) into equation~(\ref{eq:F-linear}), and we obtain
\begin{displaymath}
  F^{\hbox{\tiny exp}}_\phi(\beta) = \frac 
  {\left[
      \begin{array}{c}
        r\bigg(\beta^3 
        + \left((1-2\paccav)\cos\pmomtheta + 3\right) r \beta^2 \\
        + \left(2(1-2\paccav)\cos\pmomtheta
          + (1-\paccav)\xivv^2 + 3\right) r^2 \beta \\
        + \left(\bigl((1-\paccav)^2 \xivv^2 + 1 - 2\paccav\bigr)\cos\pmomtheta
          + (1-\paccav)\xivv^2 + 1\right) r^3\bigg)
      \end{array}
    \right]}
  {\left[
      \begin{array}{c}
        \beta^4 + \left((1 - 2\paccav)\cos\pmomtheta + 3\right) r \beta^3
        + \left(2(1-2\paccav)\cos\pmomtheta + 3 + \xivv^2\right) r^2 \beta^2 \\
        + \left(\bigl((1-\paccav)\xivv^2 + 1 - 2\paccav\bigr)\cos\pmomtheta
          + (1+\paccav)\xivv^2 + 1\right) r^3 \beta \\
        + \left(\paccav(1-\paccav)\xivv^2 \cos\pmomtheta +
          \paccav\xivv^2\right) r^4
      \end{array}
    \right]}.
\end{displaymath}
Evaluating this at \(\beta=0\) gives the integrated autocorrelation function
\begin{displaymath}
  A^{\hbox{\tiny exp}}_\phi = \frac
  {\bigl((1-\paccav)^2 \xivv^2 + 1 - 2\paccav\bigr)\cos\pmomtheta
    + (1-\paccav)\xivv^2 + 1}
  {\paccav\xivv^2 \bigl((1-\paccav)\cos\pmomtheta + 1\bigr)}.
\end{displaymath}
For HMC where \(\pmomtheta=\pi/2\) we have
\begin{displaymath}
  F^{\hbox{\tiny exp}}_\phi(\beta; \pmomtheta=\pi/2) = \frac
  {\Bigl(\beta^2 + 2r\beta + \bigl((1-\paccav)\xivv^2 + 1\bigr)r^2\Bigr) r}
  {\beta^3 + 2r\beta^2 + (\xivv^2 + 1)r^2\beta + \paccav r^3\xivv^2},
\end{displaymath}
with the corresponding integrated autocorrelation function being
\begin{displaymath}
  A^{\hbox{\tiny exp}}_\phi(\pmomtheta=\pi/2) =
  \frac{1-\paccav}\paccav + \frac1{\paccav\xivv^2}.
\end{displaymath}
In the limit of unit acceptance rate, \(\paccav=1\), we obtain
\begin{displaymath}
  F^{\hbox{\tiny exp}}_\phi(\beta; \paccav=1) = \frac
  {\bigl(\beta + (1-\cos\pmomtheta)r\bigr)r}
  {\beta^{2} + (1-\cos\pmomtheta)r\beta + r^2\xivv^2}
\end{displaymath}
and
\begin{displaymath}
  A^{\hbox{\tiny exp}}_\phi(\paccav=1) = \frac{1 - \cos\pmomtheta}{\xivv^2}.
\end{displaymath}
Finally, for HMC in the limit of unit acceptance rate
\begin{equation}
  F^{\hbox{\tiny exp}}_\phi(\beta; \pmomtheta=\pi/2, \paccav=1) =
  \frac{(r + \beta)r}{\beta^2 + r\beta + r^2\xivv^2},
  \label{eq:F-exp-acc1-linear}
\end{equation}
and
\begin{displaymath}
  A^{\hbox{\tiny exp}}_\phi(\pmomtheta=\pi/2, \paccav=1) = \frac1{\xivv^2}.
\end{displaymath}

\subsubsection{Quadratic Operators}

The Laplace transform of the quadratic single mode autocorrelation function
for exponentially distributed trajectories is obtained by substituting the
explicit form for \(G^{\pm}\) into equation~(\ref{eq:F-quadratic}), and we
obtain
\begin{displaymath}
  F^{\hbox{\tiny exp}}_{(\phi^2)_c}(\beta) = \frac
  {r\left[
      \begin{array}{c}
        \beta^4
        + \Bigl(-(\cos\pmomtheta)^2 + (1-2\paccav)\cos\pmomtheta
        + 4\Bigr)r\beta^3 \\
        + \left(
          \begin{array}{c}
            (-1 + 2\paccav)(\cos\pmomtheta)^3 - 3(\cos\pmomtheta)^2 \\
            + (3 - 6\paccav)\cos\pmomtheta + (4 - 2\paccav)\xivv^2 + 6
          \end{array}
        \right)r^2\beta^2 \\
        + \left(
          \begin{array}{c}
            (4\paccav-2)(\cos\pmomtheta)^3
            + \bigl((4\paccav-4)\xivv^2 - 3\bigr)(\cos\pmomtheta)^2 \\
            + \bigl(2(-1+\paccav)(\paccav-2)\xivv^2 + 3
            - 6\paccav\bigr)\cos\pmomtheta \\
            + (8-4\paccav)\xivv^2 + 4
          \end{array}
        \right)r^3\beta \\
        + \left(
          \begin{array}{c}
            \bigl(- 4(-1 + \paccav)^2\xivv^2 - 1
            + 2\paccav\bigr)(\cos\pmomtheta)^3 \\
            + \bigl((4\paccav - 4)\xivv^2 - 1\bigr)(\cos\pmomtheta)^2 \\
            + \bigl(2(-1 + \paccav)(\paccav - 2)\xivv^2
            + 1 - 2\paccav\bigr)\cos\pmomtheta \\
            + (4 - 2\paccav)\xivv^2 + 1
          \end{array}
        \right)r^4
      \end{array}
    \right]}
  {\left[
      \begin{array}{c}
        \beta^5
        + \Bigl(-(\cos\pmomtheta)^2 + (1 - 2\paccav)\cos\pmomtheta
        + 4\Bigr)r\beta^4 \\
        + \left(
          \begin{array}{c}
            (-1 + 2\paccav)(\cos\pmomtheta)^3 - 3(\cos\pmomtheta)^2 \\
            + (3 - 6\paccav)\cos\pmomtheta + 6 + 4\xivv^2
          \end{array}
        \right)r^2\beta^3 \\
        + \left(
          \begin{array}{c}
            (4\paccav - 2)(\cos\pmomtheta)^3
            + \bigl((2\paccav - 4)\xivv^2 - 3\bigr)(\cos\pmomtheta)^2 \\
            + \bigl((4 - 4\paccav)\xivv^2 + 3 - 6\paccav\bigr)\cos\pmomtheta \\
            + (8 + 2\paccav)\xivv^2 + 4
          \end{array}
        \right)r^3\beta^2 \\
        + \left(
          \begin{array}{c}
            (-2(-1 + \paccav)(\paccav - 2)\xivv^2 - 1 + 2\paccav)
            (\cos\pmomtheta)^3 \\
            + (-4\xivv^2 - 1)(\cos\pmomtheta)^2 \\
            + \bigl(-2(\paccav + 2)(-1 + \paccav)\xivv^2 + 1 - 2\paccav\bigr)
            \cos\pmomtheta \\
            + (4\paccav + 4)\xivv^2 + 1
          \end{array}
        \right)r^4\beta \\
        + \left(
          \begin{array}{c}
            2\paccav(-1 + \paccav)\xivv^2(\cos\pmomtheta)^3
            - 2(\cos\pmomtheta)^2\paccav\xivv^2 \\
            - 2\paccav(-1 + \paccav)\xivv^2\cos\pmomtheta + 2\paccav\xivv^2
          \end{array}
        \right)r^5
      \end{array}
    \right]}.
\end{displaymath}
Evaluating this at \(\beta=0\) gives the integrated autocorrelation function
\begin{displaymath}
  A^{\hbox{\tiny exp}}_{(\phi^2)_c} = \frac
  {\left[
      \begin{array}{c}
        \Bigl(4(1 - \paccav)^2\xivv^2 + 1 - 2\paccav\Bigr)(\cos\pmomtheta)^3
        + \Bigl(4(1-\paccav)\xivv^2 + 1\Bigr)(\cos\pmomtheta)^2 \\
        + \Bigl(2(1-\paccav)(\paccav-2)\xivv^2 - 1 +
        2\paccav\Bigr)\cos\pmomtheta + 2(\paccav-2)\xivv^2 - 1
      \end{array}
    \right]}
  {2\paccav\xivv^2(\cos\pmomtheta - 1)(\cos\pmomtheta + 1)
    \Bigl((1 - \paccav)\cos\pmomtheta + 1\Bigr)}.
\end{displaymath}
For HMC where \(\pmomtheta=\pi/2\) these equations simplify to
\begin{displaymath}
  F^{\hbox{\tiny exp}}_{(\phi^2)_c}(\beta; \pmomtheta=\pi/2) = \frac
  {\Bigl(\beta^2 + 2r\beta + \bigl(2(2 - \paccav)\xivv^2 + 1\bigr)r^2\Bigr)r}
  {\beta^3 + 2r\beta^2 + (4\xivv^2 + 1)r^2\beta + 2\paccav r^3\xivv^2}
\end{displaymath}
and
\begin{displaymath}
  A^{\hbox{\tiny exp}}_{(\phi^2)_c}(\pmomtheta=\pi/2) =
  \frac{2-\paccav}\paccav + \frac1{2\paccav\xivv^2}.
\end{displaymath}
In the limit of unit acceptance rate, \(\paccav=1\), we obtain
\begin{displaymath}
  F^{\hbox{\tiny exp}}_{(\phi^2)_c}(\beta; \paccav=1) = \frac
  {\left[
      \begin{array}{c}
        \beta^2
        + \bigl(-\cos\pmomtheta + 2 - (\cos\pmomtheta)^2\bigr)r\beta \\
        + \bigl((\cos\pmomtheta)^3 - (\cos\pmomtheta)^2 - \cos\pmomtheta +
        2\xivv^2 + 1\bigr)r^2
      \end{array}
    \right]r}
  {\left[
      \begin{array}{c}
        \beta^3
        + \bigl(-\cos\pmomtheta + 2 - (\cos\pmomtheta)^2\bigr)r\beta^2 \\
        + \bigl((\cos\pmomtheta)^3 - \cos\pmomtheta + 1 + 4\xivv^2
        - (\cos\pmomtheta)^2\bigr)r^2\beta \\
        + \bigl(-2(\cos\pmomtheta)^2\xivv^2 + 2\xivv^2\bigr)r^3
      \end{array}
    \right]}
\end{displaymath}
and
\begin{displaymath}
  A^{\hbox{\tiny exp}}_{(\phi^2)_c}(\paccav=1) = - \frac
  {(\cos\pmomtheta)^3 - (\cos\pmomtheta)^2 - \cos\pmomtheta + 2\xivv^2 + 1}
  {2\xivv^2\bigl((\cos\pmomtheta)^2 - 1\bigr)}.
\end{displaymath}
Finally, for HMC in the limit of unit acceptance rate
\begin{equation}
  F^{\hbox{\tiny exp}}_{(\phi^2)_c}(\beta; \pmomtheta=\pi/2, \paccav=1) =
  \frac {\bigl(\beta^2 + 2r\beta + (1 + 2\xivv^2)r^2\bigr)r} {\beta^3 +
    2r\beta^2 + (4\xivv^2 + 1)r^2\beta + 2r^3\xivv^2},
  \label{eq:F-exp-acc1-quadratic}
\end{equation}
and
\begin{displaymath}
  A^{\hbox{\tiny exp}}_{(\phi^2)_c}(\pmomtheta=\pi/2, \paccav=1) =
  1 + \frac1{2\xivv^2}.
\end{displaymath}

\subsection{Fixed Length Trajectories}
\label{sec:pr-fixed}

In this section we consider the case of fixed length trajectories,
\(P_R(\trjlen) = \delta(\trjlen-\trjlenav)\). In this case we find that
\begin{displaymath}
  G^{+}_{j,k} = \paccav G_{j,k} \qquad\mbox{and}\qquad
  G^{-}_{j,k} = (1 - \paccav) G_{j,k},
\end{displaymath}
without making any approximations, and from equation~(\ref{eq:g-def}) we
obtain
\begin{displaymath}
  G^{\hbox{\tiny fix}}_{j,k} = e^{-\beta\trjlenav} (\cos\xivv)^j
  (\sin\xivv)^k.
\end{displaymath}

\subsubsection{Linear Operators}

The Laplace transform of the linear single mode autocorrelation function for
fixed length trajectories \(F^{\hbox{\tiny fix}}_\phi(\beta)\) is obtained by
substituting the explicit form for \(G^{\pm}\) into
equation~(\ref{eq:F-linear}),
\begin{displaymath}
  \frac
  {(-1 + 2\paccav)\cos\pmomtheta e^{-2\beta\trjlenav}
    - (\paccav\cos\xivv + 1 - \paccav) e^{-\beta\trjlenav}}
  {(1 - 2\paccav)\cos\pmomtheta e^{-2 \beta\trjlenav}
    + \Bigl((\paccav + \paccav\cos\xivv - 1)\cos\pmomtheta
    + \paccav\cos\xivv + 1 - \paccav\Bigr) e^{-\beta\trjlenav}
    - 1}.
\end{displaymath}
Evaluating this at \(\beta=0\) gives the integrated autocorrelation function
\begin{displaymath}
  A^{\hbox{\tiny fix}}_\phi = -\frac
  {(1-2\paccav)\cos\pmomtheta + \paccav\cos\xivv + 1 -\paccav}
  {\paccav(\cos\pmomtheta + 1)(\cos\xivv - 1)}.
\end{displaymath}
For HMC where \(\pmomtheta=\pi/2\) we have
\begin{displaymath}
  F^{\hbox{\tiny fix}}_\phi(\beta; \pmomtheta=\pi/2) = \frac
  {(\paccav\cos\xivv + 1 - \paccav) e^{-\beta\trjlenav}}
  {1 - (\paccav\cos\xivv + 1 - \paccav) e^{-\beta\trjlenav}},
\end{displaymath}
with the corresponding integrated autocorrelation function being
\begin{displaymath}
  A^{\hbox{\tiny fix}}_\phi(\pmomtheta=\pi/2) =
  \frac{\paccav\cos\xivv + 1 - \paccav}{\paccav(1 - \cos\xivv)}.
\end{displaymath}
In the limit of unit acceptance rate, \(\paccav=1\), we obtain
\begin{displaymath}
  F^{\hbox{\tiny fix}}_\phi(\beta; \paccav=1) = - \frac
  {\cos\xivv e^{-\beta\trjlenav} - \cos\pmomtheta e^{-2\beta\trjlenav}}
  {(\cos\pmomtheta \cos\xivv + \cos\xivv) e^{-\beta\trjlenav}
    - \cos\pmomtheta e^{-2\beta\trjlenav} - 1},
\end{displaymath}
and
\begin{displaymath}
  A^{\hbox{\tiny fix}}_\phi(\paccav=1) = \frac
  {\cos\pmomtheta - \cos\xivv}
  {\cos\pmomtheta\cos\xivv + \cos\xivv - \cos\pmomtheta - 1}.
\end{displaymath}
Finally, for HMC in the limit of unit acceptance rate
\begin{equation}
  F^{\hbox{\tiny fix}}_\phi(\beta; \pmomtheta=\pi/2, \paccav=1) =
  \frac{\cos\xivv e^{-\beta\trjlenav}}{1 - \cos\xivv e^{-\beta\trjlenav}},
  \label{eq:F-fixed-acc1-linear}
\end{equation}
and
\begin{displaymath}
  A^{\hbox{\tiny fix}}_\phi(\pmomtheta=\pi/2, \paccav=1) =
  \frac{\cos\xivv}{1 - \cos\xivv}.
\end{displaymath}

\subsubsection{Quadratic Operators}

The Laplace transform of the quadratic single mode autocorrelation function
for fixed length trajectories is obtained by substituting the explicit form
for \(G^{\pm}\) into equation~(\ref{eq:F-quadratic}), and we obtain
\begin{displaymath}
  F^{\hbox{\tiny fix}}_{(\phi^2)_c}(\beta) = - \frac
  {\left[
      \begin{array}{c}
        (- \paccav(\cos\xivv)^2 - 1 + \paccav) e^{-\beta\trjlenav} \\
        - e^{-3\beta\trjlenav} (- 1 + 2\paccav)(\cos\pmomtheta)^3 \\
        + e^{-2\beta\trjlenav}
        (- 2\paccav + 1 + 2\paccav(\cos\xivv)^2)(\cos\pmomtheta)^2 \\
        + e^{-2\beta\trjlenav}(\paccav + \paccav(\cos\xivv)^2 -
        1)\cos\pmomtheta
      \end{array}
    \right]}
  {\left[
      \begin{array}{c}
        \Bigl(
        (- \paccav(\cos\xivv)^2 - 1 + \paccav) (\cos\pmomtheta)^2 \\
        + \bigl(- 2\paccav(\cos\xivv)^2 + 1\bigr) \cos\pmomtheta
        - \paccav(\cos\xivv)^2 - 1 + \paccav
        \Bigr) e^{-\beta\trjlenav} \\
        + \Bigl(- e^{-2\beta\trjlenav} + e^{-2\beta\trjlenav} \paccav
        + e^{-2\beta\trjlenav} \paccav(\cos\xivv)^2
        + e^{-3\beta\trjlenav} - 2e^{-3\beta\trjlenav}\paccav\Bigr)
        (\cos\pmomtheta)^3 \\
        + e^{-2\beta\trjlenav}
        \bigl(-2\paccav + 1 + 2\paccav(\cos\xivv)^2\bigr)
        (\cos\pmomtheta)^2 \\
        + e^{-2\beta\trjlenav}
        \bigl(\paccav + \paccav(\cos\xivv)^2 - 1\bigr)
        \cos\pmomtheta + 1
      \end{array}
    \right]}.
\end{displaymath}
Evaluating this at \(\beta=0\) gives the integrated autocorrelation function
\begin{displaymath}
  A^{\hbox{\tiny fix}}_{(\phi^2)_c} = - \frac {(1 -
    2\paccav)(\cos\pmomtheta)^2 + 2\cos\pmomtheta\paccav(\cos\xivv)^2 -
    \paccav(\cos\xivv)^2 - 1 + \paccav} {(\cos\pmomtheta - 1)(\cos\pmomtheta +
    1)(\cos\xivv - 1)(\cos\xivv + 1)\paccav}.
\end{displaymath}
For HMC where \(\pmomtheta=\pi/2\) these equations simplify to
\begin{displaymath}
  F^{\hbox{\tiny fix}}_{(\phi^2)_c}(\beta; \pmomtheta=\pi/2) = \frac
  {(\paccav(\cos\xivv)^2 + 1 - \paccav) e^{-\beta\trjlenav}}
  {1 - \bigl(\paccav(\cos\xivv)^2 + 1 - \paccav\bigr)
    e^{-\beta\trjlenav}},
\end{displaymath}
and
\begin{displaymath}
  A^{\hbox{\tiny fix}}_{(\phi^2)_c}(\pmomtheta=\pi/2) = \frac
  {\paccav(\cos\xivv)^2 + 1 - \paccav}
  {\paccav(1 - \cos\xivv)(1 + \cos\xivv)}.
\end{displaymath}
In the limit of unit acceptance rate, \(\paccav=1\), we obtain
\begin{displaymath}
  F^{\hbox{\tiny fix}}_{(\phi^2)_c}(\beta; \paccav=1) = - \frac
  {\left[
      \begin{array}{c}
        (\cos\xivv)^2 e^{-\beta\trjlenav}
        + (\cos\pmomtheta)^3 e^{-3\beta\trjlenav} \\
        - e^{-2\beta\trjlenav} (- 1 + 2 (\cos\xivv)^2)
        (\cos\pmomtheta)^2 \\
        - e^{-2\beta\trjlenav} (\cos\xivv)^2 \cos\pmomtheta
      \end{array}
    \right]}
  {\left[
      \begin{array}{c}
        \Bigl( (\cos\pmomtheta)^2 (\cos\xivv)^2 + ( - 1 + 2(\cos\xivv)^2)
        \cos\pmomtheta + (\cos\xivv)^2
        \Bigr) e^{-\beta\trjlenav} \\
        + \bigl(- e^{-2\beta\trjlenav} (\cos\xivv)^2 +
        e^{-3\beta\trjlenav}\bigr) (\cos\pmomtheta)^3 \\
        - e^{-2\beta\trjlenav} ( - 1 + 2(\cos\xivv)^2) (\cos\pmomtheta)^2 -
        e^{-2\beta\trjlenav} (\cos\xivv)^2 \cos\pmomtheta - 1
      \end{array}
    \right]},
\end{displaymath}
and
\begin{displaymath}
  A^{\hbox{\tiny fix}}_{(\phi^2)_c}(\paccav=1) = \frac
  {(\cos\pmomtheta)^2 - 2\cos\pmomtheta (\cos\xivv)^2 + (\cos\xivv)^2}
  {(\cos\pmomtheta)^2 (\cos\xivv)^2 - (\cos\pmomtheta)^2 - (\cos\xivv)^2 + 1}.
\end{displaymath}
Finally, for HMC in the limit of unit acceptance rate
\begin{displaymath}
  F^{\hbox{\tiny fix}}_{(\phi^2)_c}(\beta; \pmomtheta=\pi/2, \paccav=1) =
  \frac {(\cos\xivv)^2 e^{-\beta\trjlenav}} {1 - (\cos\xivv)^2
    e^{-\beta\trjlenav}},
  \label{eq:F-fixed-acc1-quadratic}
\end{displaymath}
and
\begin{displaymath}
  A^{\hbox{\tiny fix}}_{(\phi^2)_c}(\pmomtheta=\pi/2, \paccav=1) =
  \frac{(\cos\xivv)^2}{1 - (\cos\xivv)^2}.
\end{displaymath}

\section{Autocorrelations and Costs for HMC with \(\pacc\approx1\)}
\label{sec:costs-for-pacc-equals-1}

In this section we calculate autocorrelations for the special case
\(\pmomtheta=0\) (HMC) and in the approximation where the acceptance
rate is unity, \(\paccav=1\). We shall consider more general cases in
the following section.

We begin with some general observations: \label{sec:laplace-meaning}
\begin{itemize}
  
\item \emph{Integrated autocorrelation time.} It is immediately obvious that
  the integrated autocorrelation time may be obtained from the Laplace
  transform (\ref{eq:laplace-def}) by evaluating it at \(\beta=0\).
  
\item \emph{Exponential autocorrelation time.} For an ergodic Markov process
  we can write the typical asymptotic form of the autocorrelation function as
  \begin{displaymath}
    C_\Omega(\t) \asymp \hbox{\rm constant}\times e^{-\t/\texp} \qquad
      \hbox{\rm for \(\t\to\infty\)}.
  \end{displaymath}
  This exponential autocorrelation time is a different quantity from \(\nexp\)
  of Section~\ref{sec:simple-markov-processes}, which was defined in terms of
  Markov steps. \(\texp\) can be extracted from \(F_\Omega(\beta)\) by
  considering its analytic structure in the complex \(\beta\) plane. Since
  \(\texp\) governs the most slowly decaying exponential, it follows from the
  definition of the Laplace transform (\ref{eq:laplace-def}) that
  \(F_\Omega(\beta)\) will have its rightmost pole at \(\Re\beta=-1/\texp\).
  
\item \emph{Dynamical critical exponent.} One of the most relevant measures of
  the effectiveness of an algorithm for studying continuum physics is the
  exponent \(z\) relating the cost \(\cost\) to the correlation length \(\xi\)
  of the system, \(\cost\propto\xi^z\).
  
\item \emph{Optimal choice for \(\pmomtheta\).} For the GHMC algorithm we
  should minimise the cost by varying both \(\trjlen\) and \(\pmomtheta\). For
  the case of quadratic operators with exponentially distributed trajectory
  lengths, for instance, the optimal choice of parameters when \(\pacc=1\) is
  to take \(\pmomtheta\to0\) and \(\trjlen=\half\pmomtheta^2\to0\). However,
  this ignores the fact that the cost does not decrease when we take
  \(\trjlen\) smaller than the \(\dt\) required to obtain a reasonable
  Metropolis acceptance rate. If we choose \(\trjlenavopt=\dt\) (L2MC) and the
  corresponding value for \(\pmomthetaopt\) we find that the cost is less than
  for the HMC case, but only by a constant factor. As the cost is only defined
  up to a implementation dependent constant factor anyhow we may conclude that
  generalised HMC does not appear to promise great improvements over HMC. The
  situation is more complex in the real world where \(\pacc\neq1\), which is
  explored in Section~\ref{sec:costs-for-pacc-neq-1}.

\end{itemize}

\subsection{Exponentially Distributed Trajectory Lengths}

\subsubsection{Linear Operators}

The least uninteresting linear operator in free scalar field theory is the
\emph{magnetisation} \(M=\sum_x \phi(x)\). From
equations~(\ref{eq:complex-finite-fft}) and~(\ref{eq:frequency-definition})
this is simply the zero-momentum mode in Fourier space, \(\omega=m\), and is
thus expected to evolve most slowly in fictitious time.

\begin{example}
  From equation~(\ref{eq:linear-op-definition}), with \(\Omega_{pq}
  =\delta_{p0}\), and equation~(\ref{eq:F-exp-acc1-linear}), the Laplace
  transform of the autocorrelation function for the magnetisation is
  \begin{displaymath}
    F_M(\beta) = {r(r+\beta)\over \beta(r+\beta) + m^2}.
  \end{displaymath}
  
  As explained previously, the exponential autocorrelation time
  \(\texp=-1/\Re\betaexp\) corresponds to the rightmost pole in \(F_M\), and
  this occurs when
  \begin{displaymath}
    \betaexp^\pm = -{r\over2} \pm \sqrt{\left({r\over2}\right)^2-m^2},
  \end{displaymath}
  and therefore
  \begin{displaymath}
    \texp = \left\{
      \begin{array}{ll}
        2\trjlenav\over1-\sqrt{1-(2m\trjlenav)^2}
        & \hbox{\rm if \(\trjlenav\leq{1\over2m}\)} \\
        2\trjlenav
        & \hbox{\rm if \(\trjlenav\geq{1\over2m}\),}
      \end{array}
    \right.
  \end{displaymath}
  where we have used the fact that \(r=1/\trjlenav\). We observe that
  the exponential autocorrelation time is minimised when the average
  trajectory length is chosen to be \(\trjlenav=1/2m\). Note that
  \(M\) only couples to the zero momentum mode, and its relaxation
  rate determines \(\texp\) in this case.
  
  The integrated autocorrelation function \(A_M\) is given by
  \begin{displaymath}
    A_M = F_M(0) = \left({1\over m\trjlenav}\right)^2,
  \end{displaymath}
  and we can minimise the cost of computing the magnetisation by making use of
  equation~(\ref{eq:minimise-cost}). The optimal trajectory length is
  \(\trjlenavopt=\sqrt2/m\), which corresponds to the minimum integrated
  autocorrelation function value \(A_M(\trjlenavopt)=1/2\). Note that the
  optimal trajectory length does not minimise the exponential autocorrelation
  time --- they differ by a factor of \(2\sqrt{2}\).
  
  The correlation length for this system is \(\xi\propto1/m\), and our
  result indicates that the optimal trajectory length is
  \(\trjlenavopt\propto\xi^z\) with a {\em dynamical critical
  exponent\/} \(z=1\). Indeed, if we were to choose an average
  trajectory length \(\trjlenav\propto\xi^\alpha\) then we would find
  that the cost per effectively independent configuration would grow as
  \begin{displaymath}
    \cost \propto \left\{
      \begin{array}{ll}
        \xi^\alpha & \hbox{\rm for \(\alpha\geq1, \xi\to\infty\)} \\
        \xi^{2-\alpha} & \hbox{\rm for \(\alpha\leq1, \xi\to\infty\).}
      \end{array}
    \right.
  \end{displaymath}
  Keeping the average trajectory length fixed as the correlation length
  \(\xi\) increases, {\em i.e.,} choosing \(\alpha=0\), thus leads to a
  dynamical critical exponent~\(z=2\).

  We can calculate the autocorrelation function \(C_M(\t)\) in closed form by
  inverting the Laplace transform. If we expand \(F_M\) in partial fractions
  \begin{displaymath}
    F_M(\beta) =
    {1+\Delta\over\Delta(1-\Delta+2\trjlenav\beta)}
    - {1-\Delta\over\Delta(1+\Delta+2\trjlenav\beta)},
  \end{displaymath}
  where \(\Delta\equiv\sqrt{1-(2m\trjlenav)^2}\), then it is easy to verify
  that the autocorrelation function is
  \begin{eqnarray*}
    C_M(\t)
    &=& {1\over2\trjlenav\Delta} \left\{
      (1+\Delta) e^{-(1-\Delta)\t/2\trjlenav}
      - (1-\Delta) e^{-(1+\Delta)\t/2\trjlenav}
    \right\} \\
    &=& \frac1\trjlenav e^{-\t/2\trjlenav} \left\{
      \cosh\left({\Delta\t\over2\trjlenav}\right)
      + \frac1\Delta \sinh\left({\Delta\t\over2\trjlenav}\right)
    \right\}
  \end{eqnarray*}
  when \(0<\Delta<1\). When \(\Delta^2<0\) we have
  \begin{displaymath}
    C_M(\t)
    = \frac1\trjlenav e^{-\t/2\trjlenav} \left\{
      \cos\left({\Delta'\t\over2\trjlenav}\right)
      + \frac1{\Delta'} \sin\left({\Delta'\t\over2\trjlenav}\right)
    \right\},
  \end{displaymath}
  where \(\Delta'\equiv i\Delta=\sqrt{(2m\trjlenav)^2-1}\). We observe that
  there are oscillations in the autocorrelation function for long trajectories
  where \(\trjlenav>1/2m\). For the critical case \(\Delta=0\) we have
  \begin{displaymath}
    C_M(\t) = \frac1\trjlenav \left(1+{\t\over2\trjlenav}\right)
    e^{-\t/2\trjlenav}.
  \end{displaymath}
  
  Finally, the autocorrelation function expressed in terms of Markov steps is
  \(C_M(n) = e^{-n/T_0}\), where the exponential autocorrelation time \(\nexp
  = T_0
  = 1/\ln[1+(m\trjlenav)^2]\).

\subsubsection{Quadratic Operators}

\end{example}
\begin{example} We obtain the Laplace transform of the connected
  autocorrelation function\footnote{This is a synonym for the autocorrelation
    function for \(M^2_c\), the connected part of \(M^2\).} for \(M^2\) by
  setting \(\Omega_{pq} = \delta_{p0}\delta_{q0}\) in
  equation~(\ref{eq:quadratic-op-definition}). From
  equation~(\ref{eq:F-exp-acc1-quadratic})
  \begin{equation}
    F_{M^2_c}(\beta) = {r^3 + 2r^2\beta + r\beta^2 + 2m^2r
      \over \beta^3 + 2r\beta^2 + (r^2 + 4m^2)\beta + 2m^2r}.
    \label{F_M2}
  \end{equation}
  The integrated autocorrelation function is thus
  \begin{displaymath}
    A_{M^2_c} = F_{M^2_c}(0) = 1 + \frac12\left({1\over m\trjlenav}\right)^2.
  \end{displaymath}
  Minimising the cost by means of equation~(\ref{eq:minimise-cost}),
  we obtain \(\trjlenavopt = 1/\left(\sqrt3m\right)\) and thence
  \(A_{M^2}(\trjlenavopt)=5/2\).  Again, the
  dynamical critical exponent is $z = 1$.  The optimum trajectory length is
  different from that for \(M\), which is to be expected.

  For exponentially distributed trajectory lengths the Laplace
  transform of the autocorrelation function is a rational function in
  \(\beta\) with the numerator of lower degree than the denominator
  (see, eg, equations~(\ref{eq:F-exp-acc1-linear})
  and~(\ref{eq:F-exp-acc1-quadratic})), which implies that the
  autocorrelation function is a sum of exponentials. In this case the
  exponents are the roots of the cubic denominator, and they are either
  all real, or one is real and the other two are complex conjugates
  depending on the value of the mean trajectory length \(1/r\). This is shown
  explicitly in Appendix~\ref{sec:inverse-laplace-transforms}.

\end{example}
\begin{example}
  Following equation~(\ref{eq:F-exp-acc1-quadratic}), it is easy to write down
  the the Laplace transform of the connected autocorrelation function for the
  energy~\(E=\sum_p \half \omega_p^2 \phi_p^2\)
  \begin{equation}
    F_E(\beta) = \frac1N \sum_{p=1}^N
    {r^3 + 2r^2\beta + r\beta^2 + 2\omega_p^2r
      \over \beta^3 + 2r\beta^2 + (r^2 + 4\omega_p^2)\beta + 2\omega_p^2r}.
    \label{F_E}
  \end{equation}
  The integrated autocorrelation function for the energy is
  \begin{displaymath}
    A_E =  F_E(0)
    = 1 + {1\over2\trjlenav^2} \sum_{p=1}^N {1\over\omega_p^2}
    \equiv 1 + {1\over2\trjlenav^2} {1\over N}\sigma_{m^2}^{(-1)},
  \end{displaymath}
  and the optimal trajectory length is
  \(\trjlenavopt=\sqrt{\sigma_{m^2}^{(-1)}/3}\), leading to an integrated
  autocorrelation function value of \(A_E(\trjlenavopt)=5/2\).
  
  For one dimensional free field theory it remains to evaluate the spectral
  sum \(\sigma_{m^2}^{(-1)}\), details of which are discussed in
  Appendix~\ref{spectral-sum-for-one-dimensional-free-field-theory}. In the
  physically interesting limit of small \(m\) and large \(V\), we find
  \(\sigma_{m^2}^{(-1)}\approx\half m\), and thus
  \(\trjlenavopt\approx1/\sqrt{6m}\). Hence the dynamical critical exponent
  for the energy is~\(\half\).
  
  For two dimensional free field theory we get \(z{=}0\) up to logarithmic
  corrections (see
  Appendix~\ref{spectral-sum-for-two-dimensional-free-field-theory}).
  
\end{example}

\subsection{Fixed Length Trajectories}
\label{fixed-length-trajectories}

\subsubsection{Linear Operators}

From Section~\ref{sec:pr-fixed} the Laplace transform of the autocorrelation
function for the general linear operator of
equation~(\ref{eq:linear-op-definition}) is easily expressed in terms of
equation~(\ref{eq:F-fixed-acc1-linear}). The exponential autocorrelation time
\(\texp\) is related to the rightmost pole of the Laplace transform of the
autocorrelation function,
\begin{displaymath}
  \texp = -1/\Re\betaexp.
\end{displaymath}
Equation~(\ref{eq:F-fixed-acc1-linear}) has poles in the complex \(\beta\)
plane for
\begin{displaymath}
  \beta = {1\over\trjlenav} \ln\cos\omega_p\trjlenav,
\end{displaymath}
and hence
\begin{displaymath}
  \Re\beta = {1\over\trjlenav} \ln|\cos\omega_p\trjlenav|,
\end{displaymath}
which leads to
\begin{displaymath}
  \texp = \max_p{\trjlenav\over\Bigl|\ln|\cos\omega_p\trjlenav|\Bigr|}.
\end{displaymath}
This diverges when \(\omega_p\trjlenav/\pi\in\Z\) for any~\(p\), which just
reflects the fact that the {\hmc} algorithm is not ergodic for these cases, as
was first pointed out by Mackenzie~\cite{mackenzie89b}. Perhaps this is most
simply understood by considering the trajectory of the harmonic oscillator
with frequency \(\omega_p\) in the \((\omega_p\phi_p,\pi_p)\) phase space. The
{\md} trajectories are circular arcs subtending an angle of
\(\omega_p\trjlenav\) about the origin, and the momentum refreshment
corresponds to a change of the \(\pi_p\) coordinate leaving the
\(\omega_p\phi_p\) coordinate unchanged. If \(\omega_p\trjlenav\) is an
integer multiple of \(\pi\) the value of \(\phi_p\) can at most change sign,
and thus this mode certainly cannot explore the whole of its phase space.

\begin{example}
  For the magnetisation the integrated autocorrelation function is
  \begin{displaymath}
    A_M = F_M(0) = {\cos m\trjlenav\over1-\cos m\trjlenav}.
  \end{displaymath}
  If we minimise the cost we find that the optimal trajectory length
  corresponds to \(m\trjlenav\) being an odd multiple of \(\pi\), and the
  worst case occurs when \(m\trjlenav\) is an even multiple of \(\pi\). Both
  cases just reflect the non-ergodic nature of the updating scheme discussed
  above: the fact that the optimal ``ergodic'' update occurs when
  \(m\trjlenav\) is taken very close to an odd multiple of \(\pi\) is just an
  ``accidental'' consequence of the fact that \(\langle M\rangle=0\).
\end{example}

\subsubsection{Quadratic Operators}

\begin{example}
  In the case of the quadratic operator \(M^2_c\) we find from
  equation~(\ref{eq:F-fixed-acc1-quadratic}) that
  \begin{displaymath}
    F_{M^2_c}(\beta) =
      {\cos^2m\trjlenav \over e^{\beta\trjlenav} - \cos^2m\trjlenav},
  \end{displaymath}
  and thus \(A_{M^2_c} = F_{M^2_c}(0) = \cot^2m\trjlenav\) for fixed length
  trajectories. The optimal trajectory length \(m\trjlenavopt\approx1.3866\).
  As is to be expected, the non-ergodicity at \(m\trjlenav/\pi\in\Z\)
  manifests itself as a divergence in \(A_{M^2_c}\).
\end{example}

\begin{example}
  For the energy of one dimensional free field theory we obtain
  \begin{displaymath}
    F_{E}(\beta) = \frac1N \sum_p {\cos^2\omega_p\trjlenav \over
      e^{\beta\trjlenav} - \cos^2\omega_p\trjlenav},
  \end{displaymath}
  and therefore \(A_E = F_E(0) = \frac1N \sum_p \cot^2\omega_p\trjlenav\)
  which diverges whenever \(\omega_p\trjlenav/\pi\in\Z\) for any~\(p\).
\end{example}

\section{Comparison of Costs for \(\pacc\ne1\)}
\label{sec:costs-for-pacc-neq-1}

We wish to compare the performance of the HMC, L2MC and GHMC algorithms for
one dimensional free field theory. To do this we shall compare the cost of
generating a statistically independent measurement of the magnetisation \(M\)
and the magnetic susceptibility \(M^2_c\), choosing the optimal values for the
angle \(\pmomtheta\) and the average trajectory length \(\trjlenav\).

Following equation~(\ref{eq:minimise-cost}) we can minimise the cost with
respect to \(\pmomtheta\) without having to specify the form of the refresh
distribution.

The next step is to minimise the cost with respect to the average trajectory
length \(\xivv=m\trjlenav\). Strictly speaking we should note that the
acceptance probability \(\paccav\) is a function of \(\trjlenav\), but to a
good approximation we may assume that \(\pacc\) depends only upon the
integration step size \(\dt\) \emph{except} in the case of very short
trajectories, such as Langevin-type algorithms (see
Figure~\ref{fig:acc-logdh}). The exact solution of the problem would clearly
be highly transcendental. For this reason we shall find that although L2MC is
a special case of GHMC and thus can never be cheaper, our ``optimal''
solution\footnote{\emph{I.e.}, the solution obtained by neglecting the
  dependence of \(\paccav\) on \(\trjlenav\) when optimising the parameters
  \(\pmomtheta\) and \(\trjlenav\).} for very short trajectories (acceptance
probabilities very close to unity) will in fact cost more than L2MC.
Fortunately, this is irrelevant because the minimum cost for L2MC is far
greater than the minimum for GHMC, the latter occurring for long trajectory
lengths where our approximations are very good.

Another implicit approximation we make is that we treat \(\dt\) and
\(\trjlenav\) as independent parameters, although the trajectory length must
be an integer multiple of the step size; again this is a very good
approximation except for near to the Langevin limit. Of course we are also
ignoring subtleties such as that a multistep leapfrog integration is cheaper
than a series of single steps, that acceptance tests are a significant
fraction of the cost for very short trajectories, and that HMC requires less
memory than GHMC because the old momenta need not be saved. All of these facts
would disfavour L2MC, but we shall see that it is not competitive even without
these factors being taken into account.

\subsection{Linear Operators}

Applying equation~(\ref{eq:minimise-cost}) to equation~(\ref{eq:F-linear}) we
find that the cost is minimised by choosing\footnote{Setting \(\pmomtheta=0\)
  corresponds to never refreshing the momenta, and thus to a non-ergodic
  algorithm. This is just a peculiarity of linear operators in free field
  theory, and we can instead consider choosing \(\pmomtheta\) to be very small
  but non-zero to circumvent this difficulty.} \(\pmomthetaopt=0\), and at
this value the Laplace transform of the single mode autocorrelation function
becomes
\begin{displaymath}
  F_M(\beta; \pmomthetaopt) =
    -{\frac {-{G^+_{1,0}}^2 + {G^-_{0,0}}^2 - {G^+_{0,1}}^2
      + G^+_{1,0} + G^-_{0,0}} { - 1 + 2G^+_{1,0} - {G^+_{1,0}}^2 
      + {G^{-}_{0,0}}^2 - {G^+_{0,1}}^2}}.
\end{displaymath}

\subsubsection{Exponentially Distributed Trajectory Lengths}

The integrated autocorrelation function for exponentially distributed
trajectory lengths is obtained by substituting the results~(\ref{eq:g-result})
for the integrals of equation~(\ref{eq:g-def}) into \(F_M(\beta;
\pmomtheta=\pmomthetaopt)\) and setting \(\beta=0\). The cost becomes
\begin{displaymath}
  \cost^{\hbox{\tiny exp}}_M (\dt, \pmomthetaopt, \xivv) =
    \frac{V}{m\,\dt}\left[- {\frac{(\paccav-2)\xivv}{\paccav}}
      + \frac{4(\paccav-1)}{\paccav(\paccav-2)\xivv}\right].
\end{displaymath}
Minimising this with respect to \(\xivv\) we find that
\begin{displaymath}
  \opt{\xivv} = {\sqrt{1-\paccav}\over 1-\half \paccav}
\end{displaymath}
and that the cost at the optimal parameters is
\begin{displaymath}
  \cost^{\hbox{\tiny exp}}_M (\dt, \pmomthetaopt, \opt{\xivv}) =
  \frac {4V(1-\paccav)}{m\,\dt\,\paccav\sqrt{1-\paccav}}
\end{displaymath}
which corresponds to a dynamical critical exponent \(z=1\).

For the lowest order leapfrog integration scheme where we must scale
\(\dt\propto V^{-\quarter}\) to keep \(\paccav\) constant we thus find
\(\cost^{\hbox{\tiny exp}}_M \propto V^{\rational54}\) as expected.

\subsection{Quadratic Operators}

We can make the minimisation problem for quadratic operators manifestly
algebraic by writing \(c\equiv\cos\pmomtheta\) in
equation~(\ref{eq:F-quadratic}). The condition for the cost to be a minimum is
that \(\opt{c}\) is a root of the polynomial
\begin{eqnarray}
  && {G^{+}_{1,1}}^2
    \left(G^{-}_{0,0} + G^{+}_{0,2} + G^{+}_{2,0}\right)^2 \opt{c}^4
    \nonumber\\
  && - \left(2\,{G^{+}_{1,1}}^2 - G^{+}_{0,2} G^{+}_{2,0}
    + {G^{+}_{0,2}}^2 + G^{+}_{0,2} G^{-}_{0,0}\right )^2{\opt{c}}^3
    \nonumber\\
  && + \left(-2\,{G^{+}_{0,2}}^2 G^{-}_{0,0} - 2\,{G^{+}_{0,2}}^3
    - 2\, {G^{+}_{1,1}}^2 G^{-}_{0,0} -2\, G^{+}_{2,0} {G^{+}_{1,1}}^2
    + 2\, {G^{+}_{0,2}}^2 G^{+}_{2,0} -6\, {G^{+}_{1,1}}^2 G^{+}_{0,2}
    \right)\opt{c}^2
    \nonumber\\
  && - {G^{+}_{0,2}}^2\opt{c} + {G^{+}_{1,1}}^2
\end{eqnarray}
lying in the interval \([-1,1]\).

\subsubsection{Exponentially Distributed Trajectory Lengths}

\begin{figure}
  \begin{center}
    \epsfxsize=0.9\textwidth \leavevmode\epsffile{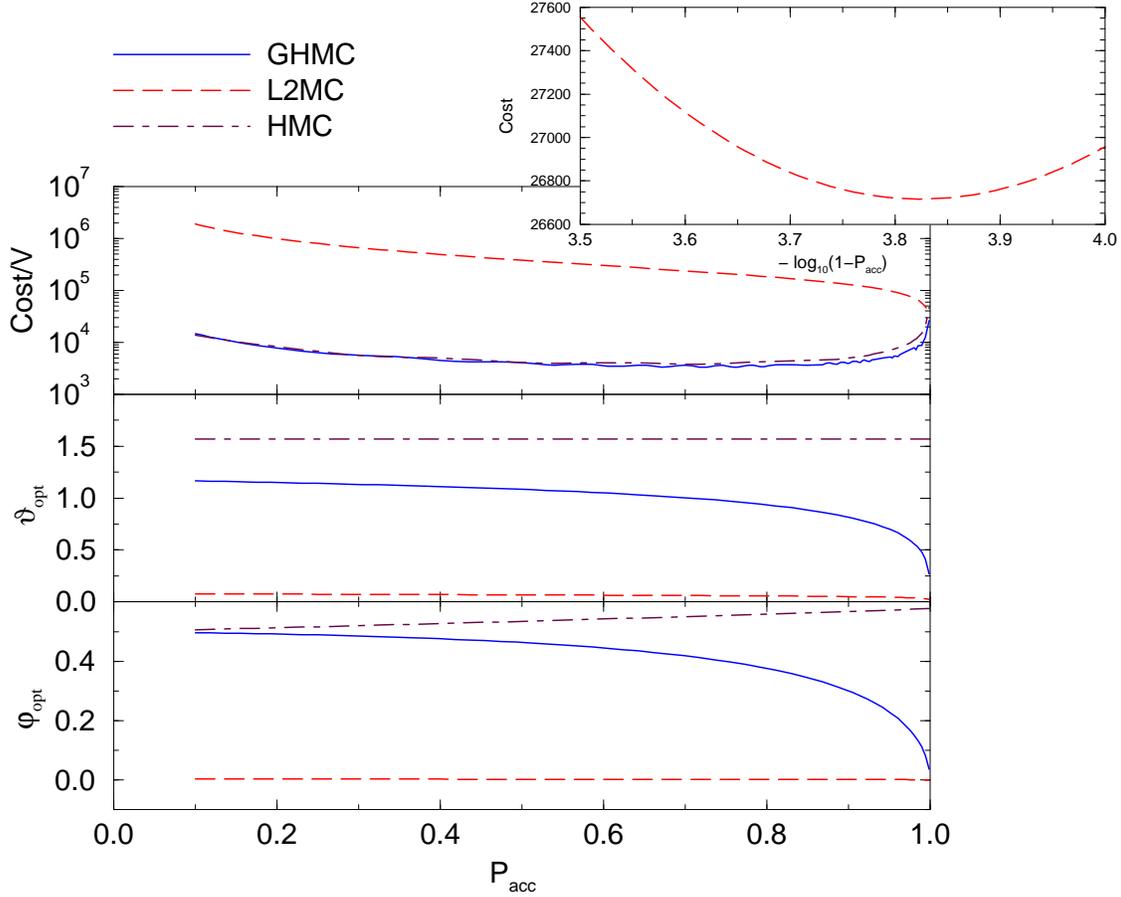}
  \end{center}
  \caption[laplace-opt.eps]{Cost as a function of average Metropolis
    acceptance rate for the GHMC algorithm compared to HMC and L2MC for free
    field theory using the lowest order leapfrog integration scheme. The
    operator under consideration is the ``magnetic susceptibility'', i.e., the
    connected quadratic operator depending only on the lowest frequency mode.
    The corresponding parameters, the momentum mixing angle \(\opt\pmomtheta\)
    and the average trajectory length measured as a fraction of the
    correlation length \(\opt\xivv=\opt\trjlen/\xi\) are also shown, all as a
    function of the acceptance rate \(\paccav\). The inset graph shows the
    region where the acceptance rate is very close to unity which is where the
    L2MC algorithm has its minimum cost.}
  \label{fig:laplace-opt}
\end{figure}

Just as in equations~(\ref{eq:minimise-cost}) the extrema of the cost occur on
the ideal defined by the polynomials \(d\cost/dc\) and \(d\cost/d\xivv\).
Finding the Gr\"obner basis with respect to the purely lexicographical
ordering with \(c<\xivv\) we find the point \((\opt{c}, \opt{\xivv})\) at
which the cost is minimal is defined by the equations
\begin{eqnarray}
  0 & = & (-4+3\paccav){\opt{c}}^4 + 4(1-2\paccav)(1-\paccav){\opt{c}}^3
    \nonumber\\
  & & + 16(1-\paccav)\opt{c}^2 + 4\opt{c} + \paccav - 4 
    \label{eq:cx2defpol2} \\[2ex]
    0 & = &
    4\paccav^2(2-\paccav)(4-\paccav)(\paccav^2-6\paccav+6){\opt{\xivv}}^2
    \nonumber\\
    & & +( - 4 + 3\paccav)({\paccav}^3 + 2{\paccav}^2 - 10\paccav +
    8)\opt{c}^5
    \nonumber \\
    & & + 2 ( - 1 + \paccav)(4\paccav^4 + 6\paccav^3 - 47{\paccav}^2 +
    68\paccav-32)\opt{c}^4
    \nonumber\\
    & & + ( - 38\paccav^4 + 108\paccav^3 - 16\paccav^2 - 152\paccav +
    96)\opt{c}^3
    \nonumber\\
    & & + ( - 16\paccav^5 + 56\paccav^4 - 20\paccav^3 - 168\paccav^2 +
    248\paccav - 96){\opt{c}}^2
    \nonumber\\
    & & + (35\paccav^4 - 110\paccav^3 + 54\paccav^2 + 88\paccav - 64)\opt{c}
    \nonumber\\
    & & + 8\paccav^5 - 60\paccav^4 + 126\paccav^3 - 62\paccav^2 - 48\paccav +
    32
    \label{eq:xix2defpol}
\end{eqnarray}
Using Sturm sequences we may easily show that equation~(\ref{eq:cx2defpol2})
has exactly one real root in \([0,1]\) and none in \([-1,0]\), and obviously
equation~(\ref{eq:xix2defpol}) has exactly one positive solution for
\(\opt{\xivv}\). The cost at the point \((\opt{c},\opt{\xivv})\) is given by
\begin{displaymath}
\cost^{\hbox{\tiny exp}}_{M^2_c} (\dt, \pmomthetaopt, \xivv) = \frac {
  V\left(
    \begin{array}{c}
      (7\paccav-3\paccav^2-4)\opt{\xivv}^2\opt{c}^3 
      + (-2\paccav+1)\opt{c}
      + 1 \\
      {}
      + (2\paccav-1)\opt{c}^3
      - \opt{c}^2 + (-\paccav+4)\opt{\xivv}^2 \\
      {}
      + (\paccav^2-5\paccav+4)\opt{c}\opt{\xivv}^2 
      + (-4+3\paccav)\opt{c}^2\opt{\xivv}^2
    \end{array}
  \right)}
{
  \left(
    \begin{array}{c}
      \paccav\dt m(\paccav-1)\opt{c}^3\opt{\xivv} 
      + \opt{\xivv}\paccav\dt m \\
      {}
      - \opt{\xivv}\paccav\dt m{\opt{c}}^2
      - \paccav\dt m(-1+\paccav)\opt{c}\opt{\xivv}
    \end{array}
  \right)}
\end{displaymath}
This solution is a function of \(\dt\) and \(\paccav\) which are not
independent variables, and using the results~(\ref{eq:pacc-formula}),
(\ref{eq:dh-formula}) and~(\ref{eq:hmc0-sigma}) we can compute the cost as a
function of \(\paccav\) as shown in Figure~\ref{fig:laplace-opt}.

\subsubsection{HMC}

The {\hmc} algorithm corresponds to setting \(\pmomtheta=\pi/2\), and thus we
find that the optimal trajectory length in this case is
\(\opt\xivv=1/\sqrt{4-\paccav}\), corresponding to a cost
\begin{displaymath}
  \opt\cost = {2V\sqrt{4-\paccav}\over\paccav\dt m}.
\end{displaymath}
This is also shown in Figure~\ref{fig:laplace-opt}.

\subsection{Fixed Length Trajectories}

For fixed length trajectories we shall only analyse the case of L2MC for which
the trajectory length \(\xivv=m\dt\). In this case \(\opt{c}\) satisfies
\begin{displaymath} 
  \frac
    {\paccav\opt{c}^2(\cos\opt\xivv)^2
      + 3\opt{c}^2\paccav + \paccav(\cos\opt\xivv)^2
      - \paccav - 2\opt{c}^2 - 4\opt{c}\paccav(\cos\opt\xivv)^2 + 2}
    {\paccav\left({\opt{c}}^{2}(\cos\opt\xivv)^2 - \opt{c}^2 
      - (\cos\opt\xivv)^2 + 1\right)} = 0.
\end{displaymath}
This, together with the corresponding cost, is also plotted in
Figure~\ref{fig:laplace-opt}. From this figure it is clear that the minimum
cost occurs for \(\paccav\) very close to unity, where the scaling variable
\(x=V\dt^6\) is very small. We may then express \(\opt{c}\) as
\begin{displaymath}
  \opt{c} = 1 - m\left(x\over V\right)^{1/6}
     + \half m^2\left(x\over V\right)^{1/3}
     - {11\over24}m^3\left(x\over V\right)^{1/2} +\ldots,
\end{displaymath}
and likewise
\begin{displaymath}
  \paccav = 1 - \sqrt{5\over8\pi}\left(1 + {9m^2\over20}
    + {39m^4\over800} + O(m^6)\right) \sqrt{x} + O(x^{3/2}).
\end{displaymath}
From these relations we find that the minimum cost for L2MC is
\begin{displaymath}
  \opt\cost = 2\left(10\over\pi\right)^{1/4} V^{5/4} m^{-3/2}.
\end{displaymath}
This result tells us that not only does the tuned L2MC algorithm have a
dynamical critical exponent \(z=3/2\), but also it has a volume dependence of
exactly the same form as HMC. We may understand why this behaviour occurs
rather than the naive \cite{horowitz90a} \(V^{7/6}m^{-1}\) by the
following simple argument.

If \(\paccav<1\) then the system will carry out a random walk backwards and
forwards along a trajectory because the momentum, and thus the direction of
travel, must be reversed upon a Metropolis rejection. A simple minded analysis
is that the average time between rejections must be \(O(1/m)\) in order to
achieve \(z=1\). This time is approximately
\begin{displaymath}
  \sum_{n=0}^\infty \paccav^n (1-\paccav) n\dt = {\paccav\dt\over1-\paccav}
  = {1\over m}.
\end{displaymath}
For small \(\dt\) we have \(1-\paccav = \erf\sqrt{kV\dt^6}\propto
\sqrt{V\dt^6}\) where \(k\) is a constant, and hence we must scale \(\dt\) so
as to keep \(V\dt^4/m^2\) fixed. Since the L2MC algorithm has a naive
dynamical critical exponent \(z=1\), this means that the cost should vary as
\(\cost\propto V(V\dt^4m^{-2})^{1/4} / m\dt = V^{5/4} m^{-3/2}\).

\section{Autocorrelations and Frequency of Measurement}
\label{sec:autocorrelations-and-frequency-of-measurement}

In this final section we perform an elementary analysis of the general problem
of determining the optimal frequency for making measurements of observables on
a Markov chain.

Suppose we wish to measure the expectation value \(\langle\Omega\rangle\) of
an operator \(\Omega\) by means of a Markov process (not necessarily HMC). If
the cost\footnote{\emph{I.e.}, the cost measured in units of computer time.}
of making one Markov step is \(\cost_S\), and the cost of making one
measurement of \(\Omega\) is \(\cost_M\), how often should we make
measurements in order to minimise the cost of the calculation? Due to the
presence of correlations between successive measurements, the answer is not
quite trivial.

Consider a sequence of \(N_u\) {\em uncorrelated} measurements of \(\Omega\).
The sample variance \(\bar V^u_\Omega\) is related to the intrinsic variance
\(V_\Omega\) in the distribution of \(\Omega\) in the usual way
\begin{equation}
  \bar{V}^u_\Omega \equiv
  \left\langle \left(\bar\Omega-\langle\bar\Omega\rangle\right)^2
  \right\rangle
  =
  { \left\langle \left(\Omega-\langle\Omega\rangle\right)^2
    \right\rangle               \over
    N_u}
  \equiv
  { V^u_\Omega \over N_u}. \label{eq:uncorr}
\end{equation}
For the general case of \(N_c\) successive correlated measurements, from
equation (\ref{eq:correlated-variance}), the sample variance \(\bar V_\Omega\)
is
\begin{equation}
  \bar{V}^u_\Omega \equiv
  \left\langle \left(\bar\Omega-\langle\bar\Omega\rangle\right)^2
  \right\rangle
  =  (1 + 2A_\Omega) \ 
  { \left\langle \left(\Omega-\langle\Omega\rangle\right)^2
    \right\rangle         \over
    N_c}
  =
  (1 + 2A_\Omega) \ { V_\Omega \over N_c}       \label{eq:corr}
\end{equation}
so that on average \(1+2A_\Omega\) correlated measurements are needed to
reduce the variance by the same amount as a single truly independent
measurement. If the cost of measuring \(\Omega\) is small (large), then it
should be beneficial to make more (less) than one measurement per
decorrelation time.

Let \(\steps\) be the total number of Markov steps required to generate one
independent sample, and let \(\gap\) be the number of Markov steps between
each measurement of \(\Omega\), so that the number of measurements performed
per independent sample is \(\steps/\gap\). If we wish to make sufficient
measurements to generate \(N\) independent samples, from
equations~(\ref{eq:uncorr}) and (\ref{eq:corr}) we have
\begin{equation}
  \frac{1 + 2A_\gap}{\steps/\gap} = 1, \label{eq:step-gap}
\end{equation}
where \(A_\gap\) is the integrated autocorrelation function for measurements
separated by \(\gap\) Markov steps. The total {\em cost} \(\cost\) of the
entire process (steps plus measurements) is clearly
\begin{equation}
  \cost = N \left(\cost_S \steps +
    \cost_M\frac{\steps}{\gap}\right).
      \label{eq:total-cost}
\end{equation}
Eliminating \(\steps\) from equations~(\ref{eq:step-gap}) and
(\ref{eq:total-cost}) we obtain
\begin{displaymath}
  \cost = N \left(\cost_S + \frac{\cost_M}{\gap}\right)
    \left(1 + 2A_\gap\right) \gap.
\end{displaymath}
The integrated autocorrelation function \(A_\gap\) can be written in terms of
the autocorrelation function \(C_\Omega(\gap l)\) of
section~(\ref{sec:simple-markov-processes}). Assuming
equation~(\ref{step-exp-autocorrelation-time}) we have
\begin{displaymath}
  A_\gap  = \sum_{l=1}^{\infty} C_\Omega(\gap l)
  = \sum_{l=1}^{\infty} e^{-Kl/\nexp}
  = \frac{1}{e^{K/\nexp} - 1}
\end{displaymath}
and so
\begin{displaymath}
  \cost = \left(\gap\cost_S + \cost_M\right)
  \left(
    \frac{e^{K/\nexp} + 1}{e^{K/\nexp} - 1}
  \right)\, .
\end{displaymath}
Minimising with respect to \(\gap\) gives
\begin{equation}
  {\rm sinh} x = x + \alpha
  \label{eq:optimal-gap}
\end{equation}
where
\begin{displaymath}
  x \equiv \frac{K}{\nexp}        \qquad  \mbox{and} \qquad
  \alpha \equiv \frac{\cost_M}{\gap\cost_S}.
\end{displaymath}
For small \(\alpha\) where measurements are ``cheap'' we find 
\begin{displaymath}
  K_{\rm opt} = \nexp \left( \frac{6\cost_M}{\cost_S\nexp}\right)^{1/3},
\end{displaymath}
while for large \(\alpha\) where they are ``expensive'', we obtain
\begin{displaymath}
  K_{\rm opt} = \nexp \ln \left( \frac{2\cost_M}{\cost_S\nexp}\right),
\end{displaymath}
the slow logarithmic increase of \(\gap_{\rm opt}\) being due to the
exponential decay in the autocorrelation function \(C_\Omega(l)\). The
crossover point, \(\gap=1\), occurs when \(\alpha\approx0.1752\).

\section{Conclusions}
\label{sec:conclusions}

We have introduced some powerful techniques for analysing a wide class of
algorithms for free field theory. Whereas inventing algorithms which are
efficacious for free field theory but useless for interesting ones is a
popular but fruitless exercise, the analysis of generic algorithms for free
fields is apparently very informative. The reason that this is so is because
our algorithms are sufficiently dumb that they spent most of their time in
dealing with the trivial almost free ``modes'' of the system.

As evidence of the more general applicability of our analysis we point to the
excellent agreement with empirical Monte Carlo data of the \(\erfc\) form for
the Metropolis acceptance probability as a function of integration step size
\cite{gupta90a,Joo:1998ib}. Futhermore preliminary results for simple models
\cite{kennedy95a} indicate that their autocorrelations have at least the same
form as expected from our free field theory results.

Perhaps the most surprising result is the failure of the L2MC (Kramers)
algorithm to have superior performance to the HMC algorithm. Despite the fact
that the noise is put into the Markov process more smoothly the desirable
properties of the L2MC algorithm are upset by its ``Zitterbewegung'' caused
by its rare Metropolis rejections.

It is also somewhat unexpected that the HMC algorithm performs nearly as well
as the full GHMC algorithm, of which it is a special case, when the parameters
of both are chosen optimally. The broad minimum of the costs of these
algorithms as a function of acceptance rate has the pleasant consequence that
no very fine tuning of parameters is required.

The result that the optimal HMC trajectory length is about the same as the
correlation length of the underlying field theory is significant, in that it
indicates that the temptation to use shorter, and hence cheaper, trajectories
for systems near criticality should be avoided.

The results pertaining to higher order (Campostrini) integrators
\cite{campostrini89a,creutz89a} are of theoretical interest, but in practice
they do not seem to be very useful because the trajectories for interacting
field theories are chaotic \cite{jansen96b,kennedy96a} and apparently limited
by the intrinsic instability of the integrators \cite{Joo:2000dh}.

\section*{Acknowledgements}

We gratefully acknowledge financial support from PPARC under grant number
GR/L22744. This research was also supported in part by the U.S. Department of
Energy through Contract Nos. DE--FG05--92ER40742 and DE--FC05--85ER250000.

We would like to thank David Daniel, Robert Edwards, Philippe de Forcrand,
Alan Horowitz, Ivan Horv\'ath, Alan Irving, B\'alint Jo\'o, Julius Kuti,
Steffen Meyer, Hidetoshi Mino, Stephen Pickles, Jim Sexton, Stefan Sint, Alan
Sokal, and Zbyszek Sroczynski for useful discussions, comments, and
hospitality during the extremely prolonged gestation period of this paper.

\newpage
\appendix
\section{Inverse Laplace Transforms\\ for Autocorrelation Functions}
\label{sec:inverse-laplace-transforms}

\subsection{Exponentially-Distributed Trajectory Lengths}

In this case the Laplace transform of the autocorrelation function, \(F(\beta)
= \LT C(\t)\), is a rational function with the numerator of lower degree than
the denominator. If the denominator is square free then we have the partial
fraction expansion
\begin{displaymath}
  F = \sum_k {A_k\over\beta-B_k}.
\end{displaymath}
Since
\begin{displaymath}
  \LT e^{-\alpha\t} \equiv \int_0^\infty d\t\, e^{-\beta\t} e^{-\alpha\t}
    = {1\over\alpha+\beta} \qquad (\Re\alpha+\beta>0),
\end{displaymath}
we have
\begin{displaymath}
  \LT^{-1} F = \sum_k A_k e^{B_k\t}.
\end{displaymath}
The roots \(B_k\) of the denominator of \(F\) come in complex conjugate pairs
because the coefficients in \(F\) are real. The autocorrelation function can
therefore always be written as a sum of real exponentials (corresponding to
the real roots) and of real exponentials multiplied by cosines (corresponding
to pairs of complex conjugate roots).

If the denominator of \(F\) is not square free then the repeated roots give
terms of the form \(A/(\beta-B)^n\) in the partial fraction expansion.
Consider then
\begin{displaymath}
  \LT \t^{n-1} e^{-\alpha\t}
    \equiv \int_0^\infty d\t\, e^{-\beta\t} \t^{n-1} e^{-\alpha\t}
    = (\alpha+\beta)^{-n} \Gamma(n) \qquad (\Re\alpha+\beta>0),
\end{displaymath}
This serves to give the inverse Laplace transform in the general case.

\subsection{Fixed Length Trajectories}

Here \(F\) is a rational function of \(\zeta\equiv e^{-\beta\trjlen}\) with
the numerator being of lower or equal degree than the denominator. For
simplicity we first consider the case where the denominator is square free,
whence by partial fractions
\begin{displaymath}
  F = C + \sum_k {A_k\over\zeta-B_k}.
\end{displaymath}
Observe that
\begin{displaymath}
  \LT \delta(\t) \equiv \int_0^\infty e^{-\beta\t} \delta(\t) = 1,
\end{displaymath}
and, more generally,
\begin{displaymath}
  \LT \sum_{j=0}^\infty \mu^j \delta(\t-\trjlen j) = \sum_{j=0}^\infty \mu^j
  e^{-\beta\trjlen j} = {1\over1-\mu e^{-\beta\trjlen}} \qquad (|\mu|<1),
\end{displaymath}
so we have
\begin{eqnarray*}
  \LT^{-1} F &=& C\delta(\t) - \sum_k {A_k\over B_k}
    \sum_{j=0}^\infty B_k^{-j} \delta(\t-\trjlen j) \\
  &=& C\delta(\t) - \sum_k A_k \sum_{j=0}^\infty
    e^{-(j+1)\ln B_k} \delta(\t-\trjlen j).
\end{eqnarray*}
All the the roots must satisfy \(|B_k|>1\) for the geometric series to
converge at \(\beta=0\quad(\zeta=1)\), which is necessary for the integrated
autocorrelation function \(F(0)\) to be finite.

In the general case where the denominator of \(F\) has multiple roots the
general form of the inverse Laplace transform may be obtained from the
identity
\begin{displaymath}
  \LT \sum_{j=0}^\infty {n+j-1\choose j} \mu^j \delta(\t-\trjlen j)
    = (1 - \mu\zeta)^{-n} \qquad (|\mu\zeta|<1).
\end{displaymath}

\subsection{Example of Computation of Autocorrelation Function}

To illustrate the computation of autocorrelation functions we shall explicitly
evaluate the connected autocorrelation function for \(M^2\) for the HMC
algorithm with exponentially distributed trajectory lengths. The Laplace
transform of the autocorrelation function is
\begin{displaymath} %% include ac-f.tex
  F(\beta) =
  {\frac {{{r}}^{3}+2\,{{r}}^{2}\beta+{r}\,{\beta}^{2}+2\,{{m}}^{2}{r}}{
  {\beta}^{3}+2\,{r}\,{\beta}^{2}+\left ({{r}}^{2}+4\,{{m}}^{2}\right )
  \beta+2\,{{m}}^{2}{r}}}.
\end{displaymath}
If we write this in terms of two distinct roots \(\beta_1\) and \(\beta_2\) of
the denominator
\begin{displaymath} %% include ac-ff.tex
  F(\beta) =
  {\frac {{r}\,\left ({{r}}^{2}+2\,{r}\,\beta+{\beta}^{2}+2\,{{m}}^{2}
  \right )}{\left (\beta-\beta_{{1}}\right )\left (\beta-\beta_{{2}}
  \right )\left (2\,{r}+\beta_{{1}}+\beta_{{2}}+\beta\right )}}
\end{displaymath}
we can expand this in terms of partial fractions to give
\begin{eqnarray*} %% include ac-pff.tex
  F(\beta) &=&
  {\frac {{r}\,\left (2\,{{r}}^{4}+4\,{{r}}^{3}\beta_{{1}}+\left (2\,{
  \beta_{{1}}}^{2}-{{m}}^{2}\right ){{r}}^{2}+\beta_{{1}}{r}\,{{m}}^{2}+
  32\,{{m}}^{4}+4\,{\beta_{{1}}}^{2}{{m}}^{2}\right )}{\left (\beta-
  \beta_{{1}}\right )\left (2\,{{r}}^{4}-13\,{{m}}^{2}{{r}}^{2}+64\,{{m}
  }^{4}\right )}} \\
  &+&{\frac {{r}\,\left (
  \begin{array}{c}
    -4\,{{r}}^{3}\beta_{{1}}+\left (-2
    \,{\beta_{{1}}}^{2}-13\,{{m}}^{2}-2\,\beta_{{2}}\beta_{{1}}\right ){
    {r}}^{2} \\
    +\left (-7\,\beta_{{2}}-8\,\beta_{{1}}\right ){{m}}^{2}{r}+16
    \,{{m}}^{4} +\left (-4\,{\beta_{{1}}}^{2}-4\,\beta_{{2}}\beta_{{1}}
    \right ){{m}}^{2}
  \end{array}
  \right )}{\left (\beta-\beta_{{2}}\right )\left (2\,{
  {r}}^{4}-13\,{{m}}^{2}{{r}}^{2}+64\,{{m}}^{4}\right )}} \\
  &+&{\frac {{r}\,
  \left (\left (2\,\beta_{{2}}\beta_{{1}}+{{m}}^{2}\right ){{r}}^{2}+
  \left (7\,\beta_{{2}}+7\,\beta_{{1}}\right ){{m}}^{2}{r}+16\,{{m}}^{4}
  +4\,\beta_{{2}}\beta_{{1}}{{m}}^{2}\right )}{\left (2\,{r}+\beta_{{1}}
  +\beta_{{2}}+\beta\right )\left (2\,{{r}}^{4}-13\,{{m}}^{2}{{r}}^{2}+
  64\,{{m}}^{4}\right )}}.
\end{eqnarray*}
The inverse Laplace transform of this is immediately obvious
\begin{eqnarray*} %% include ac-ac.tex
  &&
  {\frac {{r}\,\left (2\,{{r}}^{4}+4\,{{r}}^{3}\beta_{{1}}+\left (2\,{
  \beta_{{1}}}^{2}-{{m}}^{2}\right ){{r}}^{2}+\beta_{{1}}{r}\,{{m}}^{2}+
  32\,{{m}}^{4}+4\,{\beta_{{1}}}^{2}{{m}}^{2}\right ){e^{\beta_{{1}}t}}}
  {2\,{{r}}^{4}-13\,{{m}}^{2}{{r}}^{2}+64\,{{m}}^{4}}} \\
  &&+{\frac {{r}\,
  \left (
  \begin{array}{c}
    -4\,{{r}}^{3}\beta_{{1}}+\left (-2\,{\beta_{{1}}}^{2}-13\,{{m}}
    ^{2}-2\,\beta_{{2}}\beta_{{1}}\right ){{r}}^{2} \\
    +\left (-7\,\beta_{{2}}
    -8\,\beta_{{1}}\right ){{m}}^{2}{r}+16\,{{m}}^{4}+ \left (-4\,{\beta_{{
    1}}}^{2}-4\,\beta_{{2}}\beta_{{1}}\right ){{m}}^{2}
  \end{array}
  \right ){e^{\beta_{
  {2}}t}}}{2\,{{r}}^{4}-13\,{{m}}^{2}{{r}}^{2}+64\,{{m}}^{4}}} \\
  &&+{\frac {
  {r}\,\left (
  \begin{array}{c}
    \left (2\,\beta_{{2}}\beta_{{1}}+{{m}}^{2}\right ){{r}}^{2
    }+\left (7\,\beta_{{2}}+7\,\beta_{{1}}\right ){{m}}^{2}{r} \\
    +16\,{{m}}^{
    4}+4\,\beta_{{2}}\beta_{{1}}{{m}}^{2}
  \end{array}
  \right ){e^{-\left (2\,{r}+\beta_
  {{1}}+\beta_{{2}}\right )t}}}{2\,{{r}}^{4}-13\,{{m}}^{2}{{r}}^{2}+64\,
  {{m}}^{4}}}.
\end{eqnarray*}
\begin{figure}
  \begin{center}
  \epsfxsize=0.95\textwidth \leavevmode\epsffile{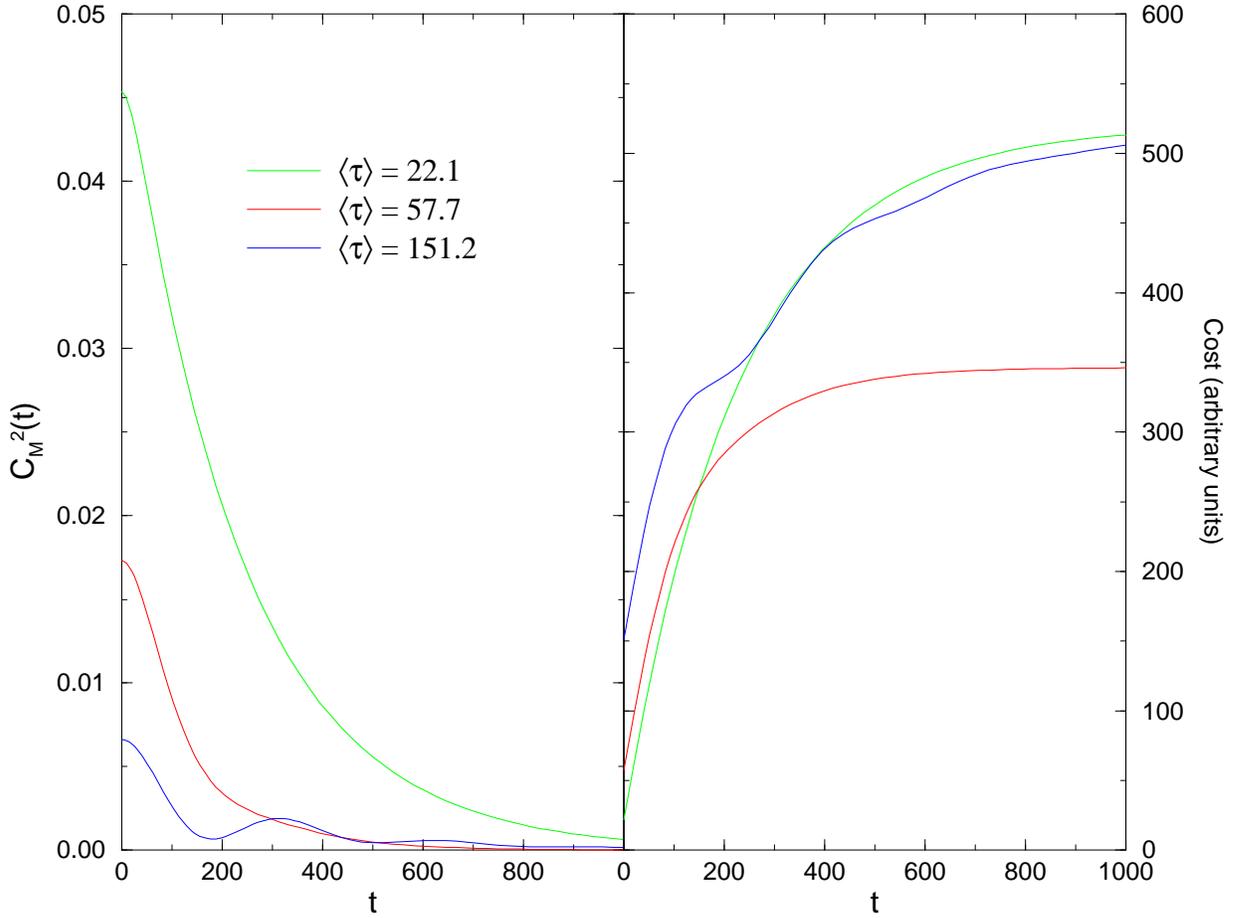}
  \end{center}
  \caption[ac-plot]{The autocorrelation function and the cumulative cost are
    shown as functions of MD time for the optimal trajectory length and those
    costing 50\% more.}
  \label{fig:ac-plot}
\end{figure}
In figure \ref{fig:ac-plot} we show this function for \(m=10^{-2}\) and the
optimal value \(\opt r=\sqrt3m\), together with the two neighbouring values of
\(r=(3\sqrt3\pm\sqrt{15})m/2\) for which the cost is 50\% greater.

\section{Evaluation of Spectral Averages}
\label{sec:spectral-sums}

\def\G{{\cal G}}                % Abstract group
\def\rep{\phi_q}                % Representation of \G

In this section we show how to evaluate the following spectral sums for
one-dimensional free field theory:
\begin{eqnarray}
  \sigma_A^{(\alpha)}
    &\equiv& {1\over V} \sum_{p\in\Z_V}
      \sin^{2\alpha}\left({\pi p\over V}\right)
  \label{eq:A} \\
  &=& {1\over2^{2\alpha}}
    \left\{
      {2\alpha\choose\alpha}
      + \sum_{k=1}^{\lfloor\alpha/V\rfloor} (-)^{kV}
        \left[{2\alpha\choose\alpha-kV} + {2\alpha\choose\alpha+kV}\right]
    \right\},
  \label{A'} \\
  \sigma_{m^2}^{(\alpha)}
    &\equiv& {1\over V} \sum_{p\in\Z_V} \omega_p^{2\alpha}
  \label{B} \\
  &=& \sum_{\beta=0}^\alpha {\alpha\choose\beta}
    m^{2(\alpha-\beta)} \left\{ {2\beta\choose\beta} +
      2\sum_{k=1}^{\lfloor\beta/V\rfloor}(-)^{kV}
        {2\beta\choose\beta-kV}\right\},
  \label{B'} \\
  \tilde\sigma_{m^2}
    &\equiv& {1\over V} \sum_{p\in\Z_V} \cos(2\trjlen\omega_p)
  \label{C} \\
  &=& \sum_{\beta=0}^\infty {(-2m^2)^\beta\over\beta!}
  \left(\prod_{\gamma=0}^{\beta-1}\tau_\gamma
    \int^{\tau_\gamma}d\tau_{\gamma+1}\right) \sum_{k\in\Z} \J{2kV}(4\trjlen),
  \label{C'} \\
  \bar\sigma_{m^2}^{(\alpha)}
    &\equiv& {1\over V} \sum_{p\in\Z_V} \omega_p^{2\alpha}
      \sin^2(\trjlen\omega_p)
  \label{D} \\
  &=& {1\over2}\sigma_{m^2}^{(\alpha)}
    - {1\over2}\left(-{1\over4}{d^2\over d\trjlen^2}\right)^\alpha
      \tilde\sigma_{m^2},
  \label{D'}
\end{eqnarray}
where
\begin{equation}
  \omega_p^2 \equiv m^2 + 4 \sin^2\left({\pi p\over V}\right).
  \label{freq}
\end{equation}

\subsection{Completeness relations}

Let \(\G\) be a group with a family of \(D\)-dimensional representations
\(\rep:\G\to(\C^D\to\C^D)\), {\em i.e.,} \(\rep(g):\C^D\to\C^D\), labelled by
some parameter \(q\); thus \(\rep(gh) = \rep(g)\rep(h)\) for \(g,h\in\G\).
Summing over group elements (or integrating with respect to Haar measure in
the continuous case) gives
\begin{displaymath}
  \sigma_q \equiv \int_\G d\mu\, \rep(g),
\end{displaymath}
and multiplying by \(\rep(h)\) for any \(h\in\G\) we find
\begin{eqnarray*}
  \rep(h)\sigma_q &=& \rep(h) \int_\G d\mu\, \rep(g)
    = \int_\G d\mu\, \rep(h)\rep(g) \\
  &=& \int_\G d\mu\, \rep(hg) = \int_\G d\mu\, \rep(g') = \sigma_q
\end{eqnarray*}
where we have made use of the (left) invariance of Haar measure \(d\mu\).
Hence either \(\rep(h)=1\quad\forall h\in\G\) or \(\sigma_q=0\). If we let
\(q=0\) label the identity representation of \(\G\), and all other \(\rep\) be
non-identity representations, we have
\begin{displaymath}
\sigma_q = \delta(q).
\end{displaymath}

\begin{example}
  \label{ZN-ZN-completeness}
  Let \(\G=\Z_V\), \(q\in\Z_V\), and \(\rep(p)=e^{2\pi ipq/V}\):
  \begin{displaymath}
    {1\over V} \sum_{p\in\Z_V} e^{2\pi ipq/V} = \delta_{q,0};
  \end{displaymath}
\end{example}

\begin{example}
  \label{Z-S1-completeness}
  Let \(\G=\Z\), \(q\in\S_1=[0,2\pi]\), and \(\rep(p)=e^{ipq}\):
  \begin{displaymath}
    {1\over2\pi} \sum_{p\in\Z} e^{ipq} = \delta(q);
  \end{displaymath}
\end{example}

\begin{example}
  \label{S1-Z-completeness}
  Let \(\G=\S_1\), \(q\in\Z\), and \(\rep(p)=e^{ipq}\):
  \begin{displaymath}
    {1\over2\pi} \int_0^{2\pi} dp\, e^{ipq} = \delta_{q,0}.
  \end{displaymath}
\end{example}

\subsection{A Poisson Resummation Formula}
\label{subsec-a-poisson-resummation-formula}

Let \(f:\S_1\to\S_1\) be a periodic function. As it is periodic it obviously
wants to be expanded in eigenfunctions of \(\S_1\):
\begin{eqnarray*}
  f(p) &=& \int_0^{2\pi} dp'\, \delta(p'-p)f(p') \qquad[p\in\S_1] \\
  &=& \int_0^{2\pi} dp'\, {1\over2\pi} \sum_{q\in\Z} e^{iq(p'-p)}f(p') \\
  &=& \sum_{q\in\Z} e^{-iqp} {1\over2\pi} \int_0^{2\pi} dp'\, e^{iqp'}f(p'),
\end{eqnarray*}
hence the ``spectral average''
\begin{eqnarray*}
  \sigma(f) &=& {1\over V} \sum_{p\in\Z_V} f\left({2\pi p\over V}\right) \\
  &=& {1\over V} \sum_{p\in\Z_V} \sum_{q\in\Z} e^{-2\pi ipq/V}
    {1\over2\pi} \int_0^{2\pi} dp'\, e^{iqp'}f(p') \\
  &=& \sum_{q\in\Z}
    \Biggl\{{1\over V}\sum_{p\in\Z_V} e^{-2\pi ipq/V}\Biggr\}
    \Biggl\{{1\over2\pi}\int_0^{2\pi} dp'\, e^{iqp'}f(p')\Biggr\} \\
  &=& \sum_{q\in\Z}
    \Biggl\{\sum_{k\in\Z}\delta_{q,kV}\Biggr\}
    \Biggl\{{1\over2\pi}\int_0^{2\pi} dp'\, e^{iqp'}f(p')\Biggr\} \\
  &=& \sum_{k\in\Z} {1\over2\pi} \int_0^{2\pi} dp'\, e^{ip'kV}f(p').
\end{eqnarray*}
As \(f\) is real valued we may take the real part of this identity to obtain
\begin{equation}
  \sigma(f) \equiv {1\over V} \sum_{p\in\Z_V} f\left({2\pi p\over V}\right)
    = \sum_{k\in\Z} {1\over2\pi} \int_0^{2\pi} dp'\, \cos(p'kV)f(p').
  \label{poisson-resummation-formula}
\end{equation}

\subsection{Multiple Angle Expansion}

The following simple identity is often useful
\begin{eqnarray}
  \sin^{2\alpha}\theta &=& \left[{1\over2i}
      \left(e^{i\theta}-e^{-i\theta}\right)
    \right]^{2\alpha} \qquad(\alpha\in\N)
    \nonumber \\
  &=& \left({1\over2i}\right)^{2\alpha}
    \sum_{q=0}^{2\alpha} {2\alpha\choose q}(-)^q e^{2i(\alpha-q)\theta}
    \nonumber \\
  &=& \left({1\over2i}\right)^{2\alpha}
    \left\{ {2\alpha\choose\alpha}(-)^\alpha +
      \sum_{q=0}^{\alpha-1} {2\alpha\choose q}(-)^q 2\cos[2(\alpha-q)\theta]
    \right\}
    \nonumber \\
  &=& {1\over2^{2\alpha}}
    \left\{ {2\alpha\choose\alpha} +
      \sum_{q'=1}^\alpha {2\alpha\choose\alpha-q'}(-)^{q'} 2\cos(2q'\theta)
    \right\}.
  \label{**}
\end{eqnarray}

\subsection{Free-field Spectral Sums}
\label{sec-free-field-spectral-sums}

Let us first consider equation (\ref{eq:A}). From equation
(\ref{poisson-resummation-formula}) we have
\begin{eqnarray*}
  && {1\over V} \sum_{p\in\Z_V}\cos\left({2\pi pq'\over V}\right)
    = \sum_{k\in\Z} {1\over2\pi} \int_0^{2\pi} dp'\,\cos(p'kV)\cos(p'q') \\
    &&\qquad = \sum_{k_in\Z} {1\over2\pi} \int_0^{2\pi} {1\over2}
      \Bigl\{\cos[p'(kV+q')] + \cos[p'(kV-q')]\Bigr\} \\
    &&\qquad = \sum_{k\in\Z} {1\over2} \Bigl\{\delta_{kV+q',0}
      + \delta_{kV-q',0}\Bigr\}.
\end{eqnarray*}
Using the multiple-angle expansion (\ref{**}) we obtain
\begin{eqnarray*}
  \sigma_A^{(\alpha)}
    &\equiv& {1\over V} \sum_{p\in\Z_V}
      \sin^{2\alpha}\left({\pi p\over V}\right) \\
    &=& {1\over V} \sum_{p\in\Z_V} {1\over2^{2\alpha}}
      \left\{
        {2\alpha\choose\alpha}
        + \sum_{q'=1}^\alpha {2\alpha\choose\alpha-q'}
          (-)^{q'} 2\cos\left({2\pi pq'\over V}\right)
      \right\} \\
    &=& {1\over2^{2\alpha}}
      \left\{
        {2\alpha\choose\alpha}
        + \sum_{q'=1}^\alpha {2\alpha\choose\alpha-q'} (-)^{q'}
          {2\over V} \sum_{p\in\Z_V} \cos\left({2\pi pq'\over V}\right)
      \right\} \\
    &=& {1\over2^{2\alpha}}
      \left\{
        {2\alpha\choose\alpha}
        + \sum_{q'=1}^\alpha {2\alpha\choose\alpha-q'} (-)^{q'}
          \sum_{k\in\Z} \left(\delta_{kV+q',0} + \delta_{kV-q',0}\right)
      \right\} \\
    &=& {1\over2^{2\alpha}}
      \left\{
        {2\alpha\choose\alpha}
        + \sum_{k=1}^{\lfloor\alpha/V\rfloor} (-)^{kV}
          \left[{2\alpha\choose\alpha-kV} + {2\alpha\choose\alpha+kV}\right]
      \right\}.
\end{eqnarray*}
From this result we may obtain an expression for the quantity
\(\sigma_{m^2}^{(\alpha)}\) defined in equation~(\ref{B})
\begin{eqnarray*}
  \sigma_{m^2}^{(\alpha)}
    &\equiv& {1\over V} \sum_{p\in\Z_V} \omega_p^{2\alpha}
      = {1\over V} \sum_{p\in\Z_V} \left[m^2+4\sin^2
        \left({\pi p\over V}\right)\right]^\alpha \\
    &=& \sum_{\beta=0}^\alpha {\alpha\choose\beta}
       {1\over V} \sum_{p\in\Z_V} m^{2(\alpha-\beta)}4^\beta\sin^{2\beta}
        \left({\pi p\over V}\right) \\
    &=& \sum_{\beta=0}^\alpha {\alpha\choose\beta}
      m^{2(\alpha-\beta)}4^\beta\sigma_A^{(\beta)} \\
    &=& \sum_{\beta=0}^\alpha {\alpha\choose\beta}
      m^{2(\alpha-\beta)} \left\{ {2\beta\choose\beta} +
        2\sum_{k=1}^{\lfloor\beta/V\rfloor}(-)^{kV}
          {2\beta\choose\beta-kV}\right\}.
\end{eqnarray*}
\begin{example}
  \begin{displaymath}
    \sigma_{m^2}^{(2)} = {1\over V} \sum_{p\in\Z_V} \omega_p^4
      = \left(6+4m^2+m^4\right) + \delta_{V,1}\left(-6-4m^2\right)
        + \delta_{V,2}\cdot2;
  \end{displaymath}
\end{example}
\begin{example}
  \begin{eqnarray*}
    \sigma_{m^2}^{(3)} &=& {1\over V} \sum_{p\in\Z_V} \omega_p^6
      = \left(m^6+6m^4+18m^2+20\right) \\
        &&\qquad + \delta_{V,1} \left(-6m^4-18m^2-20\right)
        + \delta_{V,2} \left(6m^2+12\right) - \delta_{V,3}\cdot2.
  \end{eqnarray*}
\end{example}
In order to evaluate Equation~(\ref{C}) we first investigate the simpler case
where \(m=0\)
\begin{eqnarray*}
  \tilde\sigma_0 &\equiv& {1\over V} \sum_{p\in\Z_V} \cos(2\trjlen\omega_p) \\
  &=& \sum_{k\in\Z} {1\over2\pi} \int_0^{2\pi}dp'\,\cos(p'kV)
    \cos\left[4\trjlen\sin\left({p'\over2}\right)\right] \\
  &=& \sum_{k\in\Z} {1\over\pi} \int_0^\pi d\theta\,\cos(2kV\theta)
    \cos[4\trjlen\sin\theta] \\
  &=& \sum_{k\in\Z} \J{2kV}(4\trjlen).
\end{eqnarray*}
For small \(m\) we Taylor expand \(\tilde\sigma_{m^2}\)
\begin{displaymath}
  \tilde\sigma_{m^2} = \sum_{\beta=0}^\infty {m^{2\beta}\over\beta!}
    \left.{d^\beta\tilde\sigma_{m^2}\over d(m^2)^\beta}\right|_{m^2=0}.
\end{displaymath}
Consider the quantity
\begin{eqnarray*}
  -{1\over2}{d\over dm^2}{d\over d\trjlen}{\tilde\sigma_{m^2}\over\trjlen}
    &=& -{1\over2}{d\over d\trjlen}\left({1\over\trjlen}{d\over dm^2}
      {1\over V}\sum_{p\in\Z_V}\cos(2\trjlen\omega_p)\right) \\
    &=& {d\over d\trjlen}\left({1\over\trjlen}
      {1\over V}\sum_{p\in\Z_V}\sin(2\trjlen\omega_p)\trjlen{d\omega_p\over
        dm^2} \right) \\
    &=& {d\over d\trjlen}\left(
      {1\over V}\sum_{p\in\Z_V}\sin(2\trjlen\omega_p){1\over2\omega_p}
      \right) \\
    &=& {1\over V}\sum_{p\in\Z_V}\cos(2\trjlen\omega_p)
    = \tilde\sigma_{m^2};
\end{eqnarray*}
upon integrating with respect to \(\trjlen\) we obtain
\begin{displaymath}
    {d\tilde\sigma_{m^2}\over dm^2}
    = -2\tau\int^\trjlen d\tau'\,\tilde\sigma_{m^2},
\end{displaymath}
and hence
\begin{displaymath}
  \left.{d^\beta\tilde\sigma_{m^2}\over d(m^2)^\beta}\right|_{m^2=0} =
  (-2)^\beta \trjlen \int^\trjlen d\tau_1\,\tau_1\int^{\tau_1}d\tau_2\ldots
  \tau_{\beta-1}\int^{\tau_{\beta-1}}\,d\tau_\beta\tilde\sigma_0.
\end{displaymath}
By this means we have obtained the desired result
\begin{displaymath}
  \tilde\sigma_{m^2} = \sum_{\beta=0}^\infty {(-2m^2)^\beta\over\beta!}
    \left(\prod_{\gamma=0}^{\beta-1}\tau_\gamma
      \int^{\tau_\gamma}d\tau_{\gamma+1}\right) \tilde\sigma_0.
\end{displaymath}
Equation~(\ref{D}) may be found by differentiating \(\tilde\sigma_{m^2}\) with
respect to~\(\trjlen\),
\begin{eqnarray*}
  \bar\sigma_{m^2}^{(\alpha)}
    &\equiv& {1\over V}\sum_{p\in\Z_V}
      \omega_p^{2\alpha}\sin^2(\trjlen\omega_p) \\
    &=& {1\over2V}\sum_{p\in\Z_V}\omega_p^{2\alpha}
      \Bigl\{1-\cos(2\trjlen\omega_p)\Bigr\} \\
    &=& {1\over2}\sigma_{m^2}^{(\alpha)}
      - {1\over2}\left(-{1\over4}{d^2\over d\trjlen^2}\right)^\alpha
        \tilde\sigma_{m^2}.
\end{eqnarray*}
\begin{example}
  \begin{eqnarray}
    \bar\sigma_{m^2}^{(2)}
      &=& {1\over V} \sum_{p\in\Z_V} \omega_p^4\sin^2(\trjlen\omega_p)
        \nonumber \\
      &=& {1\over2}\sigma_{m^2}^{(2)} - {1\over2}\sum_{\beta=0}^\infty
        {m^{2\beta}\over\beta!}\left(-{1\over4}{d^2\over d\trjlen^2}\right)^2
        \left.{d^\beta\tilde\sigma_{m^2}\over d(m^2)^\beta}\right|_{m^2=0}
        \nonumber \\
      &=& {1\over2}\sigma_{m^2}^{(2)} - \sum_{k\in\Z} \sum_{\beta=0}^\infty
        {(-2m^2)^\beta\over2\beta!}
        \left(-{1\over4}{d^2\over d\trjlen^2}\right)^2 \times
        \nonumber \\
      &&\qquad\qquad\qquad \times \left(\prod_{\gamma=0}^{\beta-1}\tau_\gamma
          \int^{\tau_\gamma}d\tau_{\gamma+1}\right) \J{2kV}(4\trjlen).
    \label{sigma-bar-2}
  \end{eqnarray}
  The \(\beta=0\) term in the sum is
  \begin{displaymath}
    -{1\over2}\sum_{k\in\Z} \left(-{1\over4}{d^2\over d\trjlen^2}\right)^2
      \J{2kV}(4\trjlen) = -8 \sum_{k\in\Z} \J{2kV}''''(4\trjlen).
  \end{displaymath}
  From the recurrence relation
  \begin{displaymath}
    \J{n}'={1\over2}(\J{n-1}-\J{n+1})
  \end{displaymath}
  we find that
  \begin{displaymath}
    {d^k\J{n}(z)\over dz^k} = {1\over2^k} \sum_{\gamma=0}^k
      {k\choose\gamma}(-)^\gamma\J{n-k+2\gamma}(z),
  \end{displaymath}
  which we may prove by induction:
  \begin{eqnarray*}
    && {d^{k+1}\J{n}(z)\over dz^{k+1}}
      = {1\over2^k} \sum_{\gamma=0}^k
      {k\choose\gamma}(-)^\gamma{d\J{n-k+2\gamma}(z)\over dz} \\
    &&\qquad = {1\over2^{k+1}} \sum_{\gamma=0}^k
      {k\choose\gamma}(-)^\gamma
      \Bigl(\J{n-k+2\gamma-1}(z)-\J{n-k+2\gamma+1}(z)\Bigr) \\
    &&\qquad = {1\over2^{k+1}} \sum_{\gamma=0}^k
      {k\choose\gamma}(-)^\gamma
      \Bigl(\J{n-(k+1)+2\gamma}(z)-\J{n-(k+1)+2(\gamma+1)}(z)\Bigr) \\
    &&\qquad = {1\over2^{k+1}} \Biggl\{ \sum_{\gamma=1}^k
      \left[ {k\choose\gamma}+{k\choose\gamma-1} \right] (-)^\gamma
      \J{n-(k+1)+2\gamma}(z) + \\
    &&\qquad\qquad\qquad\qquad + \J{n-(k=1)}(z) - \J{n+(k+1)}(z) \Biggr\} \\
    &&\qquad = {1\over2^{k+1}} \sum_{\gamma=0}^{k+1}
      {k+1\choose\gamma}(-)^\gamma \J{n-(k+1)+2\gamma}(z),
  \end{eqnarray*}
  using \({k\choose\gamma}+{k\choose\gamma-1}={k+1\choose\gamma}\). We
  therefore find that
  \begin{eqnarray*}
      && -8\sum_{k\in\Z} \J{2kV}''''(4\trjlen)
        = -2\sum_{k\in\Z} \sum_{\gamma=0}^4 {4\choose\gamma}
          (-)^\gamma \J{2kV-4+2\gamma}(4\trjlen) \\
      &&\qquad = -2\sum_{k\in\Z}
        \Bigl\{\J{2kV-4}(4\trjlen)-4\J{2kV-2}(4\trjlen) + \\
      &&\qquad\qquad\qquad + 6\J{2kV}(4\trjlen)-4\J{2kV+2}(4\trjlen)+
        \J{2kV+4}(4\trjlen)\Bigr\} \\
      &&\qquad = -3\J0(4\trjlen) + 4\J2(4\trjlen) - \J4(4\trjlen) -\\
      &&\qquad\qquad\qquad - 2\sum_{k=1}^\infty \Bigl\{
        \J{2kV-4}(4\trjlen) - 4 \J{2kV-2}(4\trjlen) + \\
      &&\qquad\qquad\qquad\qquad + 6 \J{2kV}(4\trjlen) - 4 \J{2kV+2}(4\trjlen)
        + \J{2kV+4}(4\trjlen) \Bigr\}.
  \end{eqnarray*}
  The \(\beta=1\) term in the sum~(\ref{sigma-bar-2}) is
  \begin{eqnarray*}
    && -{1\over2} \sum_{k\in\Z} -2m^2
      \left(-{1\over4}{d^2\over d\trjlen^2}\right)^2\trjlen
        \int^\trjlen d\tau_1\,\J{2kV}(4\tau_1) \\
    &&\qquad = m^2 \sum_{k\in\Z} {d^4\over d\hat{\trjlen}^4}
      \cdot \hat{\trjlen}\int^{\hat{\trjlen}} d\trjlen'\,\J{2kV}(\trjlen')
      \qquad ({\rm where\ } \hat{\trjlen} \equiv 4\trjlen)\\
    &&\qquad = m^2 \sum_{k\in\Z} {d^3\over d\hat{\trjlen}^3}
      \left\{ \int^{\hat{\trjlen}} d\trjlen'\,\J{2kV}(\trjlen') +
        \hat{\trjlen}\J{2kV}(\hat{\trjlen}) \right\} \\
%    &&\qquad = m^2 \sum_{k\in\Z}
%      \Bigl\{ 2\J{2kV}''(\hat{\trjlen}) +
%        \left(\hat{\trjlen}\J{2kV}'(\hat{\trjlen})
%        \right)'' \Bigr\} \\
      &&\qquad = m^2 \sum_{k\in\Z}
      \Bigl\{ 4\J{2kV}''(\hat{\trjlen})
      + \hat{\trjlen}\J{2kV}'''(\hat{\trjlen}) \Bigr\} \\ 
    &&\qquad = m^2 \sum_{k\in\Z} \Biggl\{
      \sum_{\gamma=0}^2
        {2\choose\gamma}(-)^\gamma\J{2kV-2+2\gamma}(\hat{\trjlen}) + \\
    &&\qquad\qquad\qquad
      + {\hat{\trjlen}\over8} \sum_{\gamma=0}^3 {3\choose\gamma} (-)^\gamma
        \J{2kV-3+2\gamma}(\hat{\trjlen}) \Biggr\} \\
    &&\qquad = m^2 \sum_{k\in\Z} \Bigl\{
      \J{2kV-2}(\hat{\trjlen}) - 2\J{2kV}(\hat{\trjlen}) +
        \J{2kV+2}(\hat{\trjlen}) + \\ 
    &&\qquad\qquad
      + {\hat{\trjlen}\over8}\J{2kV-3}(\hat{\trjlen})
      - {3\hat{\trjlen}\over8}\J{2kV-1}(\hat{\trjlen})
      + {3\hat{\trjlen}\over8}\J{2kV+1}(\hat{\trjlen})
      - {\hat{\trjlen}\over8}\J{2kV+3}(\hat{\trjlen}) \Bigr\} \\
    &&\qquad = m^2 \Biggl\{
      -2\J0(4\trjlen) + 2\J2(4\trjlen) + 3\trjlen\J1(4\trjlen)
      - \trjlen\J3(4\trjlen) + \\
    &&\qquad\qquad
      + \sum_{k=1}^\infty \bigl(
        2\J{2kV-2}(4\trjlen) - 4\J{2kV}(4\trjlen) + 2\J{2kV+2}(4\trjlen) + \\
    &&\qquad\qquad\qquad
        + \trjlen\J{2kV-3}(4\trjlen) - 3\trjlen\J{2kV-1}(4\trjlen) + \\
    &&\qquad\qquad\qquad
        + 3\trjlen\J{2kV+1}(4\trjlen) - \trjlen\J{2kV+3}(4\trjlen) \bigr)
      \Biggr\}.
  \end{eqnarray*}
\end{example}

\subsection{Results for Campostrini}

\begin{eqnarray*}  %% acc-sigma4.tex
  \bar\sigma^{(4)}_{m^2}
  &=& \Bigl[35 - 35\1{0} + 56\1{2} - 28\1{4} + 8\1{6} - \1{8}\Bigr] \\
  &+& \Bigl[40 - 40\1{0} + 64t\1{1} + 31\1{2} - 8\1{4} + \1{6}\Bigr] m^2 \\
  &+& \Bigl[18 + (16t^2-18)\1{0} + 48t\1{1} + 8\1{2} - \1{4}\Bigr] m^4
    + O(m^6), \\ [1ex]
%% acc-sigma6.tex
  \bar\sigma^{(6)}_{m^2}
  &=& \left[
    \begin{array}{c}
      462 - 462\1{0} + 792\1{2} - 495\1{4} \\
      + 220\1{6} - 66\1{8} + 12\1{10} - \1{12}
    \end{array}
    \right] \\
  &+& \left[
    \begin{array}{c}
      756 - 756\1{0} + 1024t\1{1} + 698\1{2} \\
      - 256\1{4} + 69\1{6} - 12\1{8} + \1{10}
    \end{array}
    \right] m^2 \\
  &+& \left[
    \begin{array}{c}
      525 + (256t^2 - 525)\1{0} + 1152t\1{1} \\
      + 316\1{2} - 74\1{4} + 12\1{6} - \1{8}
    \end{array}
    \right]m^4 + O(m^6), \\ [1ex]
%% acc-sigma8.tex
  \bar\sigma^{(8)}_{m^2}
  &=& \left[
    \begin{array}{c}
      6435 - 6435\1{0} + 11440\1{2} - 8008\1{4} \\
      + 4368\1{6} - 1820\1{8} + 560\1{10} \\
      - 120\1{12} + 16\1{14} - \1{16}
    \end{array}
    \right] \\
  &+& \left[
    \begin{array}{c}
      13728 - 13728\1{0} + 16384t\1{1} \\
      + 14075\1{2} - 6128\1{4} + 2185\1{6} \\
      - 608\1{8} + 123\1{10} - 16\1{12} + \1{14}
    \end{array}
    \right] m^2 \\
  &+& \left[
    \begin{array}{c}
      12936 + (4096t^2-12936)\1{0} \\
      + 24576t\1{1} + 9280\1{2} - 2807\1{4} \\
      + 688\1{6} - 128\1{8} + 16\1{10} - \1{12}
    \end{array}
    \right] m^4
    + O(m^6), \\ [1ex]
%% acc-sigma10.tex
  \bar\sigma^{(10)}_{m^2}
  &=& \left[
    \begin{array}{c}
      92378 - 92378\1{0} + 167960\1{2} \\
      - 125970\1{4} + 77520\1{6} - 38760\1{8} \\
      + 15504\1{10} - 4845\1{12} + 1140\1{14} \\
      - 190\1{16} + 20\1{18} - \1{20}
    \end{array}
    \right] \\
  &+& \left[
    \begin{array}{c}
      243100 - 243100\1{0} + 262144t\1{1} \\
      + 267814\1{2} - 129872\1{4} + 54252\1{6} \\
      - 19024\1{8} + 5420\1{10} - 1200\1{12} \\
      + 193\1{14} - 20\1{16} + \1{18}
    \end{array}
    \right] m^2 \\
  &+& \left[
    \begin{array}{c}
      289575 + (65536t^2-289575)\1{0} \\
      + 491520t\1{1} + 233612\1{2} - 82714\1{4} \\
      + 25164\1{6} - 6392\1{8} + 1300\1{10} \\
      - 198\1{12} + 20\1{14} - \1{16}
    \end{array}
    \right] m^4 + O(m^6).
\end{eqnarray*}

\section{Spectral sum for one dimensional free field theory}
\label{spectral-sum-for-one-dimensional-free-field-theory}

We wish to evaluate the spectral sum
\begin{displaymath}
  \sigma^{(-1)}_{m^2} \equiv {1\over V} \sum_{p\in\Z_V} \omega_p^{-2}
\end{displaymath}
for large \(V\). Using the Poisson resummation formula we obtain
\begin{displaymath}
  \sigma^{(-1)}_{m^2} = \sum_{k\in\Z} {1\over2\pi}
      \int_0^{2\pi} dp'\,\cos(p'kV) \, \omega_{Vp'/2\pi}^{-2}
    = \sum_{k\in\Z} {1\over2\pi} \int_0^{2\pi} dp'\,
      {\cos(p'kV) \over m^2 + 4\sin^2(\half p')}.
\end{displaymath}
Upon substituting \(z\equiv e^{ip'}\) we get
\begin{eqnarray*}
  \sigma^{(-1)}_{m^2} &=& \sum_{k\in\Z} {i\over4\pi}
    \oint_{|z|=1} dz\, {z^{kV} + z^{-kV} \over z^2 - (m^2+2)z + 1} \\
  &=& \sum_{k\in\Z} {i\over2\pi}
    \oint_{|z|=1} dz\, {z^{|k|V} \over (z-\mu_+)(z-\mu_-)}
  = {1\over m\sqrt{m^2+4}} \left[{1+\mu_-^V\over1-\mu_-^V}\right],
\end{eqnarray*}
where \(\mu_\pm\equiv1 + \half m^2 \pm\half m\sqrt{m^2+4}\). For \(m\ll1\) we
have \(\mu_-=1-m+O(m^2)\), and thus
\begin{displaymath}
  \sigma^{(-1)}_{m^2} = {\coth{\half mV}\over m\sqrt{m^2+4}}
\end{displaymath}
which is correct to all orders in the large \(V\) asymptotic expansion.

\section{Spectral sum for two dimensional free field theory}
\label{spectral-sum-for-two-dimensional-free-field-theory}

We want to evaluate the spectral sum
\begin{displaymath}
  \sigma \equiv {1\over L_xL_y}
    \sum_{p_x\in\Z_{L_x} \atop p_y\in\Z_{L_y}} \omega_p^{-2}
\end{displaymath}
where
\begin{displaymath}
  \omega_p^2 \equiv m^2 + 4\left(\sin^2{p_x\pi\over L_x}
    + \sin^2{p_y\pi\over L_y}\right).
\end{displaymath}
Using Poisson resummation we find that for \(L_x, L_y\to\infty\)
\begin{displaymath}
  \sigma = {1\over(2\pi)^2} \int_0^{2\pi} dp'_x\,dp'_y\,
    \left[m^2 + 4\left(\sin^2{p'_x\over2}
      + \sin^2{p'_y\over2}\right)\right]^{-1},
\end{displaymath}
so upon substituting \(z\equiv e^{ip'_x}\) we obtain
\begin{eqnarray*}
  \sigma &=& {i\over4\pi^2} \int_0^{2\pi} dp'_y
  \oint_{|z|=1} {dz\over z^2 - \Bigl(m^2 + 4\sin^2(\half p'_y)\Bigr)z + 1} \\
  &=& {i\over4\pi^2} \int_0^{2\pi} dp'_y \oint_{|z|=1}
  {dz\over(z-\mu)(z-1/\mu)}
\end{eqnarray*}
where
\begin{displaymath}
  \mu\equiv {m^2\over2} + 2\sin^2{p'_y\over2} + 1
    - \sqrt{\left({m^2\over2} + 2\sin^2{p'_y\over2} + 1\right)^2 - 1}.
\end{displaymath}
Evaluating the contour integral gives
\begin{displaymath}
  \sigma = -{1\over2\pi} \int_0^{2\pi}
  dp'_y\,\left(\mu-{1\over\mu}\right)^{-1} = {1\over4\pi} \int_0^{2\pi}
  dp'_y\, \left[ \left({m^2\over2} + 2\sin^2{p'_y\over2} + 1\right)^2 - 1
  \right]^{-\half}.
\end{displaymath}
We now make the substitution \(z'\equiv e^{ip'_y}\), which leads to
\begin{eqnarray*}
  \sigma &=& -{i\over2\pi} \oint_{|z'|=1} {dz' \over
    \sqrt{\left[(m^2 + 4)z' - z'^2 - 1\right]^2 - 4z'^2}} \\
  &=& -{i\over2\pi} \oint_{|z'|=1} {dz' \over
    \sqrt{\left[z'^2-(m^2+2)z'+1\right] \left[z'^2-(m^2+6)z'+1\right]}} \\
  &=&
  -{i\over2\pi}\oint_{|z'|=1}{dz'\over\sqrt{(z'-a)(z'-1/a)(z'-b)(z'-1/b)}},
\end{eqnarray*}
where
\begin{eqnarray*}
  a &\equiv& {m^2\over2} + 3 - \sqrt{\left({m^2\over2} + 3\right)^2 - 1}
    = (3-2\sqrt2) - {1\over8}(3\sqrt2-4)m^2 + \cdots \\
  b &\equiv& {m^2\over2} + 1 - \sqrt{\left({m^2\over2} + 1\right)^2 - 1}
    = 1 - m + \half m^2 + \cdots.
\end{eqnarray*}
Since \(1/a>1/b>1>b>a>0\) the contour integral may be shrunk to be the
integral along the branch cut from \(a\) to \(b\),
\begin{displaymath}
  \sigma = {1\over\pi} \int_a^b {dz'\over\sqrt{(z'-a)(b-z')(1/a-z')(1/b-z')}}.
\end{displaymath}
If we change variables to
\begin{displaymath}
  x \equiv {(a+1)(z'-1)\over(a-1)(z'+1)}
\end{displaymath}
we obtain
\begin{displaymath}
  \sigma = - {2\sqrt{ab}\over\pi(a+1)(b-1)}
    \int_1^{1/k'} {dx\over\sqrt{(x^2-1)(1-k'^2x^2)}}
\end{displaymath}
where \(k'\equiv{(a-1)(b+1)\over(a+1)(b-1)}\). With the substitution
\(\xi\equiv{1\over k}\sqrt{1-{1\over x^2}}\) we find that
\begin{displaymath}
  \int_1^{1/k'} {dx\over\sqrt{(x^2-1)(1-k'^2x^2)}}
    = \int_0^1 {d\xi\over\sqrt{(1-\xi^2)(1-k^2\xi^2)}} = K(k)
\end{displaymath}
where \(K\) is the {\em complete elliptic integral} and \(k^2+k'^2=1\); thus
\begin{displaymath}
  \sigma = - {2\sqrt{ab}\over\pi(a+1)(b-1)} K(k).
\end{displaymath}
This may be further simplified using Landen's transformation
\begin{displaymath}
  K\left(2\sqrt\eta\over1+\eta\right) = (1+\eta)K(\eta)
\end{displaymath}
with \(\eta\equiv{a-b\over1-ab}\), which leads to
\begin{displaymath}
  \sigma = {2\sqrt{ab}\over\pi(1-ab)} K\left({b-a\over1-ab}\right),
\end{displaymath}
using the fact that \(K\) is an even function. Expanding in powers of \(m\) we
obtain
\begin{displaymath}
  \sigma = {1\over\pi}\Bigl(1+O(m)\Bigr)K\Bigl(1-\sqrt2m+O(m^2)\Bigr)
    = {1\over4\pi}\ln\left({32\over m^2}\right) + O(m\ln m),
\end{displaymath}
because
\begin{displaymath}
  K(k) = K(\sqrt{1-k'^2}) \asymp \ln{4\over k'} \qquad\mbox{as \(k'\to0\)}.
\end{displaymath}

\bibliographystyle{adk-hunsrt}
\bibliography{lattice-bibliography,adk}

\end{document}